%% file: bluewind.tex
\documentclass[fleqn,usenatbib]{mnras}
\usepackage{subcaption}

\usepackage{newtxtext,newtxmath}
\usepackage[T1]{fontenc}
\usepackage{graphicx}   % Including figure files
\usepackage{amsmath}    % Advanced maths commands
\usepackage{comment}
\usepackage{epsfig}
\usepackage{xcolor}
\usepackage{xargs}
\usepackage{tabularx}
\usepackage{longtable}
\usepackage{threeparttable}
\usepackage{hyperref}
\hypersetup{colorlinks=true, citecolor=blue, linkcolor=blue}
\usepackage{xspace}
\let\oldAA\AA
\renewcommand{\AA}{\text{\oldAA}\xspace}

\newcommand{\redtxt}[1]{\textcolor{red}{#1}}

\newcommand{\hei}{He\,{\sc i}\xspace}
\newcommand{\heii}{He\,{\sc ii}}
\newcommand{\neiii}{[Ne\,{\sc iii}]}
\newcommand{\oiii}{[O\,{\sc iii}]}

\newcommand{\oi}{[O\,{\sc i}]}

\newcommand{\civ}{C\,{\sc iv}}

\newcommand{\feii}{Fe\,{\sc ii}}

\newcommand{\caii}{Ca\,{\sc ii}}

\newcommand{\nii}{[N\,{\sc ii}]}
\newcommand{\sii}{[S\,{\sc ii}]}
\newcommand{\mgii}{Mg\,{\sc ii}\xspace}

\newcommand{\ha}{H\ensuremath{\alpha}\xspace}
\newcommand{\hb}{H$\beta$\xspace}
\newcommand{\hg}{H$\gamma$\xspace}
\newcommand{\hd}{H$\delta$\xspace}
\newcommand{\heps}{H$\epsilon$\xspace}
\newcommand{\jwst}{\textit{JWST}\xspace}
\newcommand{\chandra}{\textit{Chandra}\xspace}
\newcommand{\ergs}{$\rm erg~s^{-1}$}
\newcommand{\ergscm}{$\rm erg~s^{-1}~cm^{-2}$}
\newcommand{\kms}{\ensuremath{\rm km~s^{-1}}\xspace}
\newcommandx{\fluxdcgs}[1][1=-20]{$\times 10^{[#1]}$~erg~s$^{-1}$~cm$^{-2}$~\AA$^{-1}$\xspace}
\newcommandx{\fluxcgs}[2][1=-20,2=\ensuremath{\times}]{${#2}10^{#1}$~erg~s$^{-1}$~cm$^{-2}$\xspace}
\newcommand\sbullet[1][.5]{\mathbin{\vcenter{\hbox{\scalebox{#1}{$\bullet$}}}}}

\newcommand{\mbh}{\ensuremath{M_{\sbullet[0.85]}}\xspace}

\newcommand{\target}{J1157\xspace}

\newcommand{\cloudy}{\textsc{Cloudy}\xspace}

\newcommand{\cigale}{\textsc{cigale}\xspace}

\newcommand{\EP}{\textit{Einstein Probe}}

\title[Varying Balmer-absorbed QSO]{\centering Gone with the wind? Potential gas inflow in the broad-line region of a QSO at $z=0.287$ traced by time-varying Balmer absorption
}

\author[Ji et al.]{Xihan Ji,$^{1,2}$\thanks{E-mail: \href{mailto:xj274@cam.ac.uk}{xj274@cam.ac.uk}}
Dominic J. Walton,$^{3}$
Francesco D'Eugenio,$^{1,2}$
Ignas Juodžbalis,$^{1,2}$
Vasily Belokurov,$^{4}$
\newauthor
Andrew C. Fabian,$^{4}$
Roberto Maiolino,$^{1,2,5}$
Jiamu Huang,$^{6}$
Douglas N.C. Lin$^{7,8,9}$
\\
$^{1}$Kavli Institute for Cosmology, University of Cambridge, Madingley Road, Cambridge, CB3 0HA, UK\\
$^{2}$Cavendish Laboratory, University of Cambridge, 19 JJ Thomson Avenue, Cambridge, CB3 0HE, UK\\
$^{3}$Centre for Astrophysics Research, University of Hertfordshire, College Lane, Hatfield AL10 9AB, UK\\
$^{4}$Institute of Astronomy, University of Cambridge, Madingley Road, Cambridge CB3 0HA, UK\\
$^{5}$Department of Physics and Astronomy, University College London, Gower Street, London WC1E 6BT, UK\\
$^{6}$Department of Physics, University of California, Santa Barbara, CA 93106, USA\\
$^{7}$Department of Astronomy and Astrophysics, University of California, Santa Cruz, CA 95064, USA\\
$^{8}$Department of Astronomy, Westlake University, Hangzhou, Zhejiang, 310030, China\\
$^{9}$Institute for Advanced Studies, Tsinghua University, Beijing, 100086, China
}

\begin{document} 
\label{firstpage}
\pagerange{\pageref{firstpage}--\pageref{lastpage}}
\maketitle
 
\begin{abstract}
We report spectroscopic observations of a remarkable QSO, J115724.80--000455.56 at $z=0.287$ from the Dark Energy Spectroscopic Instrument (DESI) survey. This QSO exhibits variability in its hydrogen Balmer absorption over a time interval of 10 months in the observed frame between 2021 and 2022.
The observed Balmer absorption can be decomposed into two components with distinct kinematics: a non-varying, ``stationary'' component and a varying, ``dynamic'' component.
Both absorbers have gas column densities of $N_{\rm H}\sim 10^{22}~{\rm cm^{-2}}$ and likely have densities of $n_{\rm H}\gtrsim 10^{8}~{\rm cm^{-3}}$.
We inferred small effective physical sizes of $\lesssim 10^{-3}$ pc for the absorbers, which, together with the time variability, suggests that the absorption occurs in the broad-line region (BLR).
The reshifted centroid velocities of both absorbers imply an inflow, where the inferred mass inflow rate is comparable to the single-epoch mass accretion rate of this QSO.
The 2022 epoch of this QSO further reveals a 10\% change in the column density over 2 months, indicating the gas inflow might be continuous.
We also obtained new X-ray observations using the \textit{Einstein Probe}, which reveals an X-ray-to-bolometric luminosity ratio significantly below the general AGN population.
The X-ray weakness is consistent with strong gas obscuration with $N_{\rm H}= 4.0^{+ 1.6}_{- 1.5}\times10^{23}~{\rm cm^{-2}}$.
The substantially higher gas obscuration seen in the X-rays compared to the optical possibly indicates a warm absorber inside the BLR or merger-driven enhancement of the galactic scale obscuration consistent with optical imaging.
This QSO might thus represent an active feeding phase after a recent gas-rich merger.

\end{abstract}

%
%-------------------------------------------------------------------

\section{Introduction}

One of the key observational probes of accreting supermassive black holes (BHs) is broad permitted line emission from spectroscopy, which is generally interpreted to come from fast moving clouds around the BHs known as the broad-line regions (BLRs).
BLRs are believed to exist in all active galactic nuclei (AGN) that reach a certain threshold in accretion rate \citep[with the Eddington ratio $\lambda _{\rm Edd}=L_{\rm bol}/L_{\rm Edd}>10^{-3}$,][and references therein]{ho_2008}, with physical sizes ranging from tens to hundreds of light days \citep[e.g.,][]{Kaspi_2000,Peterson_2004,Bentz_2013}.
To explain observations of broad lines, extreme physical conditions are needed, where the line-emitting clouds are typically characterized with high densities of $n_{\rm H}\gtrsim 10^{8}~{\rm cm^{-3}}$ and undergo Keplerian as well as off-orbital motions with velocities reaching thousands of \kms \citep[e.g.,][and references therein]{Elvis_2000,Gaskell_2009,Netzer_2013}.
BLR clouds are usually assumed to be predominately undergoing virialized motions, which is the basis of single-epoch BH mass measurements with broad emission lines.
However, observations of peculiar broad-line components tracing potential contributions from turbulence, outflows/inflows, or scattering processes in many AGN indicate that BLRs actually have complex, dynamical, and likely stratified structures \citep[e.g.,][]{Ferland_2004,Gaskell_2009,Vietri_2018,gravity+2026}.
In addition, even with successful interferometric measurements of BLR kinematics \citep[e.g.,][]{abuter_2024}, limited by the extremely small scales of the moving gas, it remains unclear whether a BLR is an ensemble of clouds \citep[although each cloud is possibly short lived,][]{Maiolino_2010,Wangjianmin_2017} or continuous winds \citep[e.g.,][]{Murray_1997,Elvis_2000,frado_2021}, which has been a longstanding puzzle given the smoothness of the broad-line profiles in high-resolution spectroscopic observations \citep[e.g.,][]{Arav_1998,Dietrich_1999,laor_blr_2006,abuter_2024}.
The physical picture is further complicated by potential turbulence and thermal electron scattering processes in BLRs \citep[e.g.,][]{Kaneko_1968,bottorff2000,baldwin2004,Laor_2006}.

In contrast to the broad emission lines, which are tracing ensemble motions of clouds or winds in BLRs, absorption lines provide an isolated view of the gas along the line of sight (LOS), which can potentially better resolve the BLR structure if the location of the absorbers can be determined.
A specific class of AGN, known as broad absorption line AGN (BALs), show metal absorption lines with typical widths reaching thousands to tens of thousands of \kms, has been identified and investigated extensively.
BALs roughly make up 10\% of the optically selected Quasi-Stellar Object (QSO) population at $z\lesssim 4$ \citep[][]{Trump_2006,Gibson_2009,Allen_2011}, although their fraction is found to rapidly increase to $\sim 50$\% by $z\sim 6$ \citep{Maiolino2004, Bischetti2023}.
There is an even rarer subclass of BALs that show predominantly low-ionization absorption, especially \feii\ absorption, known as iron low-ionization BALQSOs (FeLoBALs), which make up only 0.3\% of QSOs at $z\lesssim 4$ and might represent AGN obscured by high column densities of gas comparable to clouds/winds in the BLR \citep{Choi_2022,Leighly_2025}.
While BALs provide key evidence for outflows in AGN, there is no direct evidence connecting these outflows to the BLRs.
The outflow radii of (FeLo)BALs are typically estimated to be $\sim1- 10^4$ pc (i.e., larger than typical BLR sizes; \citealp{arav_2018,Choi_2022}), with ``loitering'' FeLoBALs, where low-velocity ($|dv|<2,000$ \kms) outflows dominate, having the smallest outflow radii of $r\lesssim 10$ pc \citep{Choi_baloutflowradii_2022}.
In addition, the density of the outflowing gas is typically lower than the BLR gas densities \citep[e.g.,][]{xu_bal_2019} except for FeLoBALs showing hydrogen Balmer absorption, which implies BLR-like density of $n_{\rm H}\gtrsim 10^{8}~{\rm cm^{-3}}$ \citep{Leighly_2025}.
In addition to BALs, AGN with narrow (width $<2,000$ \kms) Balmer absorption lines are also identified \citep[e.g.,][]{wangxu_2015,burke_abs_2021,Park_desilrd_2026,Shangguan_baq_2026}.
Recently, a class of compact and optically red broad-line sources known as the ``Little Red Dots'' \citep[LRDs,][]{labbe_2023,matthee2024} were identified at both high and low redshift \citep{Kokorev_lrd_2024,greene2024,Kocevski_lrd_2024,ji_lord_2025,Lin_lrdanalog_2025,linxiaojing_locallrd_2025,line_lrd2_2026,Park_desilrd_2026,Chen_2026}, and these sources uniquely show a high incidence of Balmer absorption \citep[$>20$-50\%;][]{juodzbalis_rosetta_2024,Juodzbalis2026absorption,ji_lrdbreak_2025,deugenio_qso1_2025,deugenio_lrdoutflow_2025,deugenio_irony,Matthee_2026,yanagisawa_lrdabs_2026}.
While their nature remains hotly debated, the presence of broad lines and of the same absorption signatures led to speculating about a connection between LRDs and absorption-line QSOs \citep{Leighly_2025,juodzbalis_2026}.

The Balmer-absorbed QSOs are among the best candidates to study any connection between outflows/inflows and the BLR due to the similarly required physical conditions on gas densities, gas column densities, and gas excitation \citep{juodzbalis_rosetta_2024,juodzbalis_2026}.
For example, \citet{Zhouhongyan_naturebalqso_2019} studied a BAL QSO J1035+1422 at $z=1.25$ with broad Balmer absorption lines redshifted by several thousand \kms and interpreted the absorption to trace a parsec-scale inflow within the BLR.
By studying another Balmer-absorbed QSO, \citet{Shixiheng_balqso_2016} suggested that the absorption comes from an inflow at $\sim 4$ pc from the accretion disc within the dusty torus.
\citet{Zhang_ebrbal_2015} discovered an extreme Balmer-absorbed QSO with the Balmer absorption blueshifted by $\sim 10^4$ \kms, and they inferred a distance of only $0.2$ pc from the accretion disc.
For these single-epoch studies, estimations of the physical scales of the absorbers generally rely on photoionization modeling.

Among the observations of absorption-line systems, an important link between the observed absorbers in these systems and their BLRs is time variability.
Typical dynamical timescales and crossing timescales of BLR clouds or winds can range from days to years, meaning that the absorption features can change over similar timescales if they come from passing clouds in the BLR.
Meanwhile, the ionizing photon output of the accretion disc can vary over similar or shorter timescales, which can also produce observable change in the absorption by changing the ionization structure of the absorber if it lies close to the accretion disc.

Until now, a number of works have focused on the time-domain behavior of ultraviolet (UV) absorption lines of BALs and found time variability indicative of small-scale ($<1$ pc) outflows and/or changes in the ionization structures of the outflows \citep[e.g.,][]{Barlow_1993,Lundgren_2007,Gibson_2010,Capellupo_2011,Capellupo_2012,Capellupo_2013,Grier_2015,Wheatley_bal_civ_variability}.
The time varibility of Balmer-absorbed QSOs has also been noted by several studies.
%few observations have focused on the time-domain behavior of Balmer absorption lines, in part due to their rarity \citep[although see e.g.,][for time-domain ultraviolet line studies for BALs]{Barlow_1993,Lundgren_2007,Gibson_2010,Capellupo_2011,Capellupo_2012,Capellupo_2013,Grier_2015,Wheatley_bal_civ_variability}.
%The most comprehensive study was conducted for t
The nearby Seyfert 1 NGC 4151, for example, has been known to exhibit optical absorption lines \citep{Mayall_1934,Anderson_1969}.
During 1997\,-\,2000, NGC 4151 showed strong absorption in hydrogen Balmer lines as well as \hei$\lambda 3889$ that varied with time, providing strong evidence for a BLR origin \citep{hutchings2002}.
\citet{Shixiheng_balmerqsovar_2017} found time variability in the bluesfhited Balmer absorption from \ha to \heps in the BAL QSO, PG 1411+442, between 2014 and 2017, which the authors interpreted as due to a change in the photoionization state of the outflow.
A similar conclusion was reached by \citet{Sunluming_balmerqsovar_2017}, who studied another BAL QSO with blueshifted Balmer absorption.
Recently, \citet{Shangguan_baq_2026} reported modest absorption variability in a Balmer absorption-line QSO at $z=0.2$ over 16 yr, but no variability in other Balmer absorption-line QSOs in their sample. The authors interpret the lack of variability as a result of the absorption coming from long-lasting ``gas envelopes'' surrounding the BLRs, similar to what have been proposed for LRDs in some recent works \citep[e.g.,][]{Inayoshi_gasremoval_2025}.
Within the LRD sample, no absorption variability has been reported thus far except for a marginal evidence at $z=5.1$ \citep{deugenio_lrdoutflow_2025}.
Still, most of the current studies are limited by the lack of dedicated spectroscopic monitoring of absorption-line AGN.

The recent data release from the Dark Energy Spectroscopic Instrument (DESI) survey \citep{desi_overview_2022,desidr1} provides a unique opportunity to investigate this problem, where millions of ultraviolet (UV)-to-optical spectra of galaxies and QSOs are available to the public.
Among the large spectroscopic sample, thousands of targets have observations taken at more than two epochs separated over several days to several months, naturally offering a probe of time variability.
In this work, we report multi-epoch DESI observations of a QSO, J115724.80--000455.56, at $z=0.287$, which exhibit clear absorption variability in the observed hydrogen Balmer lines.
To our best knowledge, J115724.80--000455.56 is the only source with sufficiently high signal-to-noise (S/N) spectroscopy to allow a comprehensive absorption variability measurement over several Balmer lines in the DESI sample.
%since the previous studies of NGC 4151 \citep{hutchings2002}.
The time variability of the absorption in J115724.80--000455.56 sets a clear connection between the LOS obscuration in the system and the passing clouds or winds in its BLR, and could provide insights for future time-domain investigations on absorption-line AGN.

The structure of this manuscript is as follows.
In Section~\ref{sec:data}, we describe our target selection and multi-epoch observations.
In Sections~\ref{sec:vary_abs} and \ref{stationary_abs}, we describe our analysis of the optical absorption spectrum.
In Section~\ref{xray}, we describe X-ray observations of our target.
We present imaging decomposition of the host galaxy of \target in Section~\ref{sec:host}.
We discuss the physical origin of the absorption in Section~\ref{sec:discuss} and draw our conclusions in Section~\ref{sec:conclude}.
Throughout this work, all observations are corrected for Galactic extinction using the \citet{Schlafly_2011} Milky Way dust map and the \citet{Fitzpatrick_1999} extinction curve with $R_{\rm V}=3.1$.
We assume a flat $\rm \Lambda CDM$ cosmology with $h=0.674$ and $\Omega _{\rm m} = 0.315$ \citep{planck2020}.
All magnitudes are given in the AB system.

\section{Data}
\label{sec:data}

\subsection{Target selection}

We identified the target source, J115724.80--000455.56 (here after J1157), from the DESI data release 1 (DR1; \citealp{desidr1}).
This source was included as one of the Balmer absorption sources we selected from DESI DR1.
Analyses of the full sample of these sources will be presented in future work.
We briefly summarize our sample selection in the following.

We performed a pre-selection for QSOs based on the DESI DR1 AGN/QSO value added catalog \citep[VAC,][]{Juneau_2025}, which includes simple line measurements and source identifications.
We selected sources identified as \texttt{QSO} and required detections of the broad \ha line at $>3\sigma$, resulting in 54,556 spectra.
This selection misses Balmer-absorption QSOs at $z>0.5$, where \ha is no longer covered by DESI spectroscopy.
Regardless, \ha absorption is typically the strongest absorption feature compared to other Balmer absorption due to its large oscillator strength, and we will present the complete selection including QSOs at $z>0.5$ in future work.

We refit all selected spectra using a customized module with the Penalized Pixel-Fitting (\textsc{pPXF}) code \citep{cappellari2004,cappellari2017}, which adopts an efficient maximum penalized likelihood approach.
We focused on a spectral window of 6280-6800 \AA in the rest frame and
fit \ha, \nii$\lambda \lambda 6548,6583$, \oi$\lambda \lambda 6300,6363$, \hei$\lambda 6678$, and \sii$\lambda \lambda 6716,6731$ as well as the continuum, which is assumed to be a simple power law.
Each emission line is modeled as a single Gaussian except for \ha, which is modeled as a single narrow Gaussian + two broad Gaussians, where the two broad Gaussians are forced to have the same centroids but with free widths.
The above choice is only to approximate the potentially complex shapes of BLR profiles and select sources efficiently.
All Gaussians are convolved with the instrumental line spread function (LSF) of DESI \citep{Guy_2023}, and all narrow Gaussians are forced to have the same kinematics.
We approximated the absorption as a negative Gaussian, which is not physically motivated but is robust and efficient for identification purposes.
%We present physical absorber model later in this manuscript.

We tested whether there is any Balmer absorption by evaluating the Bayesian Information Criterion \citep[BIC,][]{Schwarz_1978,liddle_2007} defined as
\begin{equation}
    BIC=\chi^2+k\,\ln n,
\end{equation}
where $k$ is the number of parameters and $n$ is the number of spectral pixels.
We calculated $\Delta BIC$ between the fits with and without Balmer absorption and included sources with $\Delta BIC=BIC_{\rm no~abs}-BIC_{\rm abs}>0$.
Usually, significant statistical improvement is defined as $\Delta BIC>10$, but given the potentially complex BLR profiles, we made a generous cut so as not to miss Balmer absorbers with highly non-Gaussian broad-line profiles.
We obtained a total of 4601 sources that passed this selection step.

Finally, we visually inspected all remaining sources to remove sources with potential confusion, including those with clear double-peak or boxy broad-line profiles indicative of disc components, centrally peaked, power law-like or exponential-like profiles that were not properly fit by the code (as an extremely strong and narrow negative Gaussian can also serve to reproduce the cusp), and potential confusion due to strong \nii$\lambda 6583$ emission.
We considered a source as a convincing candidate if it does not have evidence for any of the above issues and has visually identifiable absorption lines (usually \hb and/or \hei$\lambda 3889$) in addition to \ha.
We obtained 15 Balmer-absorption QSOs, two of which have been reported by \citet{Park_desilrd_2026} and \citet{Shangguan_baq_2026}.
The partial overlap of our sample with the DESI sample of \citet{Park_desilrd_2026} is due to the different selection criteria; \citet{Park_desilrd_2026} include 5 sources which we classified as potentially having disc-like components in \ha.
%; \citet{Park_desilrd_2026} are not included is mainly due to potential confusion with disc-like components.
We will present the full sample including also higher redshift sources and sources not classified as \texttt{QSO} in DESI in future work.

\subsection{Multi-epoch observations}

\begin{figure*}
    \centering
    \includegraphics[width=0.95\linewidth]{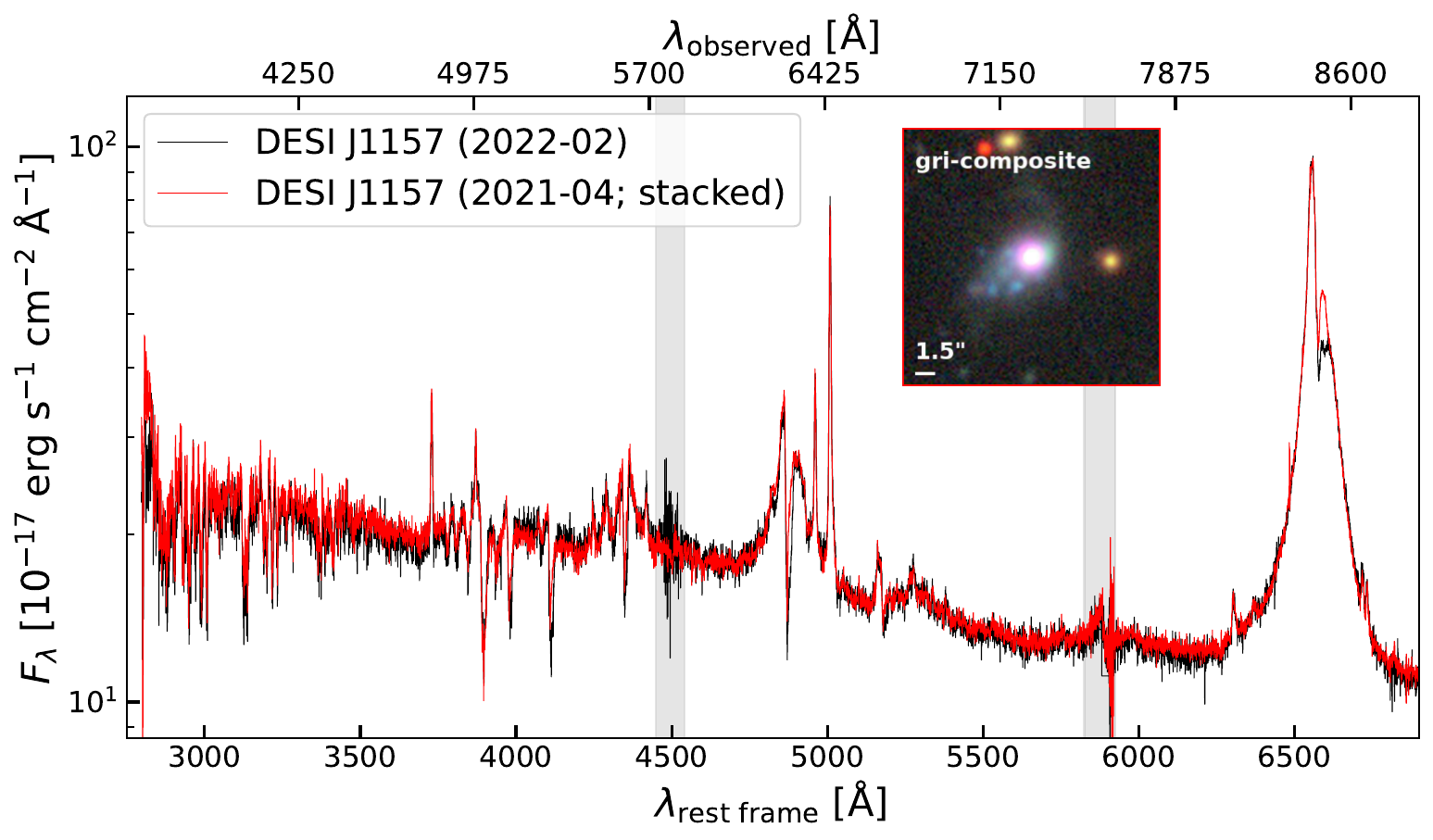}
    \includegraphics[width=0.95\linewidth]{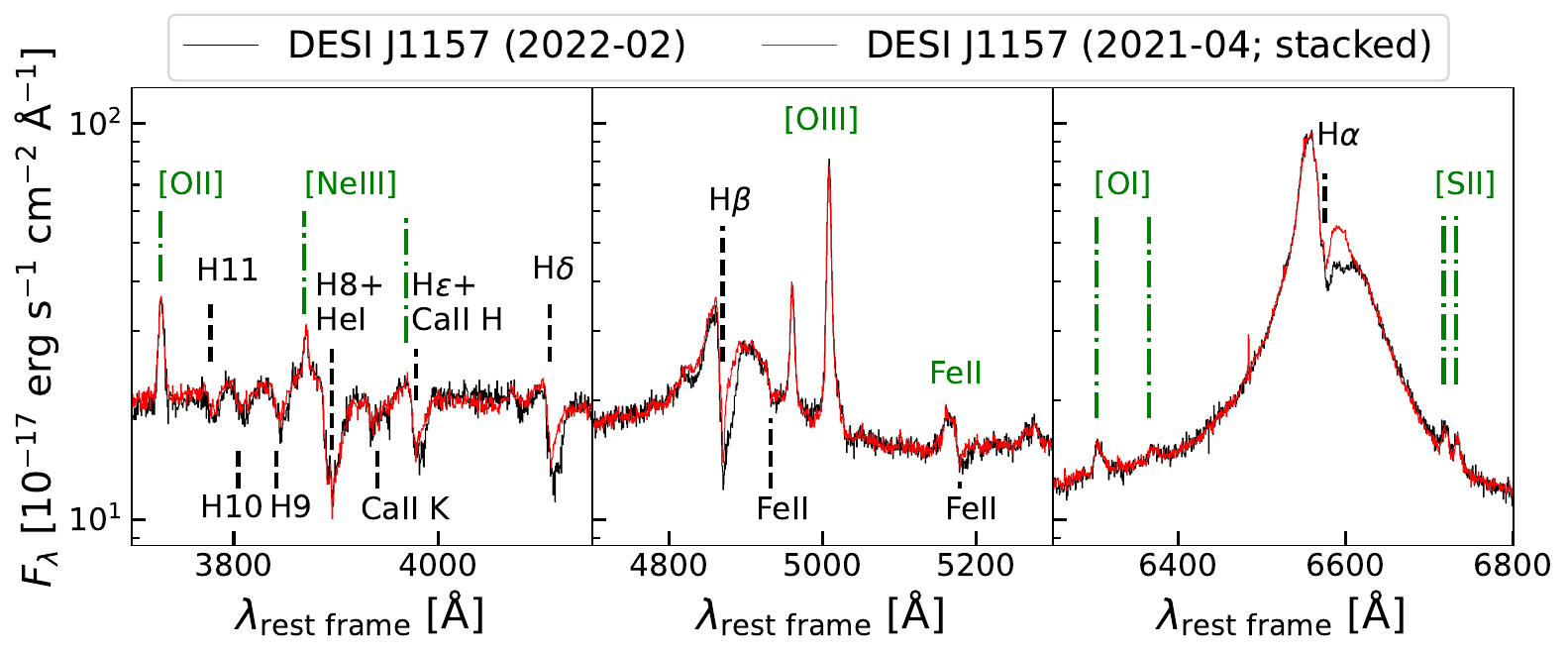}
    \caption{Comparison between the stacked spectrum of J1157 from April 2021 and the coadded spectrum from February 2022.
    In the top panel, we marked spectral regions connecting different DESI spectra (i.e., $b$ and $r$, and $r$ and $z$ arms) with gray bands, which are masked during our analyses.
    The bottom panel shows zoom-in views of part of the optical spectrum where clear absorption variation can be seen.
    The strongest variability is visible in the Balmer series, from the Balmer limit to the \ha line.
    %The erosion of the red peak of \ha mimics variation in a disc component.
    Based on the systematic change in the Balmer series, we conclude the change is likely due to a LOS obscuration by passing gas in the BLR.
    We also show the $gri$ color composite image from HSC-SSP DR3, which reveals the host and structures possibly associated with tidal features indicative of mergers.
    The bar on the image represents the on-sky fiber size of DESI.
    }
    \label{fig:full_spec}
\end{figure*}

\begin{figure}
    \centering
    \includegraphics[width=\columnwidth]{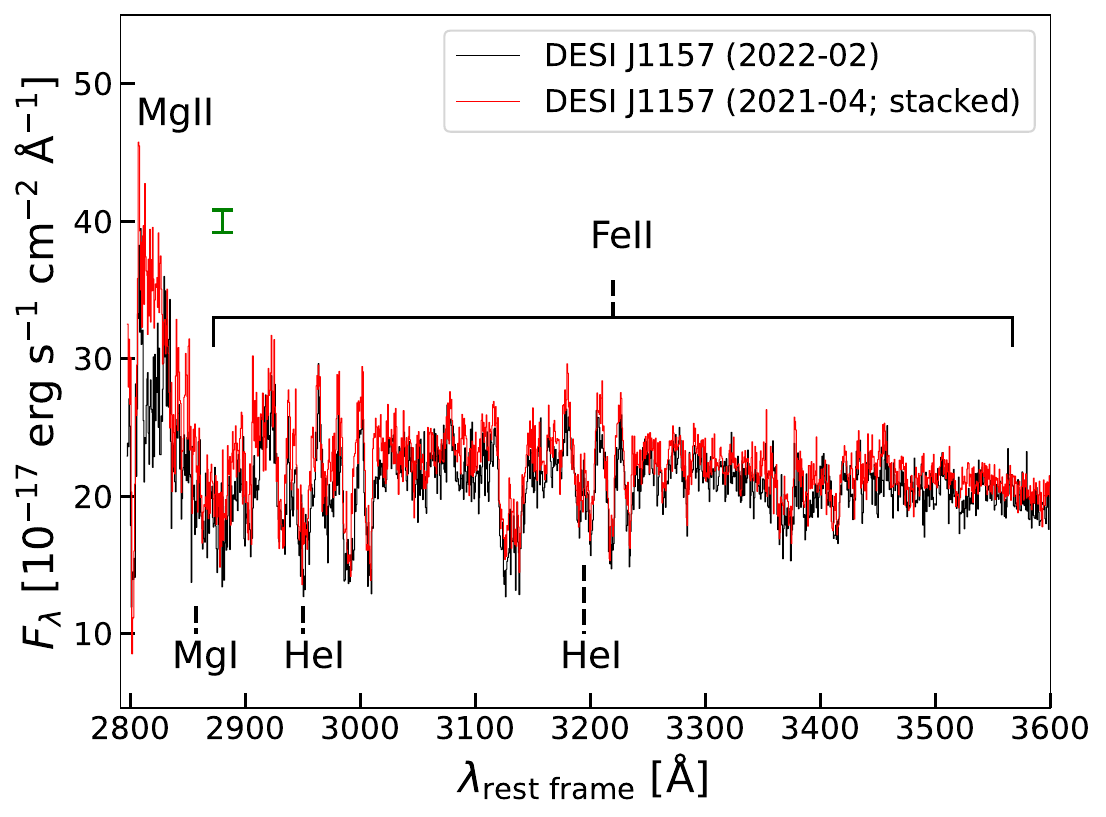}
    \caption{Continuation of Figure~\ref{fig:full_spec}, where the UV spectra are shown. The majority of the absorption is likely dominated by \feii.
    The total length of the green bar corresponds to the median $1\sigma$ uncertainty of the flux density difference between the two spectra in the plotted wavelength range.
    Comparing the 2021 and 2022 epochs, the most apparent variation is seen in the \mgii absorption. The 2022 epoch shows an additional \mgii absorption redshifted by $dv\approx 1,000$ \kms consistent with the varying Balmer absorption.
    }
    \label{fig:uv_spec}
\end{figure}

In this manuscript we focus on J1157 at $z=0.287$, the only source in our sample that has 5 coadded observations in DESI DR1, where 4 coadded observations correspond to April 2021, at 3 April, 8 April, 12 April, and 23 April, respectively, and 1 coadded observation correspond to 2 February, 2022.
{
Each of the coadded observations can be further split into individual exposures, and all exposures are distributed over 9 different nights.
}
We summarize the observations in Table~\ref{tab:target_obs}.
{
For our main analysis, we focus on coadded observations. No significant variability was found among exposures for each coadded observation, except for the February 2022 epoch, which we discuss in Section~\ref{sec:vary_abs}.
}
All observations have \texttt{coadd\_fiberstatus=0}, meaning no known fiber or observational issue.
The on-sky fiber positions of all observations are consistent within a separation of $\lesssim 0.\!\!''1$.

All epochs from 2021 show no obvious relative change in spectral shape, although the fluxes of two epochs (03 04 2021 and 08 04 2021) are systematically higher than the other epochs by roughly 10\%.
This difference could originate either from a flux calibration issue or a seeing effect, since the fluxes of narrow emission lines should not be varying over such short timescales.
To correct for this systematic difference, we renormalized the spectra from the two 2021 epochs to match the average flux of narrow emission line \oiii$\lambda 5007$ from the other epochs.
The renormalization factor we obtained is 0.88 (listed in Table~\ref{tab:target_obs}). 
We emphasize that the renormalization step is physically motivated -- without this step, directly taking the flux difference among the 2021 spectra, narrow \oiii\ emission would imply significant ``variability'' within days, which is not physical given the typical narrow-line region (NLR) sizes of hundreds to thousands of parsecs\footnote{According to \citet{Bennert_nlrsize_2006} and \citet{chenjianhang_nlrsize_2019}, the typical NLR size given $L({\rm [OIII]5007})\approx 10^{42}$ \ergs of \target would be $\gtrsim 3$ kpc in local AGN.
} and a long recombination timescale.
%(e.g., \citealp{Bennert_nlrsize_2006,chenjianhang_nlrsize_2019}; the typical NLR size given $L({\rm [OIII]5007})\approx 10^{42}$ \ergs of \target would be $\gtrsim 3$ kpc in local AGN).
Regardless, removing the two epochs of spectra to which we applied the above renormalization would not change our conclusions.
After renormalization, we took the inverse variance weighted mean of the spectra from 2021, and compared it with the spectrum from 2022.
We found that the fluxes of narrow \oiii\ are fully consistent between the stacked spectrum from 2021 and the spectrum from 2022, again suggesting a potential flux calibration difference among the 2021 spectra.

Table~\ref{tab:target_properties} summarizes some basic properties of \target, where the listed luminosities are measured from the 2021 epoch.
%We show both \ha luminosities corrected and uncorrected for the absorption.
We describe the spectral fitting in Sections~\ref{sec:vary_abs} and \ref{stationary_abs}.
%Overall, the correction does not change the derived BH properties and the uncertainties are dominated by the systematics of the BH mass calibration.
To obtain the BH mass, we used the bolometric conversion factor from \citet{sternlaor_2012} and the single-epoch virial relation from \citet{rv_bhmass_2015}.
The bolometric luminosity is corrected for dust attenuation of $A_{\rm V}=1.17$ mag based on the broad-line Balmer decrement assuming a Small Magellanic Cloud (SMC) extinction curve \citep{gordon2003}.
We also performed SED fitting of the archival photometric data of \target using the code \cigale \citep{cigale}, which gives a similar dust attenuation of $A_{\rm V} \approx 1$ mag for the AGN continuum.
We describe the SED fitting in Appendix~\ref{appendix:sed}.
For the BH mass, the propagated measurement uncertainty is much lower than 1) the systematic uncertainty of 0.3 dex associated with the virial relation; 2) the systematic uncertainty of 0.5 dex depending on whether the core of the \ha line is associated with the BLR or not. 
The Eddington ratio is obtained by taking $L_{\rm bol}/L_{\rm Edd}$ and is again dominated by the systematic uncertainty.
Based on these measurements, \target is a moderately accreting system at $\lambda _{\rm Edd}\sim 0.3$ with $M_{\rm BH} \sim 10^8 ~M_\odot$, not atypical for AGN with similar BH masses observed at $z<0.3$ \citep{schulze_2010}.
What sets \target apart from normal AGN is its absorption and associated variability, which we discuss next.

\begin{table*}
    \centering
    \caption{Summary of DESI spectroscopic observations for \target. 
    We list information for both coadded observation and individual exposures.
    The renormalization factor is applied to produce consistent narrow \oiii$\lambda 5007$ fluxes (see text).
    Removing the two spectra that require renormalization would not change our overall results and conclusions.
    The main variability is found between coadded spectra, but subtle variability is also found between certain exposures.
    }
    \begin{tabular}{l c c c c c}
         \hline
         DESI ID & Type & MJD & Date [DD MM YYYY] & Exposure time [s] & Renormalization factor  \\
         \hline
         39627787736188739 & Coadded & 59612 (mean) & 02 02 2022 (mean) &  2127 (total) & N/A \\
         & Single exposure & 59639 & 28 02 2022 & 998 & N/A \\
         & Single exposure & 59586 & 05 01 2022 & 1130 & N/A \\
         \hline
         39627787736188739 & Coadded (single night) & 59307 & 02 04 2021 &  1267 & 0.88 \\
         \hline
         39627787736188739 & Coadded (single night) & 59312 & 07 04 2021 &  1814 & 0.88 \\
         \hline
         2305843037487506305 & Coadded (single night) & 59316 & 11 04 2021 &  952 & N/A \\
         \hline
         39627787736188739 & Coadded & 59327 (mean) & 23 04 2021 (mean) &  2440 (total) & N/A \\
         & Single exposure & 59337 & 02 05 2021 & 454 & N/A \\
         & Single exposure & 59336 & 01 05 2021 & 1073 & N/A \\
         & Single exposure & 59320 & 15 04 2021 & 221 & N/A \\
         & Single exposure & 59318 & 13 04 2021 & 692 & N/A \\
         \hline
    \end{tabular}
    \label{tab:target_obs}
\end{table*}

\begin{comment}

\begin{table*}
    \centering
    \caption{Basic properties of \target.
    The BH mass and the Eddington ratios are derived based on the single-epoch virial relation (see text).
    The listed luminosities are measured values.
    The BH mass is corrected for $A_{\rm V}=0.8$ mag from the broad-line Balmer decrement assuming an intrinsic ratio of $\rm H\alpha/H\beta =2.86$ and a \citet{Fitzpatrick_1999} extinction curve with $R_{\rm V}=3.1$.
    The uncorrected BH mass is $M_{\rm BH}=10^{8}~M_\odot$.
    \redtxt{[tba: other lines and cigale]}
    }
    \begin{tabular}{l c c c c c}
    \hline
        Redshift & $\log L_{5100}$ [\ergs] & $\log L_{\rm H\alpha}$ [\ergs] & FWHM(\ha) [\kms] & $\log M_{\rm BH}$ [$M_\odot$] & $\lambda _{\rm Edd}$ \\
        %& [\ergs] & [\ergs] & [\kms] & [$M_\odot$] & \\
        \hline
        0.2872 & $44.37\pm 0.02$ & $43.37\pm 0.01$ [corrected for absorption] & $2640\pm 110$ [corrected for absorption] & $8.2\pm 0.3$ & $0.16^{\times 2}_{\div 2}$ \\
        & & $43.30\pm 0.01$ [uncorrected] & $2684\pm 110$ [uncorrected] & & \\
    \hline
    \hline
     $A_{\rm V}$ [mag] & & & FWHM(\oiii) [\kms] & & \\
     0.8 [BL] & & & & & \\
     $0.93\pm 0.02$ [\cigale] & & & & & \\
    \hline
    \end{tabular}
    \label{tab:target_properties}
\end{table*}
    
\end{comment}

\begin{table}
    \centering
    \caption{Derived properties for the Balmer-absorbed QSO \target.}
    \begin{tabular}{l c}
    \hline
    Parameter & Value \\
    \hline
    \multicolumn{2}{|c|}{Black hole/BLR}\\
    \hline
    $\log L_{5100}$ [\ergs] & $44.37\pm 0.02$ \\
    $\log L_{\rm H\alpha}$ [\ergs] & $43.298\pm 0.004$ \\
    $\log L_{\rm H\beta}$ [\ergs] & $42.630\pm 0.004$ \\
    Broad-line $A_{\rm V}$ [mag] & {$1.17\pm 0.03$} \\
    $\log L_{\rm bol}$ [\ergs] & $45.7\pm 0.4$ \citep{sternlaor_2012} \\
    FWHM(\ha) [\kms] & (2.7 [obs]\,-\,4.2 [best-fit])$\times 10^3$ \\
    $\log (\mbh/M_\odot)$ & {(8.2 [obs]\,-\,8.7 [best-fit])$\pm 0.3$} \\
    & \citep{rv_bhmass_2015} \\
    $\lambda _{\rm Edd}$ & (0.14 [best-fit]\,-\,0.38 [obs])$^{\times 2}_{\div 2}$ \\
    \hline
    \multicolumn{2}{|c|}{ISM/outflows}\\
    \hline
    $\log L({\rm [OIII]\lambda 5007})_{\rm n}$ [\ergs] & $42.047\pm 0.002$ \\
    $\log L({\rm [OIII]\lambda 5007})_{\rm outflow}$ [\ergs] & $41.64\pm 0.02$ \\
    $\rm FWHM_n$ [\kms] & $393\pm 2$ \\
    $\rm FWHM_{outflow}$ [\kms] & $865\pm 15$ \\
    \hline
    \multicolumn{2}{|c|}{SED fitting results (\cigale; Appendix~\ref{appendix:sed})}\\
    \hline
    $A_{\rm V}$ [mag] & $0.93\pm 0.02$ \\
    %$\log L_{\rm bol}$ [\ergs] & $45.06\pm 0.02$ \\
    $\log (M_{\rm \star}/M_\odot)$ & {$10.1$} \\
    \hline
    \end{tabular}
    \begin{tablenotes}
        \small
        \item $\bf Notes.$
        \item 
        The best-fit spectral model is shown in the top panel of Figure~\ref{fig:stationary_fit}.
        For the FWHM of \ha and related parameters, we list both the directly observed values without correcting for the absorption [obs], and the best-fit values [best-fit] to account for systematic uncertainties in the line decomposition.
        We used the bolometric conversion from \citet{sternlaor_2012} and the single-epoch virial black hole mass relation from \citet{rv_bhmass_2015}.
        The dust attenuation curve is from \citet{gordon2003} with $R_{\rm V}=3.1$, and we assume $\rm (H\alpha/H\beta)_{intrinsic}=3.06$ \citep{Dong_2008}.
        %The atomic data set for relevant calculations are from CHIANTI \citep[v10,][]{chianti0,chianti1}.
    \end{tablenotes}
    \label{tab:target_properties}
\end{table}

\subsection{Spectral variability}

To make a proper pixel-by-pixel comparison, we further resampled the 2021 spectrum to the wavelength grid of the 2022 spectrum using the code \textsc{spectres} \citep{spectres}.
Figures~\ref{fig:full_spec} and \ref{fig:uv_spec} show a comparison between the stacked 2021 spectrum and the 2022 spectrum for J1157.
The signal-to-noise (S/N) of each spectrum is very high, reaching a median value of 27 per pixel across the whole wavelength range in 2022 and higher in the 2021 stack.
The overall continuum is characterized roughly by a power law ($F_\lambda \propto \lambda^{\beta}$) with an index of $\beta \approx -1$ in the optical and flattens to $\beta \approx -0.2$ in the UV.
We mark major absorption lines visible in the optical spectrum in Figure~\ref{fig:full_spec}, including the Balmer series from \ha to H11, \hei$\lambda 3889$, \caii\,H and K, \feii$\lambda 4924$, and \feii$\lambda 5169$. 
The bottom panel shows three zoomed in views of the spectra around the higher-order Balmer lines until H11, the \hb+\oiii\ region, and the \ha region.
In the UV, there are also a number of absorption features likely associated with \hei, Mg\,{\sc i}, \feii, and Mg\,{\sc ii} as shown in Figure~\ref{fig:uv_spec}.
We summarize some features from the simple comparison in the following.
\begin{itemize}
    \item The absorption lines are deeper than the emission lines for lines in bluer wavelengths starting from \feii\ and \hb.
    This implies that the absorber must be absorbing the continuum.
    \item Judging from \hb, the depth of the absorption at the core is comparable to the flux density of the underlying continuum (16-$17\times 10^{-17}$ \ergscm$\,\AA^{-1}$), but the absorption core appears narrow and not obviously saturated.
    This implies that the absorber might be absorbing both the continuum and the broad lines.
    {This is further examined in Section~\ref{sec:vary_abs}.}
    This also rules out a stellar origin for the absorption, since the absorption would be strongly saturated if the broad lines are not absorbed.
    Later in Section~\ref{sec:host}, we show with imaging analyses that the host galaxy cannot contribute to more than 10\% of the optical continuum within the DESI fiber, and thus the DESI spectra are  QSO-dominated.
    \item The enhancement of the absorption in the 2022 epoch consistently happens on the red side of the original absorption, widening the overall absorption feature.
    This suggests that the variation was caused by an absorber with kinematics different from the original absorber seen during 2021.
    \item The broad-line profile, as seen in \ha and \hb (since higher-order Balmer lines are strongly absorbed) is asymmetric with a more extended blue wing.
    While from the 2022 epoch alone, the \ha profile appears to be made with a boxy disc component + a sharper core, the 2021 epoch suggests that the original profile is more similar to a BLR profile with an absorption close to the line core.
    The varying absorption erodes the red peak of \ha due to its higher LOS velocity.
    \item In addition to the absorption dominated by low-ionization atoms and ions including H\,{\sc i}, \hei, Fe\,{\sc ii}, and Ca\,{\sc ii}, there is a lack of strong high-ionization lines including \heii, [Ne\,{\sc v}], and [Fe\,{\sc vii}] in the spectrum.
    This is similar to what is observed in FeLoBALs, although the cores of the absorption (defined as the lowest flux density) seen in \target are close to the centroid of narrow lines and therefore it is more similar to the subclass of loitering FeLoBALs \citep{Choi_baloutflowradii_2022}.
    \item Unlike the Balmer absorption, the strong \hei$\lambda 3889$ absorption does not appear to vary between the 2021 and 2022 epochs. This could be an indication of the location of the absorber, which is further discussed in Section~\ref{stationary_abs}.
    \item The density of the absorber is likely high, as Balmer absorption is efficiently produced at $n_{\rm H}\gtrsim 10^8~{\rm cm^{-3}}$ \citep{juodzbalis_rosetta_2024}.
    This is also consistent with the time variability constraint.
    To have an absorption variation in $<8$ month rest-frame time, the hydrogen recombination timescale [$\tau _{\rm rec}\approx \frac{1.22\times 10^3~{\rm yr}}{n_{\rm e}/(100~{\rm cm^{-3}})}$] needs to be shorter \citep{draine2011}, meaning that $n_{\rm H}> 10^5~{\rm cm^{-3}}$.
\end{itemize}

In addition to the DESI spectra, we include a color-composite image from the third data release of the Hyper Suprime-Cam Subaru Strategic Program (HSC-SSP DR3; \citealp{Aihara_hscssp}) in Figure~\ref{fig:full_spec}. With a typical full width at half maximum (FWHM) of $\sim 0.\!\!''7$ for the point-spread function (PSF), the HSC image reveals several blue clumps in the southeast as well as diffuse shells or tails in the southeast-northwest direction.
These features possibly indicate (post) galaxy mergers, which might further be related to the inflows traced by redshifted absorption we see in the QSO spectrum.
The HSC data are described in more detail in Section~\ref{sec:host}.

Next, we describe our analysis of the absorption features.
We adopted both a parametric fitting approach and a photoionization modeling approach as detailed below.

\section{Varying absorption}
\label{sec:vary_abs}

As described in the previous section, the absorption seen in \target can be broadly decomposed into a ``dynamic'' absorber that was not present in 2021 but appeared in 2022, and a ``stationary'' absorber that was originally present in 2021.
We note that the stationary absorber is not necessarily truly stationary, but we do not have other epochs available to verify its variability or the lack thereof.

\begin{figure*}
    \centering
    \includegraphics[width=0.95\linewidth]{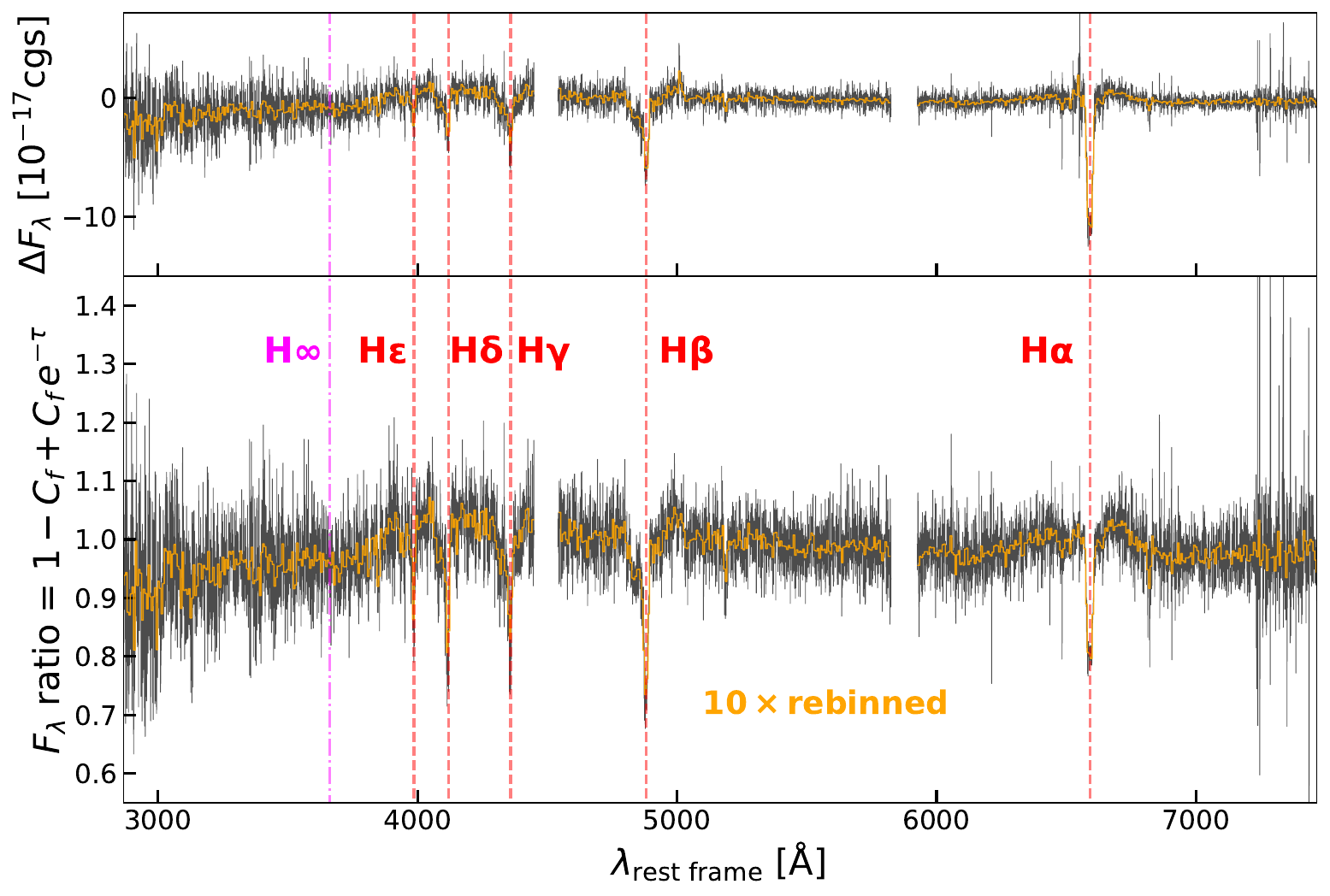}
    \caption{
    Spectral variation for J1157 revealed by DESI spectra at 2021 and 2022.
    In the top panel, we show the flux density difference (defined as $2022-2021$) in unit of $10^{-17}~{\rm erg~s^{-1}~cm^{-2}~\AA ^{-1}}$.
    In the bottom panel, we show the flux density ratio (defined as 2022/2021).
    For illustration purposes, we also plot varying spectra rebinned over every 10 wavelength pixels, which have higher S/N.
    The varying part of the spectrum shows clear Balmer absorptions from \ha towards the Balmer limit.
    We interpret the variation as a change in the LOS obscuration, where the 2022 epoch sees additional gas opacity.
    In the case of a single absorber, the 2022/2021 flux ratio reflects the absorption optical depth ($y$ axis of the bottom panel).
    }
    \label{fig:abs_var}
\end{figure*}

We start by analyzing the varying part of the absorption.
Thanks to the multi-epoch DESI observations and the high S/N of the spectra, the varying part of the absorption presents a cleaner and more easily interpretable result compared to the stationary absorption.
Figure~\ref{fig:abs_var} shows the difference spectrum between the two epochs.
The top panel shows the residual spectrum after subtracting the 2021 spectrum from the 2022 spectrum.
The major troughs visible in the difference spectrum correspond to Balmer lines from \ha to H$\epsilon$ (the higher-order Balmer absorption is generally too weak to see).
There is also an indication of a small variation in the \feii$\lambda 5169$ absorption.
The \caii\ K absorption is noisy and the \caii\ H absorption is blended with \heps.
In the UV ($\lambda _{\rm rest~frame}\lesssim 3600$ \AA), the difference spectrum is systematically lower than 0.
This might imply that there is also an intrinsic variation in the continuum between 2021 and 2022, which remains visible after our renormalization due to the fact that continuum variation of QSOs is usually wavelength dependent and stronger at shorter wavelengths \citep[e.g.,][]{burke_agnvar_2021}.
Alternatively, bound-free continuous absorption by hydrogen at $n=2$ can contribute to the UV flux variation, especially given the presence of Balmer line absorption in the optical.
Indeed, the overall varying spectrum shows a small and smooth Balmer break towards the Balmer limit, which might be associated with the absorber.
Notably, a Balmer break and line absorption with non-stellar origins are proposed to explain the spectrum of LRDs \citep{Inayoshi_maiolino_2025,ji_lrdbreak_2025,degraaff_lrd_2025,naidu_lrd_2025}, and \target potentially provides the first case showing a dynamical absorber in a QSO can produce them, which cannot come from stellar populations.

The bottom panel of Figure~\ref{fig:abs_var} shows the relative difference between the 2021 and 2022 spectra by taking the ratio of their flux densities.
Such a comparison highlights the change in the LOS obscuration. This is because if the variation results from a single absorber, then the flux density ratio should be roughly proportional to $1-C_f+C_f\exp{(-\tau )}$, where $\tau$ is the optical depth and $C_f$ is the covering factor.
Similar features including the Balmer line absorption and Balmer break can be seen in this panel.

Given the two panels, a natural question is whether the variation we see between 2021 and 2022 is due to the loss of some broad emission (top), or due to the enhancement of the absorption (bottom).
We think the latter explanation, that is, the spectral variation of J1157 is due to a change in the LOS absorption, is more plausible.
The reason is twofold.
First, should the variation be due to the partial loss of the broad-line emitting clouds undergoing orbital motions, it is difficult to explain why only clouds moving away from the LOS disappear.
Also, the flux ratio of the lost Balmer emission is highly atypical -- depending on the velocity range to integrate broad-line fluxes, the flux ratio of \ha/\hb is 1.2-1.7\footnote{Contrary to this lower-than-Case-B value, BLRs usually show Balmer decrement higher than the Case B value \citep{ilic_2012}. Extremely diffuse, density-bounded clouds can produce low Balmer decrements in SF galaxies, but the required physical conditions are quite different from those of BLRs \citep{mcclymont_db_2025}.
}.
Second, as shown in Figure~\ref{fig:full_spec}, the higher-order Balmer absorption (upper level $n > 3$) is clearly deeper than the lines and the continuum is absorbed, which would be difficult to explain with the loss of the line emitting clouds.
We note that absorption variations in BALs have been noted previously, and they cannot be explained with loss of broad emission \citep[e.g.,][]{Capellupo_2011}.
We further discuss the connection between \target and BALs in Section~\ref{sec:discuss}.
More epochs of observations will help to further verify the variability pattern.
In what follows, we focus on the varying absorption scenario and its physical interpretation.

%\section{Spectral Fitting}
%\label{sec:spectral_diagnostics}

\subsection{Partial covering model approach}

\begin{figure*}
    \centering
    \includegraphics[width=0.48\linewidth]{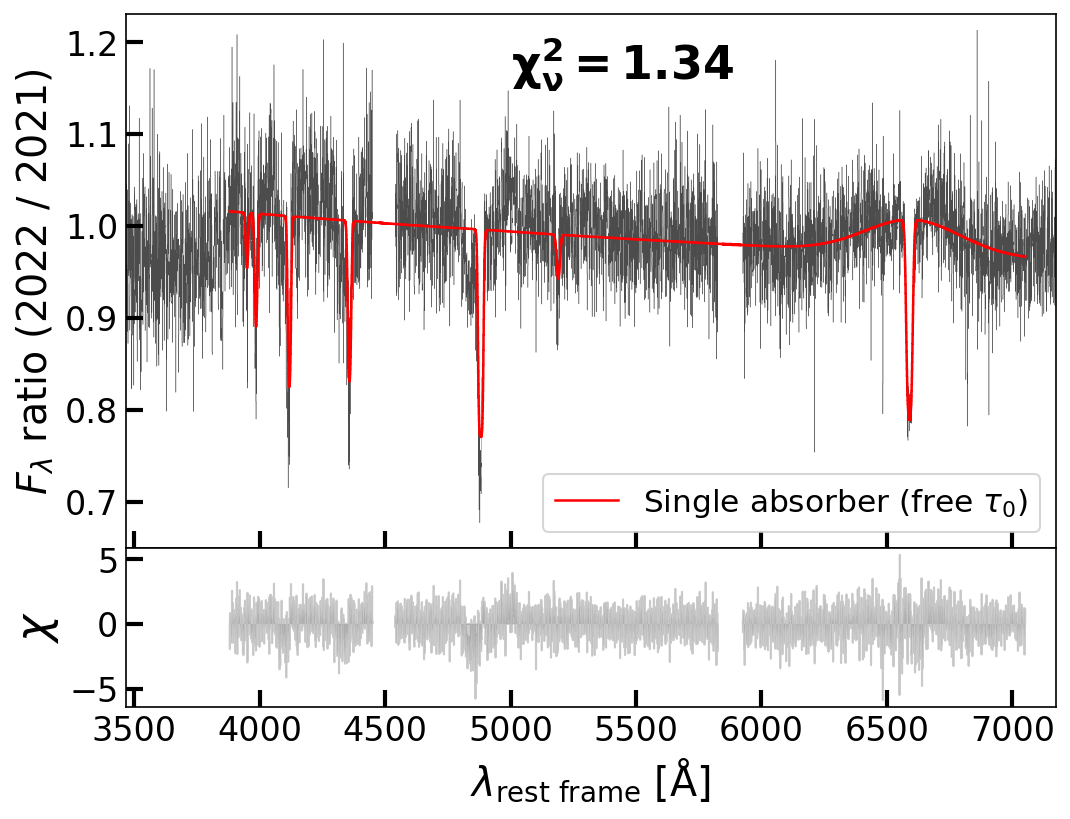}
    \includegraphics[width=0.48\linewidth]{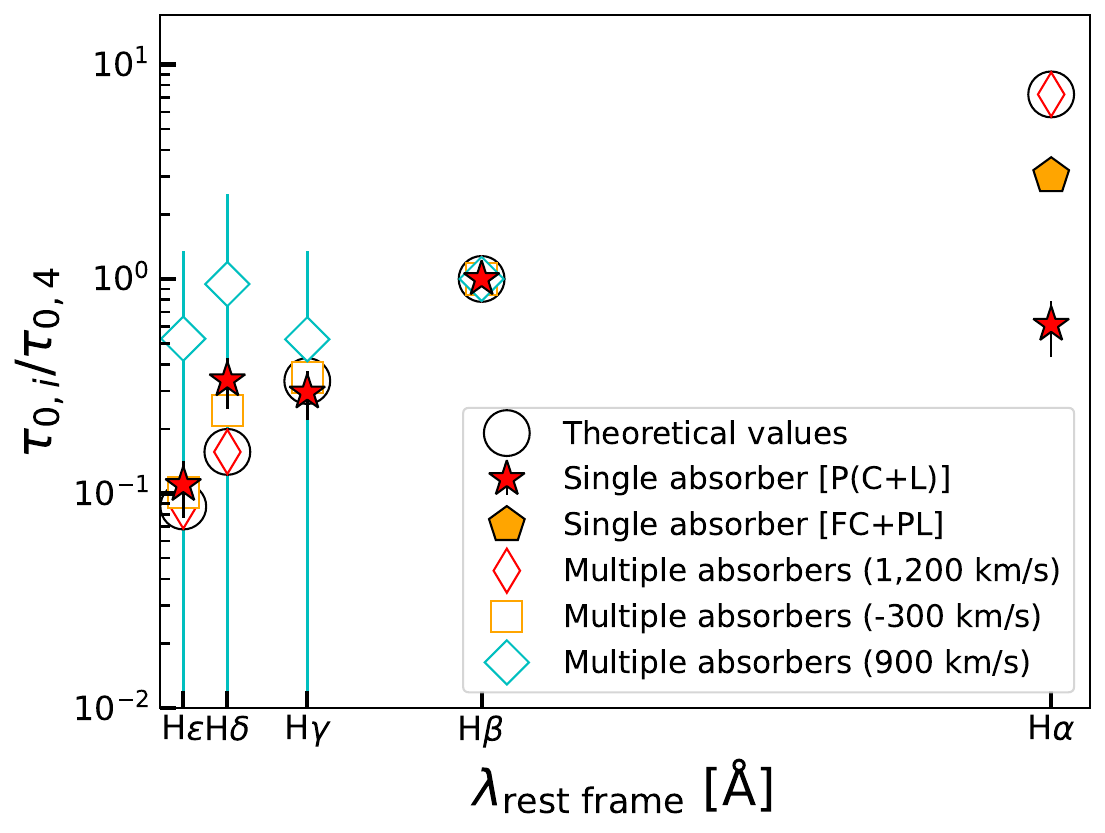}
    \caption{
    Comparison between absorber models (absorbing both broad lines and the continuum) with the varying spectrum in the optical for \target.
    The left panel shows our best fit model with a single absorber, where  we also plot normalized residuals ($\chi \equiv {\rm residual/\sigma}$) of the fit.
    Notably, the \hb absorption appears similarly strong as \ha.
    The right panel shows our best-fit line core optical depths compared to theoretical values, where the \ha optical depth relative to \hb deviates significantly from theoretical values.
    The discrepancy could imply either 1) complex partial covering where the continuum and lines see different gas covering geometries, or 2) stratified Balmer emission in the BLR, where \ha sees lower obscuration compared to other lines.
    We also show our fitted optical depths with a differential covering model (FC+PL) and multiple absorbers (see text).
    }
    \label{fig:single_abs}
\end{figure*}

In studies of BALs and LRDs, the observed absorption is typically fitted by partial covering models, which describe a screen of gas as an absorber that partially covers the central source.
We started by fitting the absorber with the following partial covering model
\begin{equation}
    f_\lambda/f_{\lambda;0}=1-C_f+C_f\exp[-\tau_0e^{-(1-\lambda/\lambda_0)^2c^2/v_0^2}],
    \label{eq:abs}
\end{equation}
where $f_\lambda/f_{\lambda;0}$ is the ratio between the transmitted flux and the initial flux, $C_f$ is the covering fraction of the absorber, $\tau_0$ is the optical depth at the line center, $\lambda_0$ is the central wavelength, $c$ is the speed of light, and $v_0$ is the velocity broadening of the absorbing gas due to a combination of thermal broadening, turbulence, and LOS motions.
The equation assumes an overall Gaussian velocity distribution for the absorber.
In principle, different forms of velocity parametrization are possible \citep{Leighly_2018} and there would also be Lorentzian (Voigt) wings in the absorption. We did not find any statistical improvement by switching to a Vogit formalism, since the damping wing parameters are generally too small compared to the widths of absorption we considered and the lines are not strongly saturated. Thus, we used Gaussian as a simple and reasonable approximation.

In a more general case where multiple kinematically distinct absorbers are present, the right hand side of the equation needs to be replaced by products of terms from different absorbers, which is $\prod\limits^{i} (1-C_{f,i}+C_{f,i}\exp[-\tau_{0,i}e^{-dv_i^2/v_{0,i}^2}])$.
%Based on this equation, the relative difference between the 2021 and 2022 epochs can be described by one or multiple kinematically distinct absorbers.
%As mentioned in the last section, this is assuming both the lines and the continuum are absorbed together.
Also, we note that here it is assumed that both the lines and the continuum are absorbed together.
We show below whether this is a plausible scenario.
This formalism ignores the contribution from the narrow Balmer lines if they overlap with the varying absorption, which would reduce the derived optical depths at the given locations.
However, the varying absorption is sufficiently redshifted compared to the redshift of the narrow lines (by $dv\sim 1,100$ \kms by comparing Figure~\ref{fig:full_spec} and Figure~\ref{fig:abs_var}), making the impact of narrow lines negligible.

\subsubsection{Single absorber fit}

We started our fit by assuming that in the flux density ratio spectrum (bottom panel of Figure~\ref{fig:abs_var}), the optical continuum is a simple power law (close to $y=x^0$ but subject to continuum variation and flux calibration) and Balmer absorptions are modeled according to Equation~\ref{eq:abs}, where we assumed a single absorber to start with for simplicity.
We fit Balmer absorptions up to H$\epsilon$ as well as \caii\,K and H and \feii$\lambda 5169$ absorptions.
All absorptions are forced to have the same kinematics ($dv$ and $v_0$) and covering factor ($C_f$), but their line core optical depths ($\tau _0$) are allowed to vary freely except for those of \caii\,K and H, which we tied based on their oscillator strengths due to their weakness.
The fit is limited to the optical ($\lambda_{\rm rest}>3860$ \AA) to avoid the Balmer break region.
All models are convolved with the DESI LSF \citep{Guy_2023}.
To account for the broad spectral features around Balmer lines possibly coming from variation of an extremely broad-line region (EBLR) or flux calibration residuals, we added broad Gaussians to \ha and \hb, where the features are most apparent.
We note that removing these Gaussians would not affect our overall results and conclusions.

To find the best-fit model, we performed Markov chain Monte Carlo calculations with the \textsc{Python} package \textsc{emcee} \citep{emcee}, doing 50,000 steps with $3k$ walkers ($k=16$ is the number of free parameters -- 2 for the power law, 10 for the absorber, and 4 for the broad Gaussians).
Flat priors are set for all parameters (in log space for the $\tau _0$ parameters).
We computed the autocorrelation time, $t_{\rm ac}$, of the chain for each parameter and set
\texttt{max}\{$2\,t_{\rm ac}$\} burn-in steps and \texttt{min}\{$0.5\,t_{\rm ac}$\}-step thinning.
We took the median values of the parameters as the best-fit values.
The final $1\sigma$ uncertainty is computed from the 68\% confidence interval of the posterior distribution.
The posterior distributions of the model parameters are shown in Appendix~\ref{appendix:posteriors}.

The left panel of Figure~\ref{fig:single_abs} shows our best-fit model, which includes Balmer absorption from \ha to H$\epsilon$.
In the normalized residuals, some clear structures especially around higher-order Balmer absorption can be seen, implying the presence of an additional absorber(s).
The right panel of Figure~\ref{fig:single_abs} compares our best-fit line core optical depths normalized by that of \hb (i.e., $\tau _{0,\,4}$) in comparison with the theoretical values.
The theoretical values are given by $\tau _{0,\,i}/\tau _{0,\,4} = (\lambda _i f_i)/(\lambda _4 f_4)$, where $i$ represents the upper level and $f_i$ is the oscillator strength.
The oscillator strengths are taken from the National Institute of Standards and Technology (NIST) Atomic Spectra Database (ASD; v5.12)\footnote{\url{https://www.nist.gov/pml/atomic-spectra-database}}.
While higher-order Balmer lines are consistent with theoretical values within $2\sigma$ measurement uncertainties, $\tau_ 0$ of \ha is significantly lower than the theoretical value, even lower than $\tau_ 0$ of \hb.
Such a behavior of lower measured $\tau _0$ in \ha is also seen in FeLoBALs and some LRDs with Balmer absorption \citep[e.g.,][]{Leighly_2025,deugenio_irony}. Below we list several possibilities to explain the deviation from theoretical values.
\begin{enumerate}
    \item The continuum, rather than the broad lines are absorbed (or the continuum is more strongly absorbed than the lines; see e.g., \citealp{Leighly_2025}).
    In this case of differential obscuration, the total flux ratio including the contribution from broad lines would bias the derived optical depth low.
    Since the broad emission of \ha is much stronger than higher order Balmer lines, this effect is strongest for \ha, which could explain why higher order Balmer absorptions appear more consistent with theoretical predictions.
    \item The LOS absorption has multiple components and is poorly approximated with a single absorber. Since the line core optical depth is most sensitive to the deepest absorption trough, this might lead to a fitting bias.
    \item The absorber is within the BLR and has spatial stratification together with line-emitting clouds. Since Balmer emission from BLRs is known to come from different radial locations based on reverberation mapping studies \citep[e.g.,][]{Kaspi_2000}, emission produced deeper in the BLR (i.e., higher-order Balmer emission) naturally has larger optical depths.
\end{enumerate}

The three scenarios are not mutually exclusive and can happen at the same time.
Regardless, the comparison in Figure~\ref{fig:single_abs} indicates that a single screen of absorption is only a rough approximation of the actual obscuration at play here, and that more complex scenarios are required to fully describe the data, which we investigate next.

\subsubsection{More complex models}
\label{more_complex_models}

%\redtxt{[tbd the structure since the stationary absorber is involved here as well]}

\begin{figure*}
    \centering
    \includegraphics[width=0.515\linewidth]{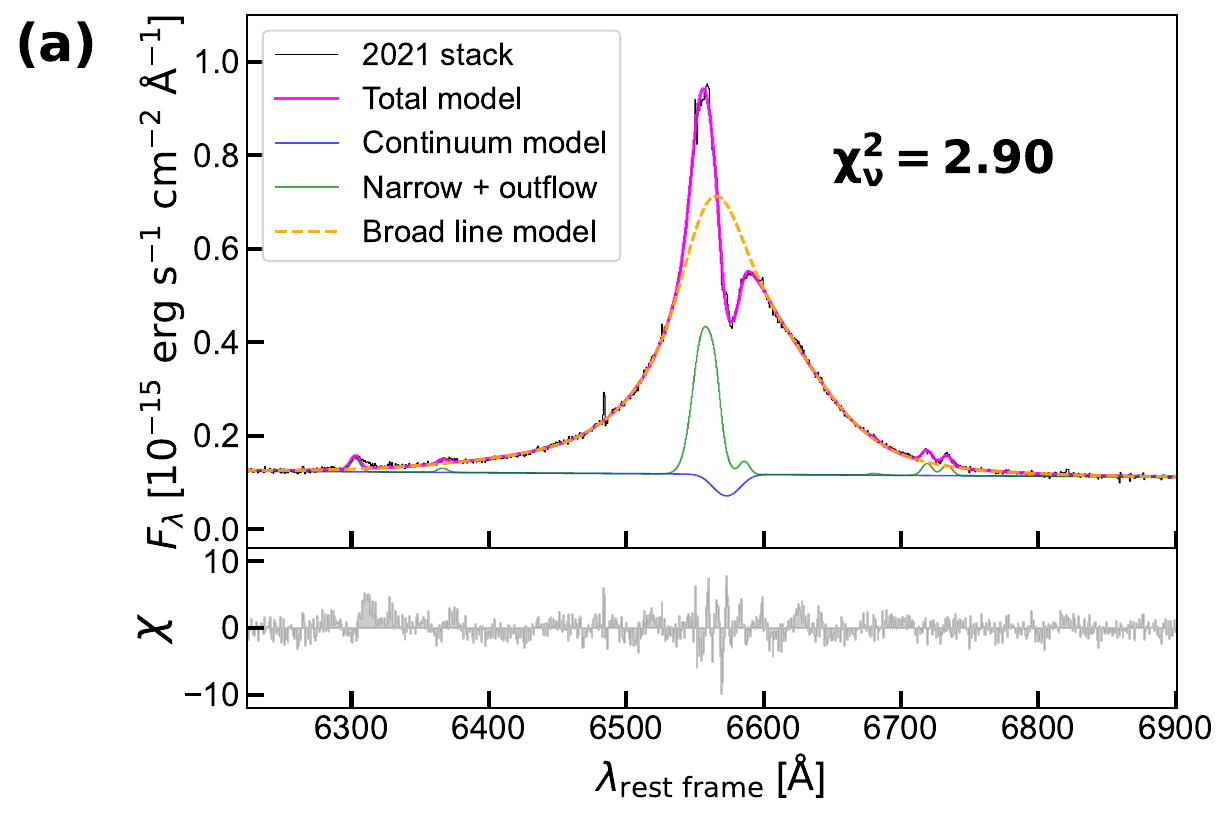}
    \includegraphics[width=0.47\linewidth]{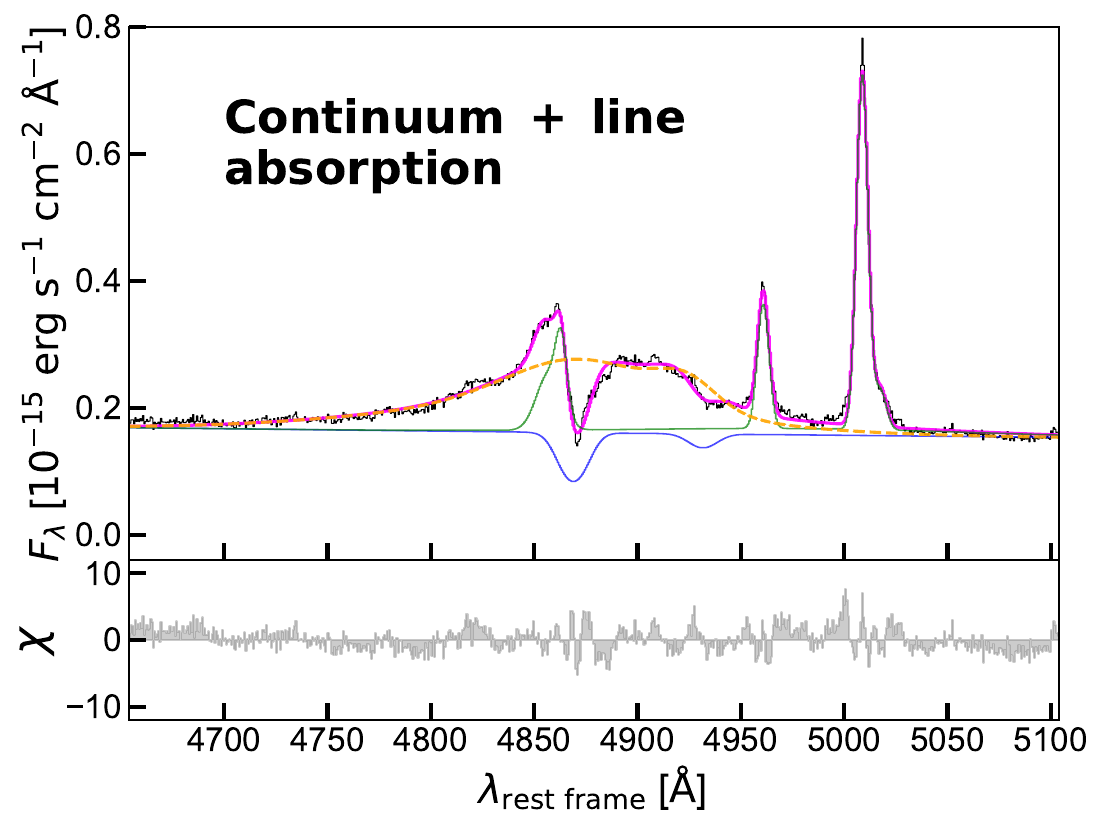}

    \includegraphics[width=0.515\linewidth]{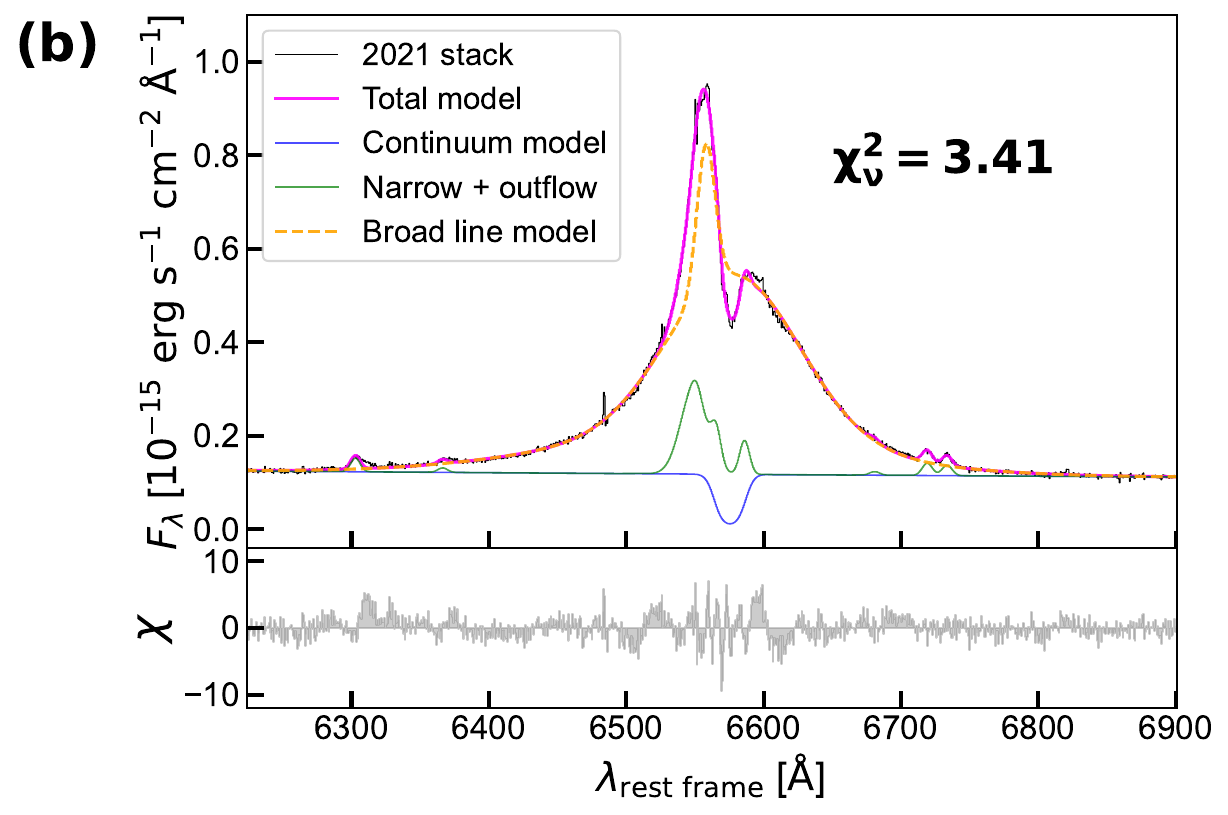}
    \includegraphics[width=0.47\linewidth]{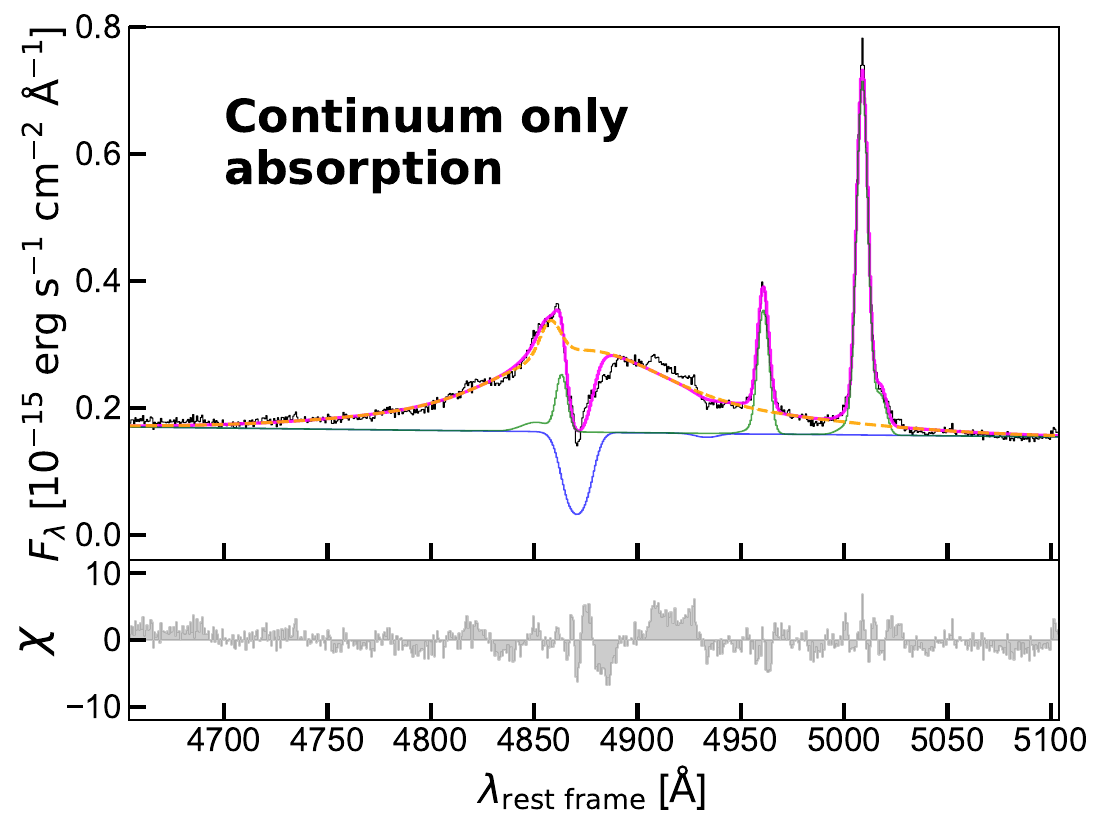}
    \caption{
    Comparison between two different model fits to spectral regions around \ha, \hb, and \oiii$\lambda \lambda 4959,5007$ in \target.
    Model (a) assumes that both broad lines and continuum are absorbed by the same absorber. Model (b) assumes that only the continuum is absorbed.
    Model (b) predicts significantly saturated absorption cores, and the overall fit is worse.
    The outflow component in Model (a) might also represent a P-Cygni emission (see text).
    A common systematic uncertainty shared by both models is the blending of emission and absorption near the line core, which requires higher resolution and/or more time-domain observations to investigate.
    %\redtxt{tbc}
    }
    \label{fig:stationary_fit}
\end{figure*}

\begin{figure}
    \centering
    \includegraphics[width=\columnwidth]{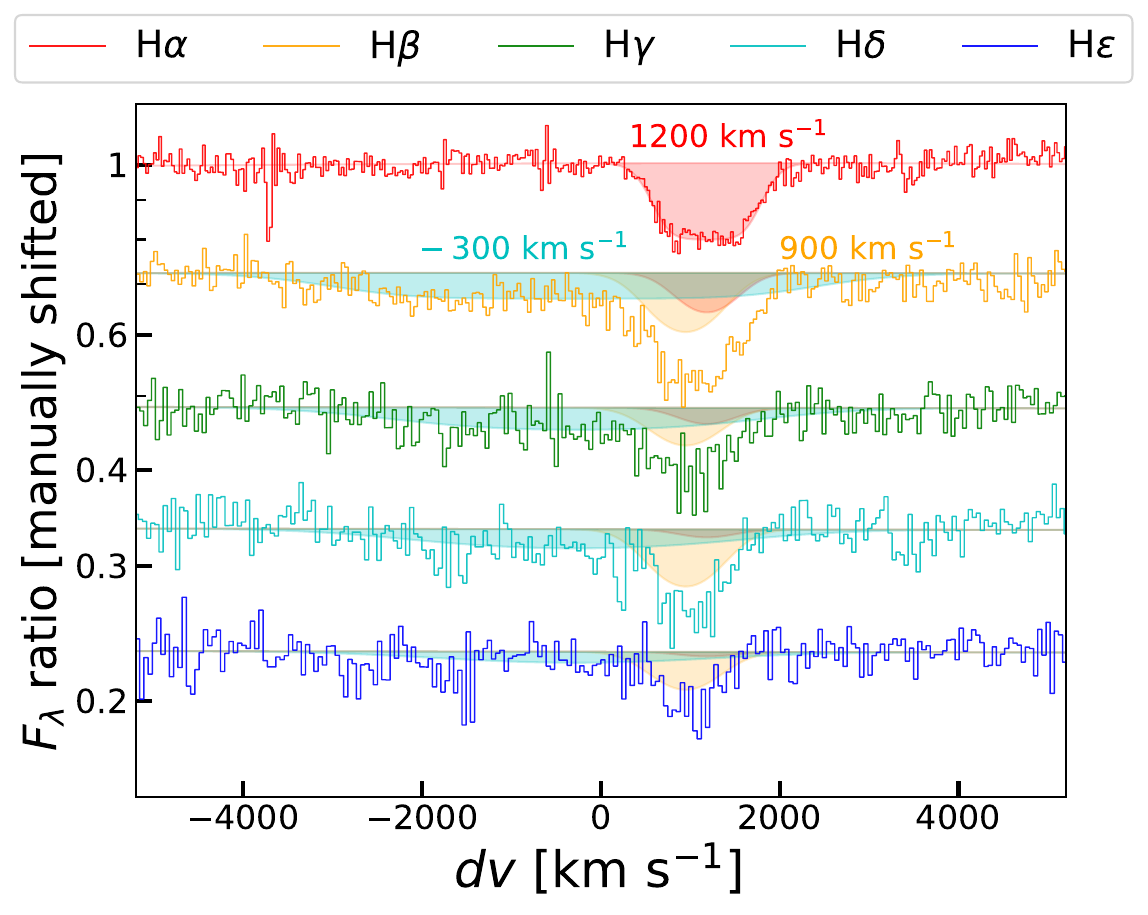}
    \caption{Comparison
    of different Balmer absorptions that vary between 2021 and 2022 in the velocity space.
    The zero-point of the velocity is set to the systemic redshift of \target determined by narrow \oiii$\lambda 5007$.
    While the minimum locations of all absorptions are consistently redshifted with $dv\approx 1,000$ \kms, the blue and red wings appear asymmetric, indicative of multiple components.
    Meanwhile, \hb absorption appears particularly strong.
    We also show a tentative decomposition indicated by the shaded regions, where $n>3$ Balmer lines are assigned with two additional absorbers assuming they come from a deeper region.
    In this case, a strongly saturated and blueshifted ($dv\sim -300$ \kms) component with low $C_f$ is found in $n>3$ Balmer lines, and the main absorption trough is still characterized by $dv\approx 1,000$ \kms.
    }
    \label{fig:vel_com}
\end{figure}

\begin{table*}
    \centering
    \caption{Derived parameters for the absorbers based on the Balmer absorption in the spectra of \target.
    The velocity shift of the absorption, $dv$, is relative to the systemic velocity set by narrow emission lines.
    The intrinsic width of the absorption is given by $v_0$.
    The column density of $n=2$ hydrogen, $N_{\rm H(n=2)}$, is derived based on the best-fit optical depth of the given line under the ``Major line tracer''.
    Parameters for the dynamic and stationary absorbers are listed separately, and different absorbing models (under ``Model'') are compared.
    For the absorber models, $P(C+L)$ means that both the continuum and broad lines are partially covered and equally absorbed; $FC+PL$ means that while the continuum is fully covered with $C_f=1$, the broad lines are partially covered by the absorber; $PC$ means that only the continuum is partially covered and absorbed, and the broad lines remain unabsorbed.
    }
    \begin{tabular}{l c c c c c c c c}
    \hline
    \multicolumn{9}{|c|}{Dynamic (varying) absorber (between 2021 and 2022 epochs)}\\
    \hline
        Model & $\chi^2_\nu$ & $BIC$ & $C_f$ & $dv$ [\kms] & $v_0$ [\kms] & $\log N_{\rm H(n=2)}$ [$\rm cm^{-2}$] & $\log N_{\rm H}$ [$\rm cm^{-2}$;  & Major line \\
        & & & & & & & model-based] & tracer \\
        \hline
        \hline
        Single absorber & 1.34 & 6668 & $0.23\pm 0.01$ & $1109 \pm 7$ & $458\pm 26$ & $14.4\pm 0.1$ & $\sim 21-23$ & \ha \\
        $[P(C+L)]$ & & & & & & $15.4 \pm 0.1$ & (w/ \hei; Section~\ref{sec:vary_abs}) & \hb \\
        \hline
        Single absorber$^*$ & 1.76 & - & $0.042^{+0.017}_{-0.013}$ & $1054 \pm 9$ & $569\pm 15$ & $14.26\pm 0.05$ & $\sim 21-23$ & \ha \\
        $[FC+PL]$ & & & & & & $14.64 \pm 0.01$ & (w/ \hei; Section~\ref{sec:vary_abs}) & \hb \\
        \hline
        \hline
        Multiple absorbers & 1.20 & 6059 & $0.206\pm 0.004$ & $1157\pm 7$ & $434\pm 15$ & $14.60\pm 0.05$ & $\sim 21-23$ & \ha \\
        $[P(C+L)]$ & & & $0.24^{+0.41}_{-0.08}$ & $919\pm 28$ & $506^{+56}_{-68}$ & $14.8\pm 0.5$ & (w/ \hei; Section~\ref{sec:vary_abs}) & \hb \\
         & & & $0.075^{+0.006}_{-0.005}$ & $-321\pm 107$ & $1910\pm 200$ & $16.09^{+0.15}_{-0.14}$ & $\gtrsim 23$ (strong saturation) & \hb \\
    \hline
    \multicolumn{9}{|c|}{Dynamic (varying) absorber (between 2022-01-05 and 2022-02-28)}\\
    \hline
    Single absorber & - & - & {$0.30^{+0.45}_{-0.25}$} & $1218 \pm 76$ & $671^{+89}_{-280}$ & $13.27^{+1.58}_{-0.41}$ & $\sim 20-22$ & \ha \\
    $[P(C+L)]$ & & & & & & & (scaled from above) & \\
    \hline
    \multicolumn{9}{|c|}{Stationary absorber (2021 epoch)}\\
    \hline
    Single absorber & 2.90 & 5612 & {$0.99^{+0.01}_{-0.02}$} & $354 \pm 13$ & $573\pm 23$ & $13.66\pm 0.04$ & $\sim 21.5-22.5$ & \ha \\
    $[P(C+L)]$ & & & & & & $14.64 \pm 0.03$ & (w/ \hei; Section~\ref{stationary_abs}) & \hb \\
        \hline
    Single absorber & 3.41 & 6553 & $0.996^{+0.003}_{-0.007}$ & $458\pm 14$ & $455\pm 10$ & $14.23\pm 0.02$ & $\sim 21.5-22.5$ & \ha \\
    $[PC]$ & & & & & & $14.94 \pm 0.02$ & (w/ \hei; Section~\ref{stationary_abs}) & \hb \\
        \hline
    \end{tabular}
    \begin{tablenotes}
%	    \centering
        \small
        \item $\bf Notes.$ $^*$ This model only fits spectral regions around \ha and \hb to reduce the number of free parameters. 
        We thus do not list its $BIC$ since it cannot be directly compared with other models.
    \end{tablenotes}
    \label{tab:abs_params}
\end{table*}

Scenario (i) can be tested by removing the contribution from the broad Balmer emission, although this step itself might suffer from large uncertainties due to the complex line shape.
We note that simply taking the flux density difference (top panel of Figure~\ref{fig:abs_var}) and adding it back to the best-fit intrinsic (unabsorbed) continuum, and then normalizing it by the intrinsic continuum cannot correctly retrieve the varying absorption.
This approach would only work when the varying absorption is sufficiently separated from the non-varying absorption since absorption (in this case where only one central ionizing source is present) is likely a multiplicative rather than additive process.
In the case of \target, the varying absorption is clearly blended with the red wing of the non-varying absorption as shown in Figure~\ref{fig:full_spec}.

To remove the impact from broad lines and non-varying absorption, we took the 2021 stack and fit the spectral region around \ha and \hb.
We included a continuum model as a simple power law, a narrow line component, an outflow component, three Gaussian components with independent kinematics to approximate the asymmetric broad-line shape, and an absorption component that only absorbs the continuum.
The narrow and outflow components were also fitted for \oiii$\lambda \lambda 4959,5007$, but the broad-line component was only fitted for \ha and \hb.
As a control fit, we also checked a model where both the broad-line component and the continuum are equally absorbed.
We note that in both models, the narrow and the outflow components are not absorbed, due to the likely much larger physical scales of these line emitting regions compared to the absorber.

We ran MCMC to get the best-fit parameters, and the best-fit models are shown in Figure~\ref{fig:stationary_fit}.
Panel (b) of Figure~\ref{fig:stationary_fit} shows the continuum-only absorption model, where the continuum absorption is nearly saturated and the apparent non-saturation in the spectrum is explained with line filling in this model.
To find the varying part of the absorption, we then took the direct flux density difference, $\Delta F_{\lambda}=F_{\lambda,2022}-F_{\lambda,2021}$, and added it to the best-fit continuum model with absorption in 2021, which gives the continuum model for the 2022 spectrum.
We did not directly model the absorbed continuum of the 2022 epoch due to the more complex shape of \ha at this epoch.
We show the reconstructed 2022 continuum model in Appendix~\ref{appendix:contabs}.
We found that the lowest location of the reconstructed absorption trough is clearly negative in this case, which is unphysical.
Masking negative pixels cannot recover the ``intrinsic'' absorption since the line core optical depth is mostly driven by the lowest location in the absorption.
Also, as shown in Figure~\ref{fig:stationary_fit}, the continuum-only absorption model [panel (b)] performs statistically worse compared to the model where both the continuum and broad lines are equally absorbed [panel (a)].
The $\chi^2_\nu$ and $BIC$ of different models are also lited in Table~\ref{tab:abs_params}.
Finally, the continuum-only absorption model also does not produce an optical depth ratio between \ha and \hb consistent with the theoretical value for the stationary absorber during the 2021 epoch. 
Based on the above reasons, we do not consider the continuum-only absorption to be a plausible solution for the optical depth discrepancy seen in the dynamic absorber, and we use the model shown in panel (a) as our fiducial model.
Notably, similar analysis has been performed for some LRDs with high S/N spectra, where negative absorption was also found if it only applies to the continuum \citep[e.g.,][]{juodzbalis_rosetta_2024,deugenio_irony,Ivey_2026}.

In principle, more complicated partial covering scenario is plausible.
For example, as discussed in \citet{Leighly_2025}, one can also assume that the continuum and broad lines see different covering factors if the absorber is in part co-spatial with the line emitting region, where the continuum is fully covered with $C_f=1$ but the lines are partially covered with $C_f<1$.
The different covering scenarios are motivated by the observation that deviations from the theoretical optical depth ratios are frequently observed in FeLoBALs.
To check this, we performed another fit of the spectrum from 2021 by fixing the continuum covering factor to 1 while setting the broad-line covering factor free.
Our best-fit model yields $C_f\approx 1$ for broad lines, again implying that for the stationary absorber traced by the 2021 epoch, broad lines have a high covering factor similar to the continuum.
The optical depth ratio between \ha and \hb still deviates significantly from the theoretical value compared to the default case where the continuum and lines are equally absorbed, indicating that this model does not solve the optical depth issue for the stationary absorber.

For the dynamic absorber, it is non-trivial to test the scenario with different $C_f$.
Still, based on the best-fit model of the stationary absorber at the 2021 epoch, we can reformulate the flux ratio in this scenario as the following
\begin{equation}
    F_{\rm  ratio} (2022/2021) = \frac{f_{\rm abs,\,c}F_{C}+f_{\rm abs,\,l}F_L}{F_C+F_L},
    \label{eq:fc_pl}
\end{equation}
where $F_C$ and $F_L$ are best-fit continuum and broad-line models from the 2021 epoch, and $f_{\rm abs}$ is the absorption factor given by Equation~\ref{eq:abs}.
Following \citet{Leighly_2025} and setting $C_f = 1$ for the continuum, with MCMC fitting, we obtained a small line covering factor of $C_f \approx 0.04$ and an optical depth ratio between \ha and \hb closer to the theoretical value, as shown in the right panel of Figure~\ref{fig:single_abs}.
This implies that differential coverage could be a plausible explanation.

We again caution that the fit of the 2021 stack to retrieve the parameters for the stationary absorber suffers from blending with emission lines.
One complicating issue is whether there is a P-Cygni-like profile contributing to the line-core emission due to infalling clouds from the other side of the accretion disc, which is a phenomenon observed in supernovae with expanding or contracting shells \citep[e.g.,][]{Dessart_2009,Fransson_sne_2014}.
In our current fit, the core emission is largely assigned to narrow and outflow components that are not absorbed, which can also be considered as a flexible ``P-Cygni fit''.
{We also checked a more constrained P-Cygni fit, where we replaced the outflow component by a Gaussian emission with a velocity offset opposite to the stationary absorption for \ha and \hb. The fit has $\chi ^2_{\nu}=3.3$ and is statistically worse compared to our fiducial fit.
The major difference compared to the fiducial fit is the covering factor, which is reduced to $C_f \approx 0.5$.
}
Future time-domain follow up to investigate the relative variations in the line core emission and absorption could help to further break the degeneracy.

%There are several problems associated with the approach of continuum-only absorption. First, given that the line core is strongly saturated and blended with in-filling emission, systematic uncertainties associated with the optical depth measurements are difficult to characterize. Second, the reconstructed continuum in the 2022 epoch contains pixels showing continuous negative flux densities, implying that the continuum-only absorption might not be a self-consistent model.

To examine Scenarios (ii) and (iii), we plot the varying absorption spectrum in the velocity space in Figure~\ref{fig:vel_com}.
Interestingly, the minimum flux ratios are similar in \ha and \hb, consistent with the fact that a single absorber fit yields comparable line core optical depths for the two lines.
There is no particularly deep trough in \hb that drives the whole fit.
In fact, the \hb absorption appears overall broader than \ha especially in the blue wing, which is also visible in higher-order Balmer lines.
The missing blue wing absorption in \ha supports Scenario (iii), where higher-order Balmer lines undergo more absorption.
As a test, we do a fit with one absorption component for \ha but three absorption components for higher-order Balmer lines.
Among the three components for higher-order Balmer lines, one is tied to \ha based on their oscillator strengths and wavelengths, and two are allowed to vary freely.
This increases the total number of parameters, and thus we enlarge the MCMC chain to 100,000 steps with an initial 20,000-step burnin.
The posterior distributions of the parameters are also shown in Appendix~\ref{appendix:posteriors}.
Figure~\ref{fig:vel_com} shows the best-fit model with multiple absorbers.
The model is statistically better than the single absorber model with $\Delta BIC \gg 10$.
The absorptions are dominated by two redshifted components with $dv\approx 1,000$ \kms.
An additional blueshifted, strongly saturated absorption with $dv\approx -300$ \kms is found in higher-order Balmer lines.
As shown in the right panel of Figure~\ref{fig:single_abs}, the optical depth ratios for higher-order Balmer lines are compatible with theoretical values, although the uncertainties are relatively large.
{The absorption is unlikely to be caused by optical \feii\ transitions, as the typically strong \feii\ group near \hb is at 4434\,-\,4684 \AA in the rest frame, corresponding to a blueshift velocity of $>10,000$ \kms with respect to \hb.
It is possible that this blueshifted component is not physically associated with the absorber that makes the major redshifted absorption trough, and it might trace, for example, a strong wind component within the BLR.
Following up \target in the UV to cover wind-sensitive lines such as \civ\ might help to verify the origin of the blueshifted Balmer component.
Regardless, for the rest of the manuscript, we still refer to this blueshifted component as part of the dynamic absorber since it varies between the 2021 and 2022 epochs.
}

Based on these model fits, we derive column densities of hydrogen at $n=2$ associated with the varying absorber with the equation
\begin{equation}
    N_{\rm H~(n=2)} = \frac{m_{\rm e}c}{4\pi e^2 f_0 \lambda _0}\tau,
\end{equation}
where $m_{\rm e}$ is the mass of electron, $c$ is the speed of light, $e\equiv q_{\rm e}/\sqrt{4\pi \epsilon _0}$ is the electron charge in Gaussian units, $f_0$ is the oscillator strength, $\lambda _0$ is the line core wavelength, and $\tau$ is the integrated optical depth in velocity space.
The values are listed in Table~\ref{tab:abs_params}.
In the varying absorber, regardless of the fitted models, the absorbing trough with $dv\sim 1,000$ \kms traced by \ha absorption has a column density of excited hydrogen $(n=2)$ of roughly $10^{14-14.5}~{\rm cm^{-2}}$, whereas \hb implies a larger column of $\sim 10^{15}~{\rm cm^{-2}}$ with an uncertainty up to 0.5 dex.
%The covering factors for these absorbers are 15\,-\,20\%.
In contrast, the blueshifted kinematic component corresponds to a small covering factor of $C_f \approx 8\%$ and a high column density reaching $10^{16}~{\rm cm^{-2}}$, which implies a strong saturation of this absorption.
We also list our measurements for the stationary absorber from the best fit model shown in panel (a) of Figure~\ref{fig:stationary_fit}.
There, the model implies that the stationary absorber is nearly fully covering the LOS.
However, the overall column densities of the stationary and varying absorbers are comparable within 1 dex, with $N_{\rm H(n=2)}\sim 10^{14-15}~{\rm cm^{-2}}$.
All these column densities only reflect the excited hydrogen at $n=2$.
To infer the total column density of hydrogen, one needs to make assumptions about the ionizing source and gas conditions, which we discuss next.

\subsection{Photoionization modeling}
\label{subsec:cloudy_models}

\subsubsection{Total gas column density}

\begin{table}
        \centering
        \caption{Input parameters for \textsc{Cloudy} absorber models.
        The models are computed to describe the dynamic absorber.
        }
        \label{tab:cloudy_models}
        \begin{tabular}{l c}
            \hline
            \hline
            Parameter & Values \\
            \hline
            $\log~Z/Z_\odot$ & $0$ \\
            \hline
            $\log U$& $-4$\,-\,$0$ (with 0.3 dex interval) \\
            \hline
            $\log (n_{\rm H}/{\rm cm^{-3}})$& 6\,-\,11 (with 0.5 dex interval) \\
            \hline
            $\log (N_{\rm H}/{\rm cm^{-2}})_{\rm stop}$ & 23 \\
            \hline
            $v_{\rm turb}$/\kms & 400 \\
            \hline
            Geometry & Open; plane-parallel\\
            \hline
            AGN SED & $M_{\rm BH}=10^8~M_\odot$, $\lambda _{\rm Edd}=0.1$\\ 
             & \citep{Pezzulli_2017} \\
            \hline
            Dust & No dust\\
            \hline
            Atomic data & CHIANTI (v7, \citealp{chianti0};\\
             & \citealp{chianti_v7})\\
            \hline
            Solar reference & \citet{grevesse2010} \\
            \hline
        \end{tabular}
\end{table}

\begin{figure*}
    \centering
    \includegraphics[width=\columnwidth]{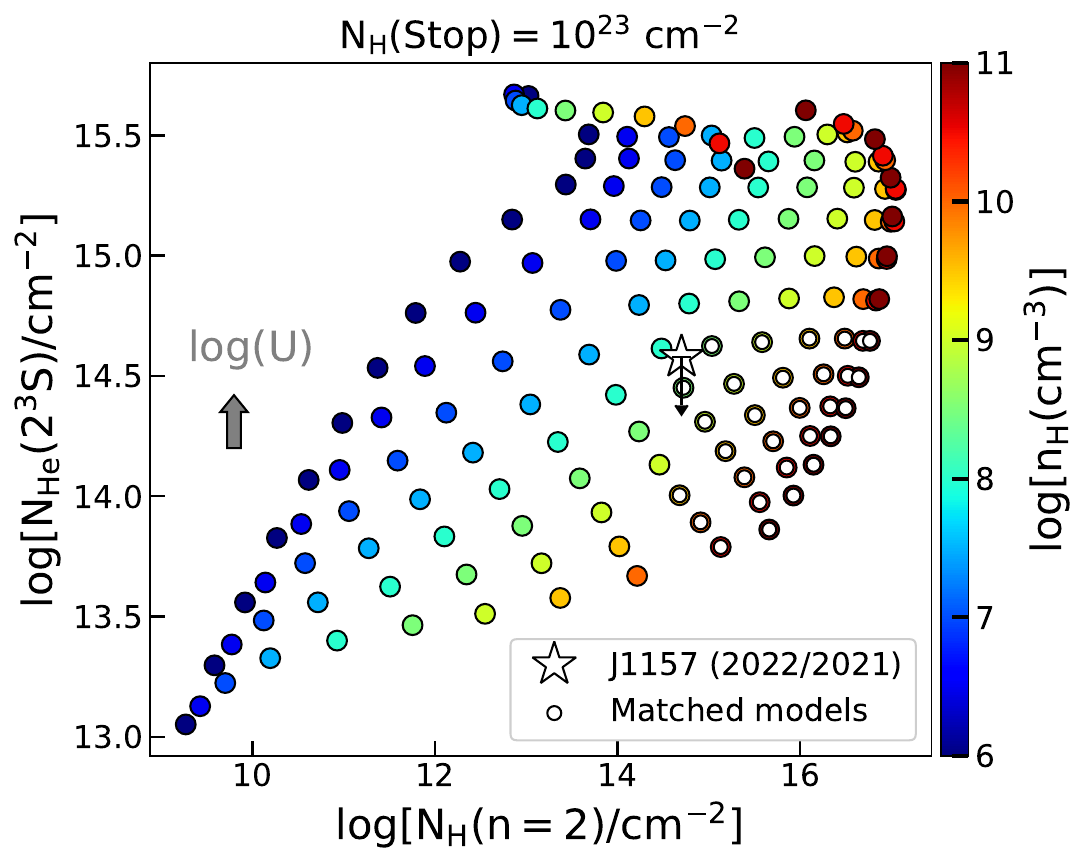}
    \includegraphics[width=\columnwidth]{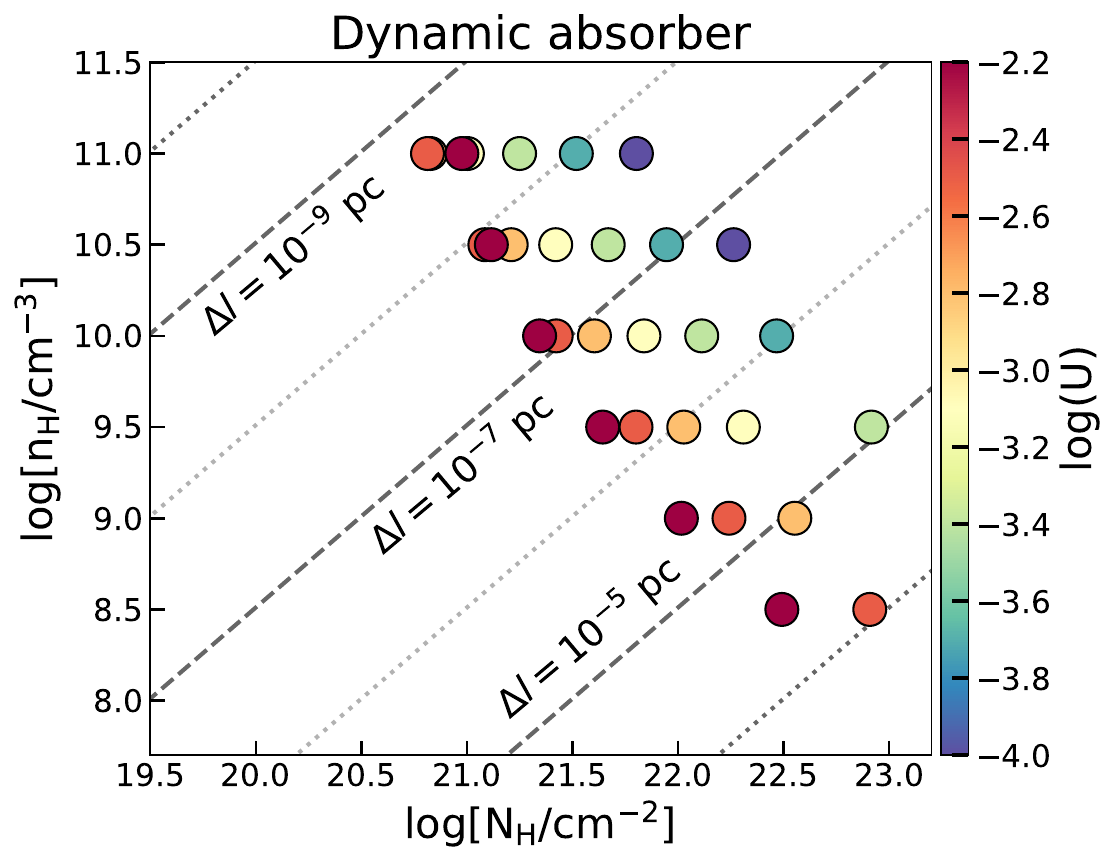}
    \includegraphics[width=\columnwidth]{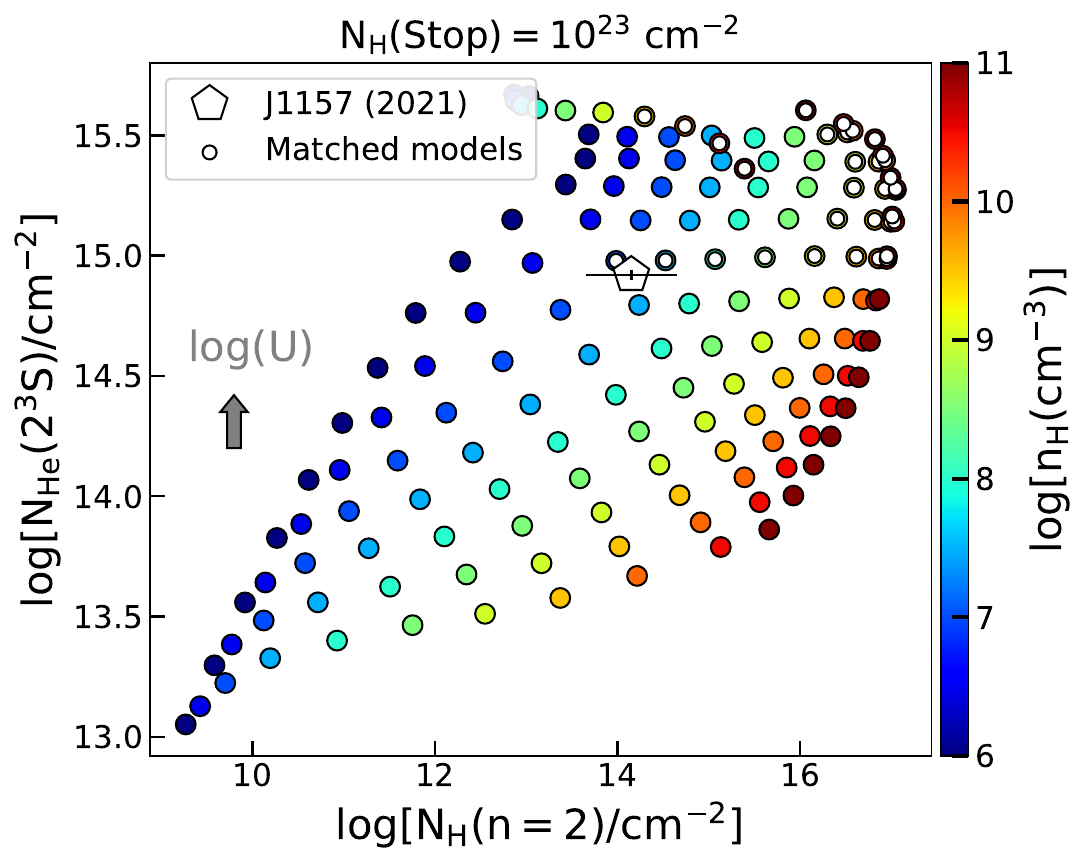}
    \includegraphics[width=\columnwidth]{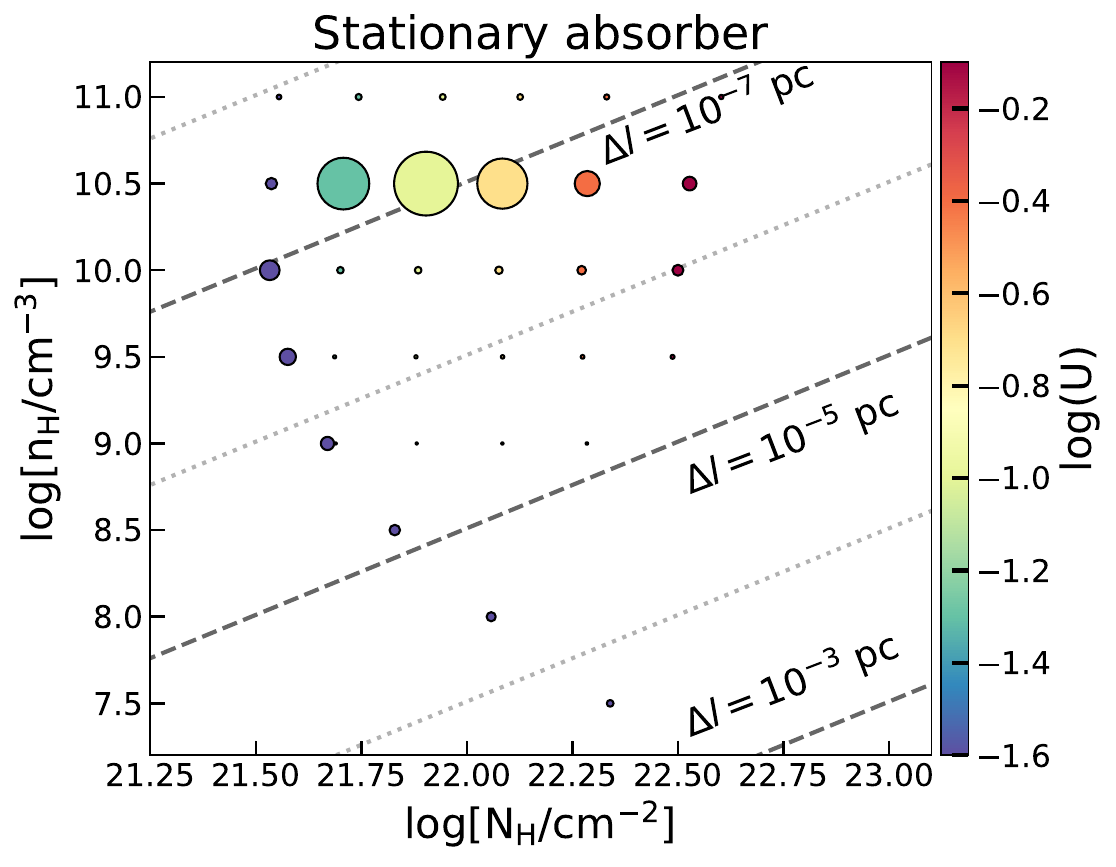}
    \caption{Possible parameter space for the dynamic and stationary absorbers inferred from \cloudy photoionization models. \textit{Top left:} model grid spanning $-4\leq \log U \leq 0$ and $10^6\leq n_{\rm H}~[{\rm cm^{-3}}]\leq 10^{11}$.
    The increasing direction of $U$ is roughly indicated by the arrow.
    All models are stopped at $\log N_{\rm H}= 23~[{\rm cm^{-2}}]$. The star symbol marks our measure column densities from absorption lines in the dynamic absorber (the \hei\ column density is an upper limit), and white circles mark individual models within the grid that contains a certain depth reproducing the measured column densities within 0.25 dex.
    \textit{Upper right:} properties of individual models that can reproduce our measured column densities for the dynamic absorber. The $x$ axis shows the depth (or equivalently, the total hydrogen column density) where the model reproduces our measurements.
    The dashed lines and dotted lines mark constant effective thickness defined as $\Delta l \equiv N_{\rm H}/n_{\rm H}$.
    \textit{Bottom:} same as the upper panel but for the stationary absorber.
    In this case, both $N_{\rm H(n=2)}$ and $N_{\rm He(2^3S)}$ are measured, and we only show matched models with $\chi ^2 < 5$.
    The sizes of the symbols in the lower right panel are inversely proportional to $\chi ^2$.
    Reproducing the measurements for the both dynamic and stationary absorber requires high densities ($n_{\rm H}\gtrsim 10^{8}~{\rm cm^{-3}}$).
    The dynamic absorber has lower ionization ($\log U \lesssim -2$) compared to the stationary absorber ($\log U \sim -1$). Both absorbers have total hydrogen column densities of $N _{\rm H}\sim 10^{21-23}~{\rm cm^{-2}}$), and the implied $\Delta l$ mostly lies between $10^{-8}-10^{-4}$ pc.
    }
    \label{fig:abs_models}
\end{figure*}

\begin{figure}
    \centering
    \includegraphics[width=\columnwidth]{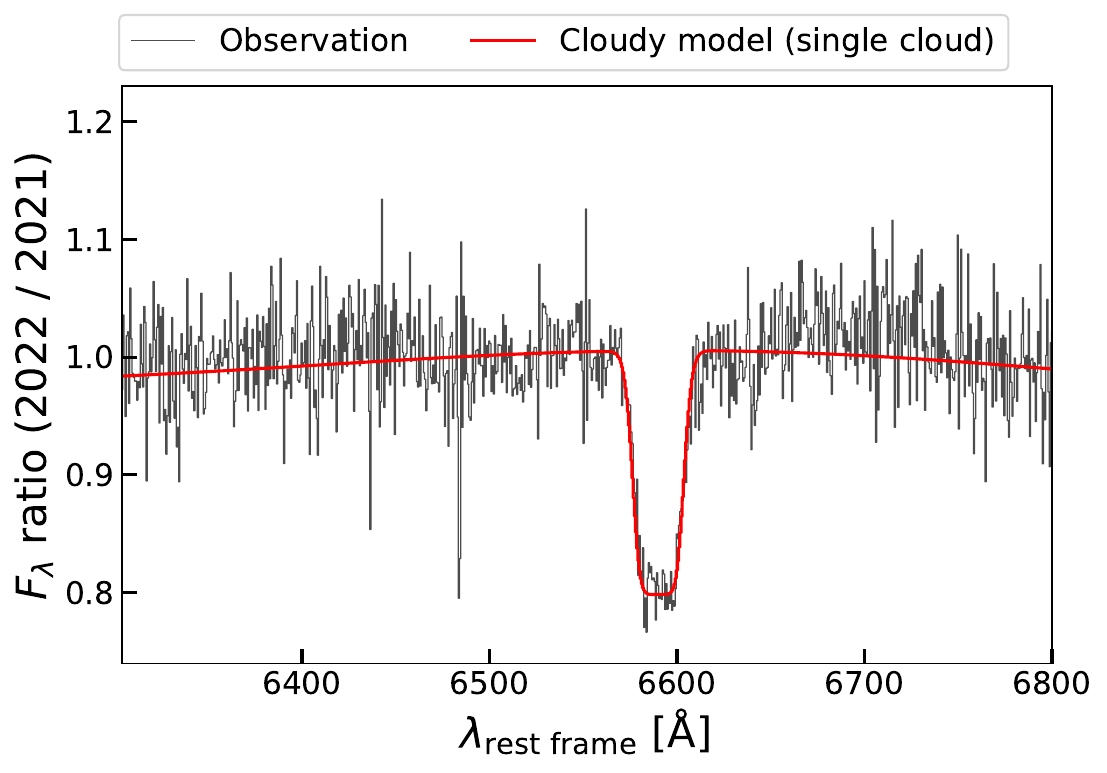}
    \caption{
    Comparison (not fitting) between a fine-grid ($\lambda /\Delta \lambda \approx 1.7\times 10^4$) \cloudy photoionization model with $N_{\rm H}=10^{21.5}~{\rm cm^{-2}}$, $\log U=-2$, and $C_f=0.2$ (see Tables~\ref{tab:abs_params} and \ref{tab:cloudy_models}) and the varying absorption spectrum of \target between 2021 and 2022.
    The model absorption matches well with the observation of \ha, although models with different column densities are needed to match all Balmer absorption lines.
    }
    \label{fig:cloudy_fine_ha}
\end{figure}

\begin{figure*}
    \centering
    \includegraphics[width=.57\linewidth]{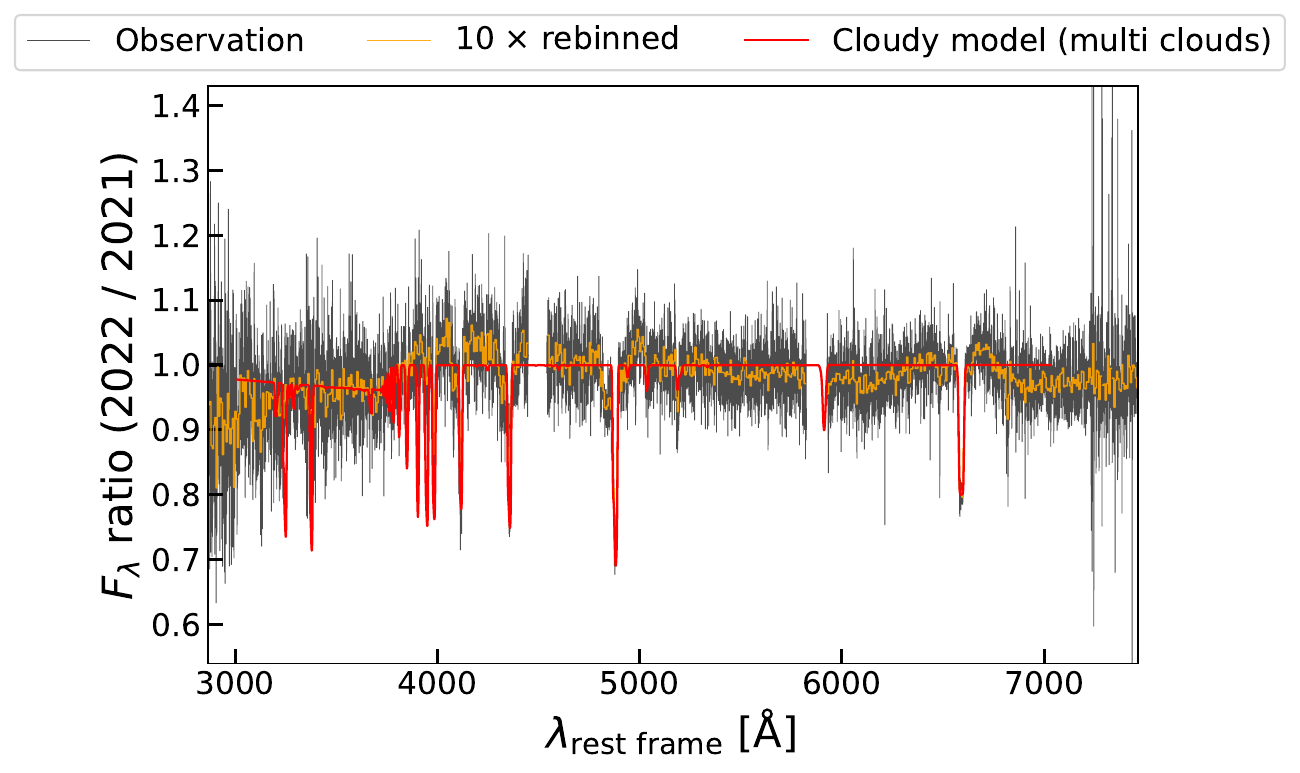}
    \includegraphics[width=.42\linewidth]{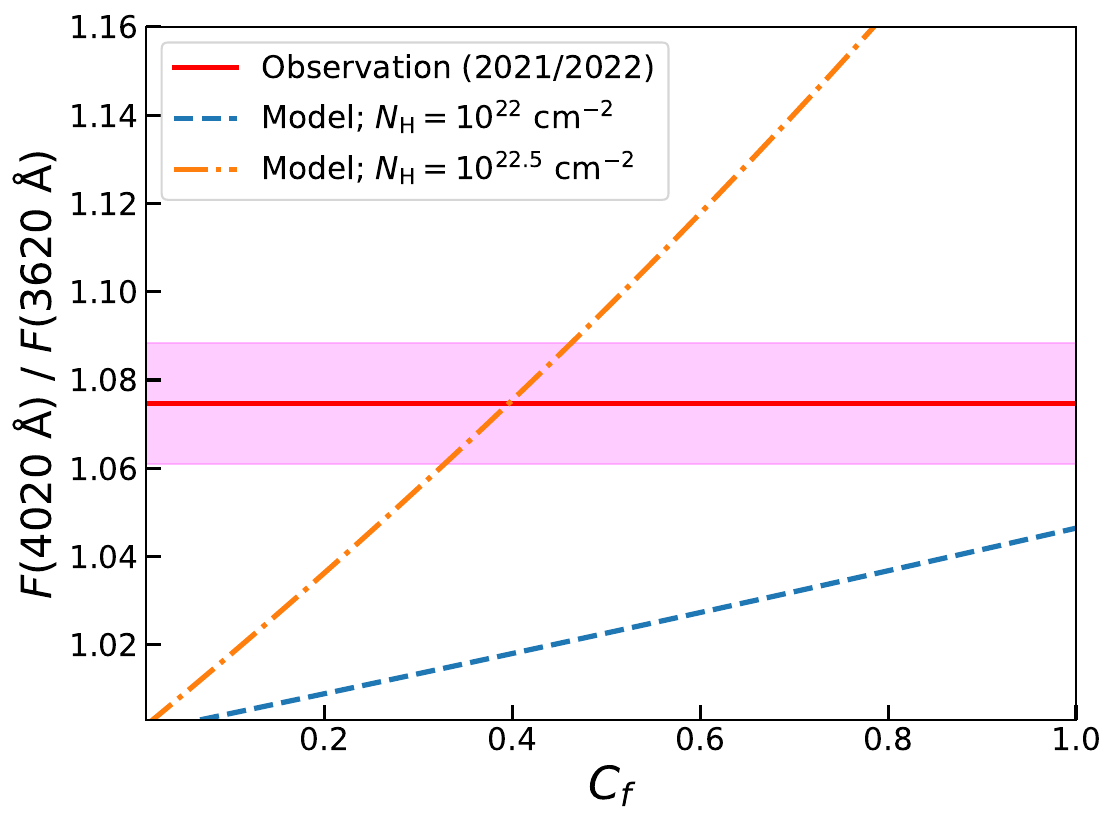}
    \caption{Comparison (not fitting) between the varying absorption spectrum and a composite \cloudy model.
    \textit{Left:}
    composite \cloudy models based on the two absorbers (see text).
    The model recovers the Balmer absorption and predicts a Balmer break roughly consistent with the observation, although the UV \feii\ absorption is clearly overestimated.
    %The plotted model is a coarse grid ($\lambda/\Delta \lambda \approx 300$) from \cloudy to allow continuous absorption, and individual absorption lines are not fully resolved.
    \textit{Right:} comparison between the observed Balmer break strength in the varying absorption spectrum (represented by the ratio on the $y$ axis) and the prediction from single-cloud photoionization models at $\log U=-2$ and $n_{\rm H}=10^{10}~{\rm cm^{-3}}$.
    To have a physical covering factor of $C_f\leq1$, an effective column density of $N_{\rm H}>10^{22}~\rm cm^{-2}$ is needed, consistent with the prediction by the composite model.
    }
    \label{fig:break_strength}
\end{figure*}

With photoionization models, we can estimate the total amount of gas by matching the model predictions to the observed spectrum.
In this section, we describe the models computed with \cloudy \citep[v17.03,][]{ferland2017,Chatzikos_2023,Gunasekera_2025}.
The modeling approach relies on assumptions about the ionizing source and the gas conditions \citep[see e.g.,][]{juodzbalis_rosetta_2024,deugenio_lrdoutflow_2025}.
We list our model assumptions in Table~\ref{tab:cloudy_models}.
In summary, we used a model AGN SED as the ionizing source, which is from \citet{Pezzulli_2017} and has a BH mass of $M_{\rm BH}=10^{8}~M_\odot$ and an Eddington ratio of $\lambda _{\rm Edd}=0.1$ matching our virial estimates from averaging the best-fit and direct measurements listed in Table~\ref{tab:target_properties}.
We assumed a wide range of gas densities of $n_{\rm H}=10^{6-11}~{\rm cm^{-3}}$.
For a BLR origin, the typical density would be $n_{\rm H}>10^{8}~{\rm cm^{-3}}$ \citep{netzer_1990}, and $n_{\rm H}\sim10^{10}~{\rm cm^{-3}}$ would be capable of efficiently producing Balmer absorption \citep{juodzbalis_rosetta_2024,Inayoshi_maiolino_2025,ji_lrdbreak_2025}.
The ionization parameter [$U\equiv \Phi_0/(n_{\rm H}\,c)$, where $\Phi_0 $ is the flux of hydrogen ionizing photons, $n_{\rm H}$ is the hydrogen volume density, and $c$ is the speed of light] at the illuminated face of the cloud is also set to a wide range of $\log U=-4$\,-\,0.
The metallicity is set to solar.
To set the velocity parameter, $v_0$, of the absorber, we used the turbulence velocity parameter $v_{\rm turb}$ in \cloudy, which changes the escape probability of line photons accordingly.
We note that in this case, $v_{\rm turb}$ represent any motions of an ensemble of absorbing clouds on small physical scales and not necessarily the micro-turbulence, which would otherwise imply large energy dissipation unless non-dissipative mechanisms are invoked \citep{bottorff2000}.
We set $v_{\rm turb}=400$ \kms, which is roughly the maximum values allowed by the observed width of the main absorption trough.
We also tested other values as detailed later.
The photoionization simulations are stopped at a maximum column density of $N_{\rm H}=10^{23}~{\rm cm^{-2}}$, and for each model, cumulative column densities of atoms and ions at different energy levels are saved.
Our goal here is to see if any of these models contain a certain depth within the cloud that can reproduce our measurements of the dynamic absorber in Table~\ref{tab:abs_params}.

The column density of hydrogen at $n=2$ alone cannot solve for $n_{\rm H}$ and $\log U$.
Adding absorption from another atom/ion with very different excitation/ionization energy would help the constraint, and \hei$\lambda 3889$ is a frequently used probe since its lower level is a metastable triplet state at $\sim 20$ eV above the ground level and He/H is less dependent on the overall metallicity compared to metals \citep{juodzbalis_rosetta_2024}.
As we mentioned in Section~\ref{sec:data}, no \hei$\lambda 3889$ absorption is seen in the dynamic absorber.
{The non detection yields an upper limit on the column density of \hei\ in the metastable triplet state $2^3S$, which we found to be $\log N_{\rm HeI^*(2^3S)}<14.6~[\rm cm^{-2}]$ at $3\sigma$.}
For the $n=2$ hydrogen column density, we took a value of $10^{14.7}~{\rm cm^{-2}}$ as the log-average of the major absorption components in the multiple absorber fit, and we note that the results would be largely the same if we took the average of other model fits.
To find all plausible models, we searched within each model whether there is any depth reproducing $N_{\rm H(n=2)}$ and compatible with the upper limit of $N_{\rm HeI^*(2^3S)}$ within 0.25 dex.
The results are shown in the upper panels of Figure~\ref{fig:abs_models}.
In the upper left panel, we plot the total $N_{\rm H(n=2)}$ and $N_{\rm HeI^*(2^3S)}$ of all models color coded by their densities, where models that can match our measurements within certain depths are marked with open circles.
The upper right panel further shows all plausible models in terms of their $n_{\rm H}$ and $N_{\rm H}$ at the depths that reproduce our measurements, and the models are color coded by their ionization parameters.
All models have $n_{\rm H}>10^{8}~{\rm cm^{-3}}$ consistent with a BLR origin, and their relatively low ionization of $\log U \lesssim -2$ implies the dynamic absorber might lie close to the outskirt of the BLR.
The column densities of most models are around $10^{21-23}~{\rm cm^{-2}}$, which, combined with $n_{\rm H}$, indicate extremely small effective thickness of $\Delta l=N_{\rm H}/n_{\rm H}<10^{-4}$ pc and are compatible with previous limits on the sizes of individual BLR clouds \citep{Maiolino_2010}.
Since the dynamic absorber varied within 8 months in the rest frame, the crossing timescale should be $l/V_{\rm absorber}\lesssim 8$ months, where $l=\Delta l/\epsilon _f$ is the physical length scale of the absorber with $\epsilon _f \in (0,1)$ being the LOS filling factor.
With a characteristic inflow velocity of $V_{\rm absorber}=|dv|+\sqrt{2}v_0 \approx 1800$ \kms for the major absorption trough, the crossing length scale is $l_{\rm cross}\sim 10^{-3}$ pc, compatible with our constraints on $l\lesssim 10^{-4}/\epsilon _f$ pc.

With the above constraints on $U$ and $n_{\rm H}$, we can also estimate the radial location of the dynamic absorber from the accretion disc using
\begin{equation}
    r_{\rm abs} = \sqrt{Q_0 /(4\pi \,c\,n_{\rm H}\,U)},
\end{equation}
where $Q_0$ is the hydrogen ionizing photon rate. We calculated the hydrogen ionizing photon rate with $Q_0=f_0L_{\rm bol}/<h\nu>$, where $f_0$ and $<h\nu>$ are the ionizing energy fraction and average energy of ionizing photons and are derived based on the model AGN SED (Table~\ref{tab:cloudy_models}).
From Figure~\ref{fig:abs_models}, if we take a model with $\log n_{\rm H}\sim 10~{\rm [cm^{-3}]}$ and $\log U=-3$, we obtained $r_{\rm abs}=1.2$ pc.
This result strongly depends on the adopted parameter especially given the wide parameter space we derived.
At $\log U=-2.2$, for example, the distance is reduced to $r_{\rm abs}=0.5$ pc.
In comparison, the size of the BLR for \target scaled from $L_{\rm bol} = 10^{45.7}$ \ergs using the single-epoch method \citep{greene_ho_2005} is roughly 0.07 pc.
{This seems to suggest that we need a very tenuous absorber with $r_{\rm abs}\gg l$ while having a considerable covering factor of $C_f\sim 0.2$.}
However, we note that the above calculation assumes no shielding of the disc radiation field, which is likely not realistic since the absorber also absorbs broad emission lines.
It is possible that the dynamic absorber only traces the neutral gas layer responsible for the Balmer absorption, and the whole absorbing gas might form a coherent structure over a larger scale and thereby achieve substantial covering of the BLR.

In addition to \hei, there are other absorption lines that can constrain the gas properties.
{In principle, since optical \feii\ absorption lines have different lower levels, their relative strengths can also be used to constrain $n_{\rm H}$ with \cloudy models.
Specifically, \feii$\lambda 5169$ and \feii$\lambda 4233$, which trace $\rm Fe^+$ levels of $b\,^4P_{5/2}$ (2.58 eV above the ground state) and $a\,^6S_{5/2}$ (2.89 eV above the ground state) respectively, have been used to constrain the gas densities in BALs \citep{Shixiheng_balqso_2016}.
In the varying spectrum shown in Figure~\ref{fig:abs_var}, we only detect \feii$\lambda 5169$ absorption.
Regardless, with the upper limit on the column density of \feii$\lambda 4233$, we found that the density constraints are compatible with those from \hei, albeit looser, which are shown in Appendix~\ref{appendix:feii_abs}.}

{Finally, we comment on the effect of $v_{\rm turb}$ in modeling.}
The value of $v_{\rm turb}$ affects the escape probability of line photons and higher (lower) $v_{\rm turb}$ generally reduces (increases) $N_{\rm (n=2)}$ at given $N_{\rm H}$ due to less (more) efficient line trapping.
For example, at $v_{\rm turb}=100$ \kms as sometimes assumed to reproduce \feii\ emission in QSOs \citep{baldwin2004}, the predicted $N_{\rm (n=2)}$ for the absorbers in \target would be reduced by a factor of 1.3. In contrast, assuming an extreme value of $v_{\rm turb}=1900$ \kms matching the blueshifted absorption component would increase $N_{\rm (n=2)}$ by roughly a factor of 1.4.
Next, we investigate whether the models reproduce the actual spectral features observed in \target.

\subsubsection{Balmer absorption and Balmer break}

As a sanity check of the models, we compare the simulated spectra from \cloudy directly with the observed absorption spectrum.
The simulated absorption spectrum is extracted by first taking the ratio between the transmitted spectrum and the incident spectrum from \cloudy, which represents the $C_f=1$ case with Equation~\ref{eq:abs}, and then correcting for $C_f$.
We then multiplied the best-fit unabsorbed spectrum for the dynamic absorber (which also includes EBLR components) with the \cloudy absorption model and obtained the final absorption spectrum.
Figure~\ref{fig:cloudy_fine_ha} shows an example of a fine-grid model ($\lambda /\Delta \lambda \approx 1.7\times 10^4$), where the model-predicted \ha absorption matches the observed one in the varying absorption spectrum well and the parameters are consistent with what we measured in Table~\ref{tab:abs_params}.

With the direct comparison, we can examine other continuum features in the observed spectrum.
As we mentioned in Section~\ref{sec:vary_abs}, there is a potential Balmer break associated with the varying absorption spectrum shown in Figure~\ref{fig:abs_var}.
A hydrogen Balmer break together with Balmer absorption is generally predicted by photoionization models as a natural consequence of bound-free absorption by $n=2$ hydrogen \citep{Inayoshi_maiolino_2025,ji_lrdbreak_2025,degraaff_lrd_2025,naidu_lrd_2025,Sneppen_lrd_2026}.
The strength of the Balmer break increases with increasing gas column densities, ionization parameter, and covering factor \citep{ji_lrdbreak_2025}.
In the left panel of Figure~\ref{fig:break_strength}, we compare the observed spectrum with a composite \cloudy model.
The model is made by combining a coarse-grid ($\lambda /\Delta \lambda\approx 300$ around \ha) model and a fine-grid ($\lambda /\Delta \lambda \approx 1.7\times 10^4$) model.
This is because in \cloudy, the continuum absorption is only saved in the coarse grid, and the fine grid saves only the line absorption (see Hazy 1, C17\footnote{\url{https://gitlab.nublado.org/cloudy/cloudy/-/wikis/home}}).
Therefore, we used the coarse grid model with all absorption lines masked as the base continuum, and the fine-grid continuum to provide the detailed absorption line shapes.
%coarse-grid ($\lambda /\Delta \lambda\approx 300$ around \ha) \cloudy model, where individual absorption lines are not  fully resolved.
%We used the coarse grid to capture the continuum absorption feature, which is not saved in the fine-grid output (see Hazy 1, C17\footnote{\url{https://gitlab.nublado.org/cloudy/cloudy/-/wikis/home}}).
%The model grid is a composite made by multiplying absorption spectra computed  using the multi-absorber parameters listed in Table~\ref{tab:abs_params}. 
The input parameters for the models are set to $n_{\rm H}=10^{10}~{\rm cm^{-3}}$ and $\log U=-2$.
We used two absorbers to describe the main absorption trough of \target given the potential presence of multiple absorbers, with $dv\sim 1,000$ \kms, $C_f=0.2$, and $N_{\rm H}=10^{21.5}~{\rm cm^{-2}}$ and $N_{\rm H}=10^{22.5}~{\rm cm^{-2}}$ for the two clouds, while ignoring the blueshifted absorber due to its strong saturation (this is further discussed in Section~\ref{xray}).
We emphasize that the above choices are not the only possibility for the dynamic absorber given the wide parameter space we derived.
The final composite model grid matches major absorption troughs in Balmer lines and produces a small Balmer break potentially tracing what is seen in the observed spectrum.
Notably, while the model roughly reproduces the optical \feii\ absorption, it overpredicts the UV \feii\ absorption, which might be related to the Fe abundance.
The detailed analysis of metal absorption is beyond the scope of this work and we leave it for future investigations.

We further quantify the Balmer break strength in the right panel of Figure~\ref{fig:break_strength}.
The Break strength is defined as the ratio at two locations on both sides of the Balmer limit avoiding absorption lines.
At each location, we set a 20 \AA-wide spectral window to calculate the mean value.
With this approach, we found a small but significant Balmer break strength of $1.075\pm 0.013$ for the varying absorber, which can only be reproduced with absorption models at $N_{\rm H}>10^{22}~{\rm cm^{-2}}$ as shown in Figure~\ref{fig:break_strength}.
%This could imply either an underestimation of the total gas column density + a strong partial covering effect where the continuum is covered more compared to lines, or additional effects that boost the apparent strength such as an intrinsic continuum variation in the UV.
Monitoring more epochs with UV-optical spectroscopy will help verify the Balmer continuum absorption, which we leave for future investigations.

{We have discussed the absorption variability over an interval of 10 months in the observed frame.
Whether such variability happens over a shorter timescale requires shorter time baselines.
As shown in Table~\ref{tab:target_obs}, the 2022 observation has two exposures separated by roughly 2 months.
These two exposures show small but significant variability of the Balmer absorption, which we discuss next.
}

\subsubsection{Short-term absorption variability in 2022}

\begin{figure*}
    \centering
    \includegraphics[width=0.49\linewidth]{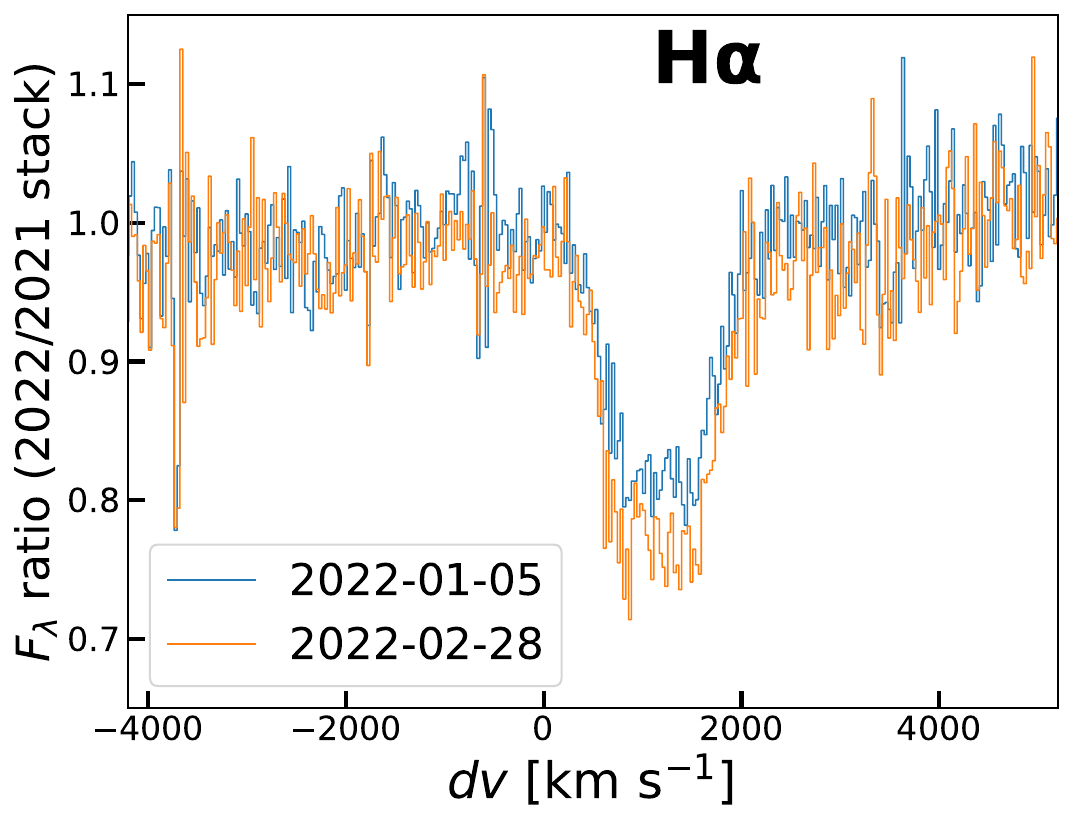}
    \includegraphics[width=0.49\linewidth]{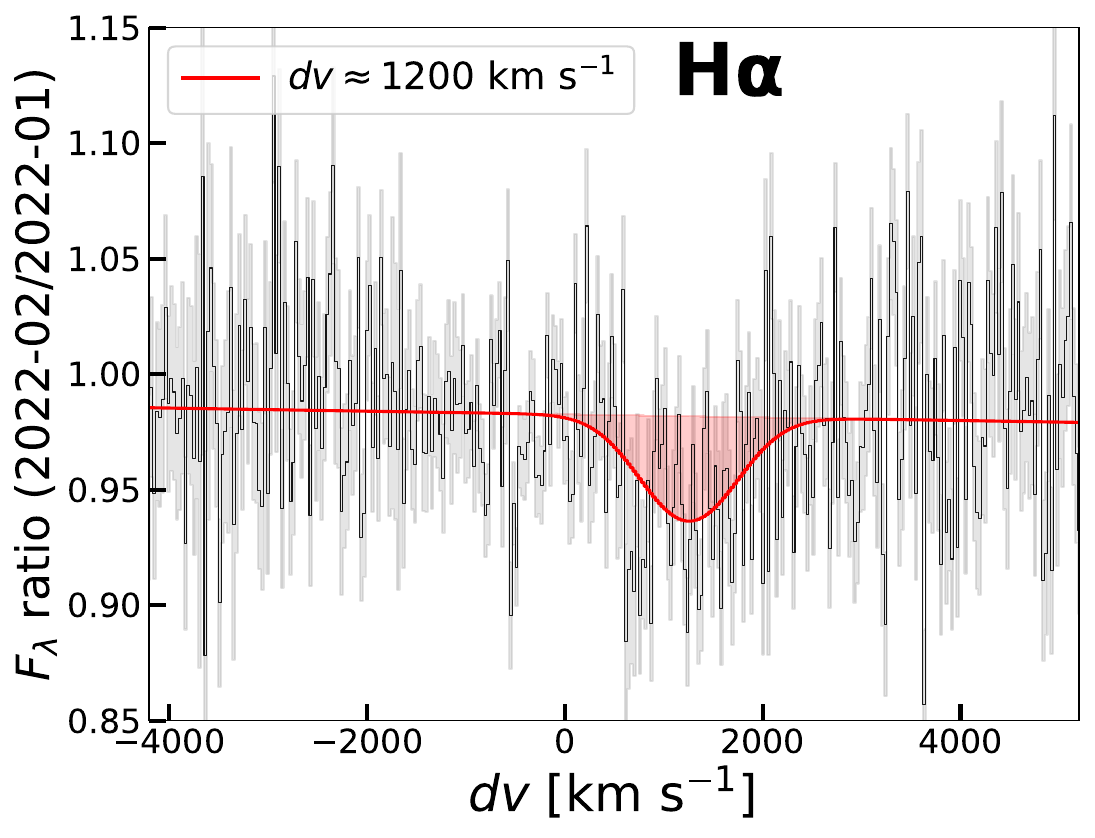}
    \caption{Absorption variability revealed by two exposures taken at 2022 separated by 54 days. \textit{Left:} Comparison between the dynamic absorbers at two epochs in 2022.
    Flux densities of both exposures are divided by $F_\lambda$ of the 2021 stack and normalized by the median flux density in 6400-6500 \AA.
    The later epoch at the $28^{\rm th}$ of February shows a slight deepening of the absorption.
    \textit{Right:} The flux density ratio between the two 2022 exposures (black) and the best-fit absorbing model (red). The \ha absorption is detected at $6\sigma$ and has a velocity offset consistent with the major dynamic absorber that occurred between 2021 and 2022.
    }
    \label{fig:exposure_var}
\end{figure*}

To investigate the short-term variability traced by the two 2022 exposures, we extracted their reduced spectra from DR1.
In the left panel of Figure~\ref{fig:exposure_var}, we compare the dynamic absorbers traced by the two exposures by taking the flux density ratios between these exposures and the stacked 2021 spectrum.
Since the absorption change is subtle, to avoid any flux calibration effects, we normalized the spectra of two 2022 exposures based on their median flux densities in the rest-frame 6400-6500 \AA.
The figure shows that the February exposure, separated from the January exposure by 54 days, shows a deepened \ha absorption.

In the right panel of Figure~\ref{fig:exposure_var}, we further quantify this absorption change by taking the flux density ratio between the two 2022 exposures.
While the noise level is high as indicated by the shaded grey region, the absorption is significantly detected with $\rm EW(H\alpha _{abs})=1.29\pm 0.20~\AA$.
In comparison, \hb absorption only shows a marginal detection at $2.4\sigma$ and no higher-order Balmer absorption is detected at $>3\sigma$.
We used the partial covering model to fit the only significantly detected \ha absorption.
The best-fit model is shown in the right panel of Figure~\ref{fig:exposure_var} and the model parameters with uncertainties from MCMC are listed in Table~\ref{tab:abs_params}.
Although the uncertainties of model parameters are large, the kinematics are broadly consistent with the main dynamic absorption between the 2021 and 2022 epochs, with a redshifted velocity of $dv\approx 1200$ \kms.
Also, the covering factor and the width of the absorption are consistent with the main dynamic absorption within $1\sigma$ uncertainties.
Based on the best-fit model, we derived a $n=2$ column density of $\log N_{\rm H(n=2)}=13.27^{+1.58}_{-0.41}~[{\rm cm^{-2}}]$.
Despite the large uncertainty, the column density is consistent with being $\sim 10\times$ smaller than that of main dynamic absorber.
Given that the main dynamic absorber occurred within a 10-month interval, and a further $\sim 10\%$ gas column density variation occurred over a 2-month interval, it might indicate that the variation was a relatively continuous process accumulated over several months.
The gas column density might therefore continue to increase after the 2022 epoch.

Thus far, we have focused our discussion on the varying absorber. Next, we discuss our measurements for the stationary absorber and compare the results with those for the varying absorber.

\section{Stationary absorber}
\label{stationary_abs}

\begin{figure}
    \centering
    \includegraphics[width=\columnwidth]{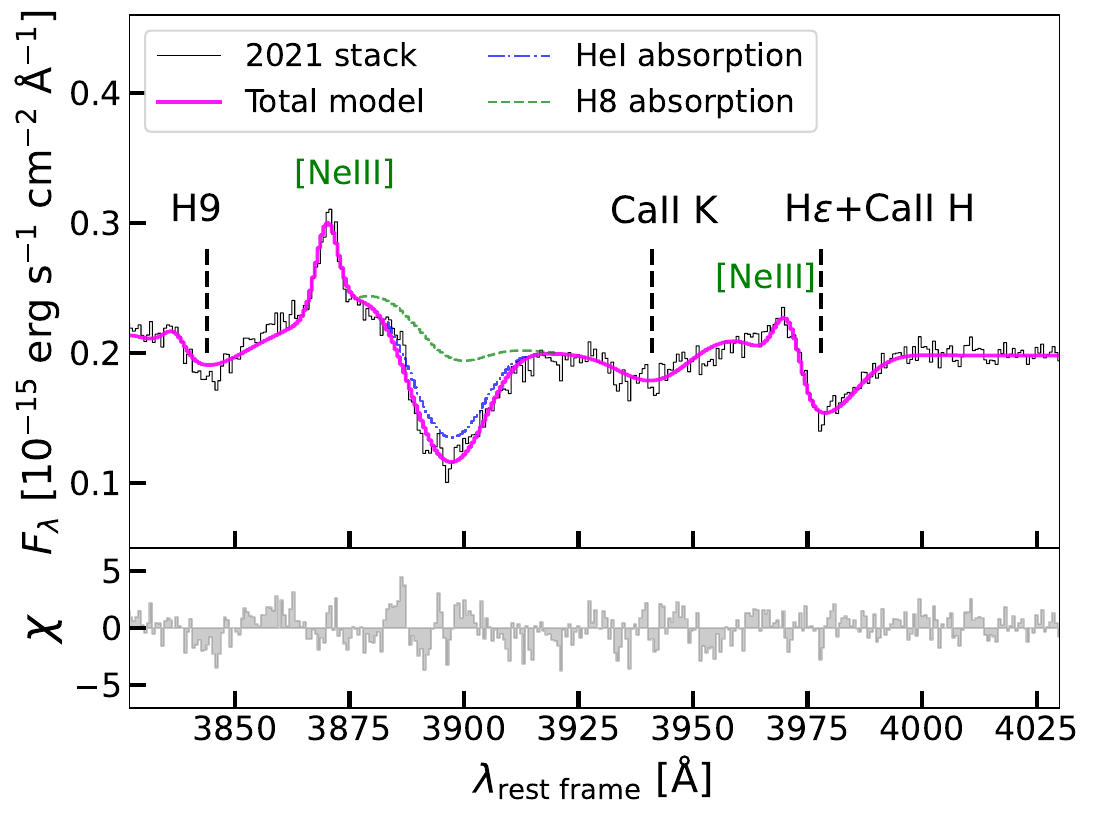}
    \caption{Best-fit single absorber model for the stacked 2021 spectrum in the region from H9 to H$\epsilon$.
    We highlight decomposition of the blended \hei+H8 absorption, where the H8 optical depth is constrained by other Balmer lines.
    Presence of the \hei absorption implies a substantial amount of excited He with $\log N_{\rm He(2^3S)}\sim ~ 15~[{\rm cm^{-2}}]$ comparable to the column density of excited hydrogen.
    }
    \label{fig:abs_sgheps}
\end{figure}

Based on the partial covering measurements in Table~\ref{tab:abs_params}, compared to the varying absorber, the stationary absorber (i.e., the absorption component that is not varying between 2021 and 2022 epochs) has comparable column densities of $n=2$ hydrogen traced by different lines.
Meanwhile, the LOS covering factor of the stationary absorber is nearly $5\times$ that of the varying absorber.
The velocity shift of the stationary absorber is 400\,-\,600 \kms less compared to the major components of the varying absorber, but still significantly redshifted compared to the systemic redshift of \target.
While one needs to be cautious as the derived parameters for the stationary absorber hinge on reliable fit of the broad lines, its physical properties, especially kinematics, are clearly different from those of the varying absorber.
The non-variability of the stationary absorber implies a higher stability of the medium with a large covering factor, in contrast to a possibly more fragmented, low-covering medium traced by the varying absorber.
Compared to other absorption-line AGN, the low covering factor of the varying absorber is more typical of FeLoBALs \citep[e.g.,][]{Leighly_2025}, whereas the high covering factor of the stationary absorber is similar to interpretations of some LRDs with high S/N absorption-line measurements \citep[e.g.,][]{juodzbalis_rosetta_2024,deugenio_irony,ji_lord_2025}.

Another noticeable feature of the stationary absorber,
as shown in Figures~\ref{fig:full_spec} and \ref{fig:abs_var}, is that it shows more absorption lines compared to the dynamic absorber.
The only identifiable absorptions in the dynamic absorber are Balmer series absorptions (including a potential Balmer break), \feii$\lambda 5169$, potential \caii\ K and H, \mgii, and some potential but weak \feii\ absorptions in the UV.
In contrast, in the stationary absorber, a prominent \hei$\lambda 3889$ absorption can be seen.
%, which traces the metastable triplet state $2^3S$.
The excitation energy of the \hei\ metastable state is roughly 20 eV from the ground state, significantly higher than the hydrogen $n=2$ excitation energy of 10 eV.
This indicates that the stationary absorber has a higher ionization parameter compared to the dynamic absorber.
%This indicates that between the 2021 and 2022 epochs, only the low-ionization absorption is varying but high-ionization (i.e., \hei) absorption remains stable.

In the literature, Balmer absorption in AGN is usually found to be accompanied by \hei absorption\footnote{In fact, \hei absorption is more frequently observed since the physical conditions to produce the line is less extreme compared to Balmer absorption (\citealp{Liu_heibal_2015,juodzbalis_rosetta_2024,wangbingjie_2024}).}.
For example, \citet{juodzbalis_rosetta_2024} studied an LRD at $z=2.26$, where \hei$\lambda 10830$ absorption implies a larger velocity offset with respect to the systemic redshift compared to the Balmer absorption.
Since the high excitation \hei likely originates in a region closer to the ionizing source, the authors infer a decelerating outflow that produces the absorption.
The earlier study by \citet{hutchings2002} noted that both \hei$\lambda 3889$ and Balmer absorptions are varying in the nearby Seyfert 1 NGC 4151, where \hei absorption had a smaller velocity offset compared to Balmer absorptions consistently over all epochs, suggesting instead an accelerating outflow origin.
Meanwhile, significant continuum variability was accompanying the line absorption variations in NGC 4151 \citep{hutchings2002}.
Notably, as shown by \citet{Choi_2022} with observations of BAL QSOs, QSOs with modest accretion rates ($\lambda _{\rm Edd}<0.3$) tend to have outflow speeds increasing with radius, whereas high-accretion rates ($\lambda _{\rm Edd}>0.3$) QSOs tend to show the opposite trend, although in these systems the outflow velocities are generally higher and reach thousands of \kms.

The non-variation of \hei in \target appears puzzling since the line should be physically closer to the ionizing source and thus response faster to any variability.
However, if the variability is not driven directly by continuum variation, as we have inferred thus far, and instead driven by varying obscuration, the non-varying \hei would imply that the obscuration change only occurs in the outer, lower-ionization/excitation region traced by hydrogen Balmer series.

To understand the ionization structure more quantitatively, we fitted the \hei\ absorption associated with the stationary absorber in the 2021 spectrum. Since \hei emission appears completely absorbed as shown in Figure~\ref{fig:full_spec}, we set priors for the fit based on our model for the stationary absorber. By adopting the $P(C+L)$ model, we jointly fit \hei, H9, H8, H$\epsilon$, and \caii\ K and H absorption as well as corresponding narrow and broad emission + \neiii$\lambda \lambda 3869,3968$ emission.
To reduce the number of parameters, we tied $\tau _0$ of Balmer absorption according to their wavelengths and oscillator strengths.
%We found that if we set $C_f$ free, we obtained $C_f\approx 0.23$ and $\log N_{\rm He(2^3S)}=13.8\pm 0.1~[{\rm cm^{-2}}]$; if 
We also constrained {$0.95<C_f<1$} according to our best-fit model of the stationary absorption based on \ha and \hb. The best-fit model is shown in Figure~\ref{fig:abs_sgheps}, and we obtained $\log N_{\rm He(2^3S)}=14.92\pm 0.04~[{\rm cm^{-2}}]$.
%As a result, the product of $C_fN_{\rm He(2^3S)}$ remains roughly constant.
%There is a large uncertainty in $\log N_{\rm He(2^3S)}$ due to a general lack of constraints on line emission over this wavelength regime.

%Regardless, considering a broad range of $\log N_{\rm He(2^3S)}\sim 13.2-13.8~[{\rm cm^{-2}}]$ can still provide useful constraints on the physical conditions of the absorber.
%As shown by \citet{juodzbalis_rosetta_2024}, if one 
Assuming that the \hei and hydrogen Balmer absorptions in the 2021 spectrum come from the same stationary absorber, we constrained $n_{\rm H}$ and $\log U$ for the stationary absorber as we did in Section~\ref{sec:vary_abs}, despite that in this case we had a detection of the \hei absorption rather than an upper limit.
%Adopting the same approach as \citet{juodzbalis_rosetta_2024}, we examined a large grid of \cloudy models with $-2.5\leq \log U \leq 0$ and $10^6\leq n_{\rm H}~[{\rm cm^{-3}}]\leq 10^{11}$, which are all stopped at $\log N_{\rm H}= 23~[{\rm cm^{-2}}]$.
%The rest of the parameters are the same as listed in Table~\ref{tab:cloudy_models} with $v_{\rm turb}=400$ \kms.
%To find all plausible models, we search within each model to find at which depths $\log N_{\rm He(2^3S)}= 14.9~[{\rm cm^{-2}}]$ and $\log N_{\rm H(n=2)}= 14.2~[{\rm cm^{-2}}]$ (i.e., average of the \ha- and \hb-based values).
%As long as the two depths agree within 0.5 dex, we consider the model as a plausible model.
To select all plausible models, we searched within each model to find at which depth the column densities of excited \hei and H are closest to our best-fit values of $\log N_{\rm He(2^3S)}= 14.9~[{\rm cm^{-2}}]$ and $\log N_{\rm H(n=2)}= 14.2~[{\rm cm^{-2}}]$ (i.e., average of the \ha- and \hb-based values).
We used $\chi ^2$ as a metric to evaluate the goodness of models, where we manually set the uncertainty of $\log N_{\rm H(n=2)}$ to 0.5 dex (i.e., the difference between the \ha- and \hb-based values divided by 2) to account for the systematic uncertainty.
Figure~\ref{fig:abs_models} shows all plausible models with $\chi ^2<5$ for the stationary absorber, where solutions with high densities ($n_{\rm H}\gtrsim 10^{8}~{\rm cm^{-3}}$) and moderate column densities ($N _{\rm H}\sim 10^{22}~{\rm cm^{-2}}$) are preferred.
As a result, the effective thickness of the stationary absorber is also extremely small, with most models having $\Delta l = N_{\rm H}/n_{\rm H}\lesssim 10^{-4}$ pc.
%Since the absorber remains stable for nearly 8 months in the rest frame, the crossing timescale should be $l/V_{\rm absorber}>8$ months.
%With a characteristic inflow velocity of $V_{\rm absorber}=|dv|+\sqrt{2}v_0 \approx 1200$ \kms, the crossing length scale is $l_{\rm cross}\sim 10^{-3}$ pc.
Combining $\Delta l$ with a crossing lengths scale of $l_{\rm cross}\sim 10^{-3}$ pc for 8 months in the rest frame, the stationary absorber probably has a filling factor of $\epsilon _f \lesssim \Delta l/l_{\rm cross} \lesssim 10^{-1}$ to remain stable.
Given the range of $\Delta l$, the filling factor can be as low as $\epsilon _f \sim 10^{-5}$,
which is still possible for BLR clouds \citep{netzer_1990}.
Alternatively, the single cloudlet approximation we adopted with \cloudy might not be exact to describe the absorber, which can undergo self-shielding for sufficiently large $C_f$ \citep{Gaskell_2009}.
%Either case, the actual physical size of the stationary absorber needs to be large enough.

The self-shielding scenario might also explain the lack of \hei$\lambda 3889$ absorption in the dynamic absorber as mentioned in Section~\ref{sec:vary_abs}.
%Without shielding, the implied ionization parameter should be sufficiently low ($\log U\ll -4$) in the varying absorber.
%If the Balmer break feature in the varying absorber is real, a single cloudlet with an extremely low ionization parameter cannot reproduce the observed strength by \cloudy models.
In the self-shielding scenario, the varying absorber is simply a nearly neutral layer of a radially extended cloud structure and therefore has little \hei\ from recombination.

\begin{comment}

In addition to the physical size, we can infer the radial location of the stationary absorber from the accretion disc based on $U$ and $n_{\rm H}$ using
\begin{equation}
    r_{\rm abs} = \sqrt{Q_0 /(4\pi \,c\,n_{\rm H}\,U)},
\end{equation}
where $Q_0$ is the hydrogen ionizing photon rate. We calculated the hydrogen ionizing photon rate with $Q_0=f_0L_{\rm bol}/<h\nu>$, where $f_0$ and $<h\nu>$ are the ionizing energy fraction and average energy of ionizing photons and are derived based on the model AGN SED (Table~\ref{tab:cloudy_models}).
    
\end{comment}
The overall higher ionization of the stationary absorber indicates that it lies closer to the accretion disc compared to the dynamic absorber.
From Figure~\ref{fig:abs_models}, the best-fit model (i.e., the largest symbol) has $\log n_{\rm H}=10.5~{\rm [cm^{-3}]}$ and $\log U=-1$, which gives $r_{\rm abs}=0.07$ pc, which is roughly the same as the BLR radius inferred from the single-epoch method for \target.
%Meanwhile, the size of the BLR for \target scaled from $L_{\rm bol} = 10^{45.7}$ \ergs using the single-epoch method \citep{greene_ho_2005} is roughly 0.07 pc.
%The above consistency again suggests that the absorber lies within the BLR.
For all models with $\chi ^2 < 5$, $r_{\rm in}$ spans a wide range from $10^{-2}$ pc to $7.7$ pc.

To better understand the structure of the absorbers, a useful and complementary probe is the X-ray.
The absorption spectrum of X-rays provides key information on small-scale LOS obscuration down to the hot corona.
Next, we discuss X-ray observations of \target.

\begin{figure*}
    \centering
    \includegraphics[width=510pt]{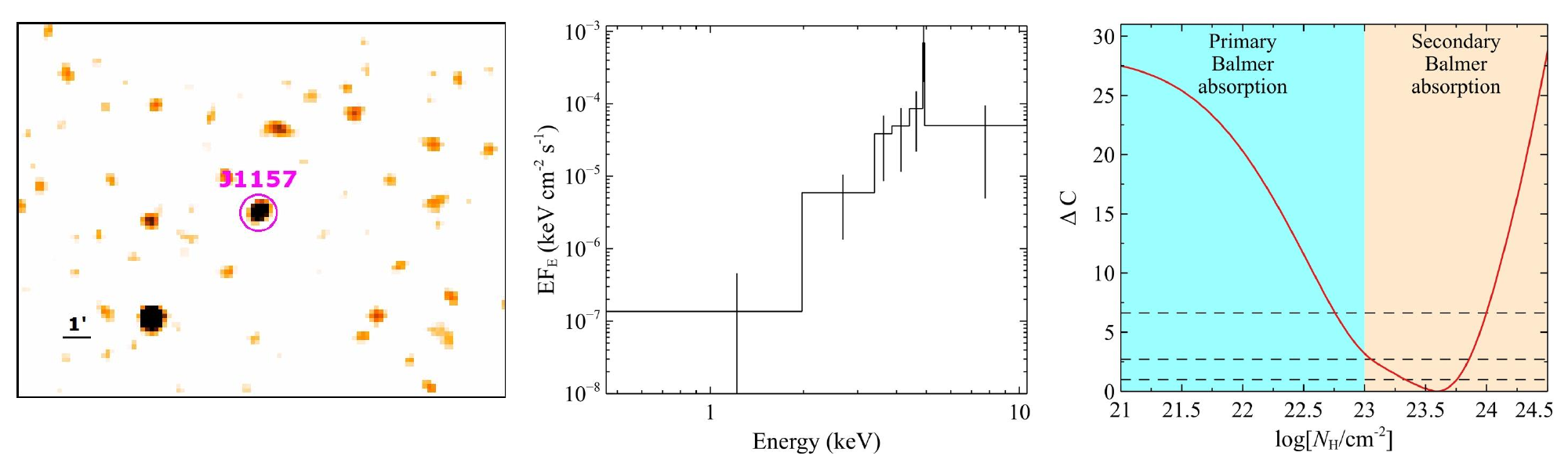}
    \caption{
    New X-ray observations of \target.
    \textit{Left:} the 2--10\,keV X-ray image of the field around \target\ from \EP/FXT (both FXT modules combined), showing the clear detection of the source ($>$3$\sigma$ significance). \textit{Middle:} the X-ray spectrum extracted from the \EP/FXT data (again, both FXT modules combined), rebinned for visual clarity and unfolded through a model that is constant with energy. Though the data are low S/N, the spectrum is clearly
    rising over the FXT bandpass; such a hard spectrum is reminiscent of heavily obscured AGN. \textit{Right:} constraints on the X-ray column density based on a simple absorbed powerlaw model for which we limited $1 \leq \Gamma \leq 3$ and assumed a fully-covering neutral absorber. The dashed lines indicate the thresholds for the 67, 90 and 99\% confidence intervals, respectively. The X-ray column is larger than the primary Balmer absorption components (cyan shading; see Table~\ref{tab:abs_params}), but consistent with the secondary (blueshifted) Balmer compoennt (orange shading).
%    \redtxt{[need to update the blue band based on Figure 7; other adjustments?]}
    }
    \label{fig:xrays}
\end{figure*}

\section{X-ray Constraints}
\label{xray}

\begin{figure}
    \centering
    \includegraphics[width=\columnwidth]{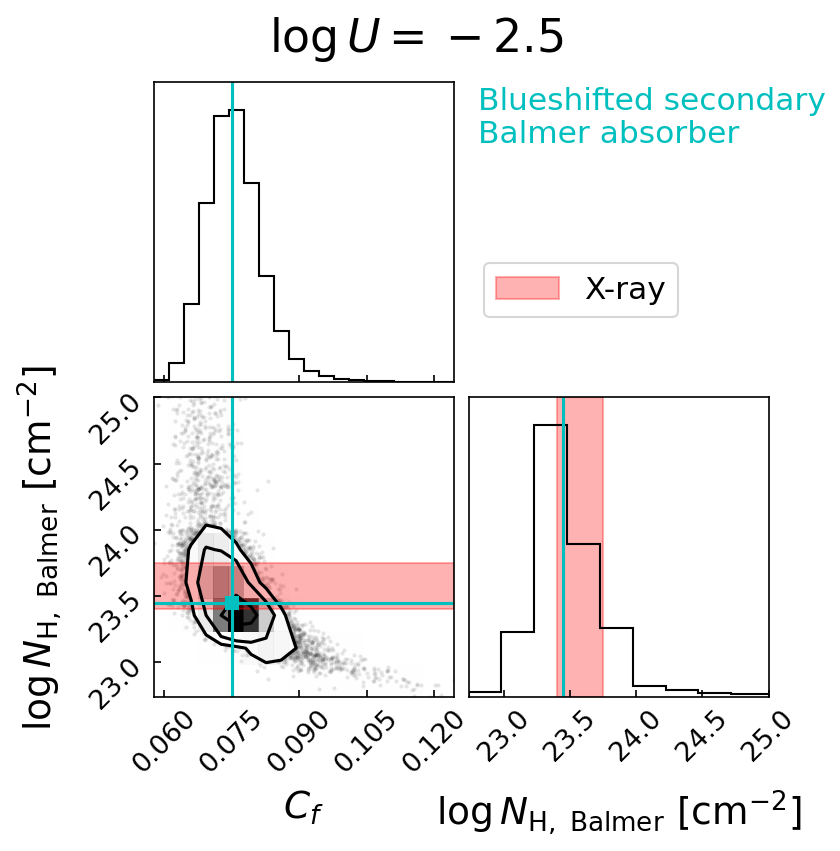}
    \caption{Posterior distributions for the covering factor and the hydrogen column density from 100,000-step MCMC runs for the blueshifted Balmer absorber within the multi-component dynamic absorber model shown in Figure~\ref{fig:vel_com}.
    The column density is converted from the $n=2$ column density with \cloudy models (see Section~\ref{subsec:cloudy_models}).
    The shaded region corresponds to the 67\% confidence interval for the X-ray-based gas column density, which overlaps with the median $N_{\rm H}$ inferred from the blueshifted Balmer absorber.
    This implies that the blueshifted Balmer absorber might trace the same gas responsible for the X-ray obscuration.
    }
    \label{fig:babs_subcorner}
\end{figure}

In order to search for X-ray emission from \target\ we observed the source with the Follow-up X-ray Telescope (FXT) on board the \EP\ (\citealt{ep,ep_2025}). The observation (OBSID 11900754048) was conducted during the course of 5\,-\,6 June 2026 for a total on-source exposure of 10.2\,ks (collected as a series of shorter snapshots). We processed the cleaned event files produced as standard by the \EP\ team, which were themselves produced from the raw events files with \textsc{fxtchain} (part of the \textsc{fxtdas} v1.30 software package; \citealt{EP_FXTDAS}) with the default settings. The 2\,-\,10\,keV (observed frame) image from both of the FXTA and FXTB modules combined, extracted using \textsc{xselect}, clearly shows a source at the known position of \target, which is plotted in the left panel of Figure \ref{fig:xrays}. 
We subsequently extracted spectra for \target\ from each FXT unit using a circular region of radius 30$''$, and estimated the background from a larger region of blank sky adjacent to the source position, again using \textsc{xselect}. Instrumental responses were generated using \textsc{fxtarfgen} and \textsc{fxtrmfgen}, and the spectra from the two FXT units were then combined using \textsc{addspec}.

Although the source is significantly detected (at $>$3$\sigma$ significance, based on \citealt{Kraft91}), the net spectrum only contains $\sim$12 counts in total. We therefore rebinned the spectrum to a minimum of 1 count per bin, and analyzed the data by reducing the Cash statistic (suitable for low-count data; \citealt{cstat}) using \textsc{xspec} (\citealt{xspec}). The FXT data cover the $\sim$0.5\,-\,10.0\,keV band, and the spectrum is shown in the middle panel of Figure \ref{fig:xrays}. Although the S/N is low, the spectrum is clearly very hard over the FXT bandpass, reminiscent of other AGN that are heavily obscured in the X-ray band (e.g., \citealt{Guainazzi02, Comastri10, Rivers15_7582, Walton18, Walton22, TorresAlba25}).

We therefore model the data over the $\sim$0.5--10\,keV band using a simple absorbed power-law model. The absorption is modeled with the \textsc{tbabs} code (\citealt{tbabs}), and is assumed to be neutral, fully covering, and to have solar abundances. We also set the redshift of the absorber to the systemic redshift of \target\ (i.e., $z = 0.287$). We also account for the expected losses due to scattering for a given column density with a \textsc{cabs} component, which has its column density linked to that of the neutral absorber. The full \textsc{xspec} model expression is thus \textsc{tbabs} $\times$ \textsc{cabs} $\times$ \textsc{powerlaw}. Our initial modeling found the photon index of the power-law continuum to be essentially unconstrained, so we limited this to the range $1.0 \leq \Gamma \leq 3.0$, suitable for sub-Eddington AGN (e.g., \citealt{Ricci17}). With this setup, we find the gas column density traced by X-rays to be $\log[N_{\rm{H}}/\rm{cm}^{-2}] = 23.6^{+0.15}_{-0.2}$ (at 67\% confidence; see the right panel of Figure \ref{fig:xrays}), {more than a factor of $\sim$10 higher than the possible column densities inferred from most models for the primary Balmer absorption components (c.f., Figure~\ref{fig:abs_models}), although marginally consistent with the highest model value at $N_{\rm H}\approx 10^{23}~{\rm cm^{-2}}$ if the gas density is $n_{\rm H}\sim 10^{8.5}~{\rm cm^{-3}}$.}

The ``observed'' 2\,-\,10\,keV (i.e., evaluated in the rest frame of \target but not corrected for gas obscuration) is
$\log[F_{\rm{2-10}} / \rm{erg~cm^{-2}~s^{-1}}] = -13.0^{+0.1}_{-0.2}$, which corresponds to a luminosity of $\log[L_{\rm{2-10}} / \rm{erg~s^{-1}}] = 43.3^{+0.1}_{-0.2}$. Taking this luminosity at face value and combining it with the 5100\AA-based bolometric luminosity reported in Table \ref{tab:target_properties} would imply a 2\,-\,10\,keV bolometric correction of $\kappa_{2-10} \sim 250$. 
This is significantly larger than what would be expected regardless of whether one adopts a scenario in which $\kappa_{2-10}$ scales with Eddington ratio or with bolometric luminosity (both of which predict $\kappa_{2-10} \sim 25$ for \target; \citealt{Duras_2020}), implying significant X-ray weakness compared to normal AGN. 
However, if we correct the observed X-rays for the losses (both photoelectric and scattering) due to the LOS gas obscuration, we find the intrinsic 2\,-\,10\,keV luminosity to be $\log[L_{\rm{int,2-10}} / \rm{erg~s^{-1}}] = 44.6^{+0.1}_{-0.7}$. This would imply an intrinsic bolometric correction of $\kappa_{2-10} \sim 15$, consistent with the expectation for normal AGN \citep{Duras_2020}. As such, in this case the observed X-ray weakness is entirely consistent with being due to the level of X-ray obscuration present.

The $\sim 1$ dex discrepancy between the optical-based $N_{\rm{H}}$ and X-ray-based $N_{\rm{H}}$ is puzzling.
We consider three possible scenarios that lead to the discrepancy: 1) since the X-ray observations were taken in a more recent epoch (2026) compared to the optical observations (2021\,-\,2022), a significant enhancement of gas obscuration might occur in the current epoch; 2) since the optical Balmer absorption traces neutral gas obscuration outside the line emitting region, there might be a warm absorber closer to the accretion disc only seen in the X-rays; 3) since the optical Balmer absorption is most sensitive to dense ($n_{\rm H}\gtrsim 10^{8}~{\rm cm^{-3}}$) and warm ($T\sim 10^4$ K) gas \citep{juodzbalis_rosetta_2024}, the X-rays might instead see extra columns of diffuse, low-density, and/or low-temperature gas distributed over a larger physical scale even beyond the BLR.

Scenario 1) can be tested with multi-epoch X-ray observations.
While the \EP\ observation discussed above is the only pointed X-ray observation of \target\ to date, the source also lies in the German half of the \textit{eROSITA} survey data (\citealt{erosita_dr1}). However, \target\ is not detected in the eRASS1 data (accumulated between Dec 2019 - June 2020). 
We therefore instead determined the upper limit on the 0.2\,-\,5.0\,keV flux in the observed frame from \textit{eROSITA} using the upper limit server released by the \textit{eROSITA} team (as detailed in \citealt{EROSITA_DR1_UL})\footnote{\url{https://erosita.mpe.mpg.de/dr1/erodat/upperlimit/single/}}; we found $F_{\rm{obs,0.2-5}} < 1.25 \times 10^{-13}$\,erg~cm$^{-2}$~s$^{-1}$. The best-fit model to the 2026 \EP\ data gives an observed flux in this band of $F_{\rm{obs,0.2-5}} \sim 4.5 \times 10^{-14}$\,erg~cm$^{-2}$~s$^{-1}$, consistent with the earlier \textit{eROSITA} limit. However, we note that if we decrease the column density observed in the X-ray band to $\log[N_{\rm{H}}/\rm{cm}^{-2}] = 22$, comparable to the primary Balmer absorption components, then \target\ \textit{should} have been detected by \textit{eROSITA}, assuming $\Gamma = 2$ and no intrinsic variation in flux between the \textit{eROSITA} and \EP\ epochs. This suggests that the large X-ray column seen in the \EP\ data probably does not represent a transient enhancement in $N_{\rm{H}}$ in the current epoch, although further monitoring will be needed to conclusively rule out this scenario.

Scenario 2) would imply again differential gas obscuration where the Balmer-absorbing gas only traces part of the X-ray obscuring gas.
As shown in Figure~\ref{fig:vel_com}, in the varying Balmer absorber, there exists a low-covering yet strongly saturated component with a velocity offset of {$dv\sim -300$} \kms.
The fact that this absorption is kinematically distinct from the main absorption trough and only visible in higher-order Balmer lines with upper levels $n>3$ imply a potentially different spatial location and strong optical-depth effects.
%This is also consistent with its low covering factor of $C_f \approx 8\%$, since it becomes difficult for an absorber to cover BLR clouds if it lies at a small radial distance.
An immediate question is whether this Balmer absorber is related to the X-ray obscuration.
In Figure~\ref{fig:babs_subcorner}, we compare the gas column density inferred from the blueshifted Balmer absorption with that inferred from the X-rays.
We show posterior distributions of both $C_f$ and $N_{\rm H}$ since they are correlated.
As in Section~\ref{subsec:cloudy_models}, $N_{\rm H}$ is calculated from $N_{\rm H(n=2)}$ based on photoionization models.
We consider models with $v_{\rm turb}=1900$ \kms, $n_{\rm H}=10^{10}~{\rm cm^{-3}}$, and different ionization parameters.
We only computed models up to $N_{\rm H}=10^{24.5}~{\rm cm^{-2}}$, but extrapolated the $\log N_{\rm H}$-$\log N_{\rm H(n=2)}$ relation linearly at higher column densities\footnote{At higher column densities, scattering optical depths might start to become important and more sophisticated radiative transfer treatment is needed, which is beyond the capability of \cloudy modeling here.}.

We found that at $\log U \sim -2.5$,
the median column density inferred from Balmer absorption is $\log N_{\rm H} = 23.4~{\rm [cm^{-2}]}$ (not affected by the extrapolation), which is compatible with the X-ray-based column density within 90\% confidence interval as shown in Figure~\ref{fig:babs_subcorner}.
At $\log U \gtrsim -2$, the median column density would be lower than the X-ray-based range; at $\log U \lesssim -3$, the median column density would be higher than the X-ray-based range.
The range of the ionization parameter is consistent with that of the redshifted absorption trough in the dynamic absorber inferred in Section~\ref{sec:vary_abs}.
The consistent column densities at $\log U\sim -2.5$ indicate that a high column density from a blueshifted absorber is a plausible solution, although the above calculation relies on our model assumptions.
%Regarding the dependency on $U$, at $\log U=-3$, the measured $N_{\rm H(n=2)}$ cannot be achieved; at $\log U=-1$, the median column density for the blueshifted absorber is reduced to $\log N_{\rm H} = 22.9~{\rm [cm^{-2}]}$ outside the 90\% confidence interval constrained by X-rays.
%Due to the absence of varying \hei absorption, the dynamic absorber should have a lower ionization compared to the stationary absorber, favoring $\log U<-1$.

Finally, in Scenario 3), the X-ray obscuring gas is distributed on a physical scale large than the Balmer absorber and potentially beyond the BLR, and likely has a low ionization parameter and/or a low density.
As a result, according to \cloudy models shown in the left panels of Figure~\ref{fig:abs_models}, not enough hydrogen can be excited to $n=2$, and the gas would become invisible to the optical Balmer absorption.
Two questions need to be addressed in this scenario: 1) how (or whether) the large-scale and low-density gas manage to produce a high covering factor in X-rays; 2) how such a large column density of gas is accumulated on a large physical scale.
To answer Question 1), deeper X-ray observations are needed to determine $C_f$. Regardless, if $C_f$ is much lower than unity, $N_{\rm H}$ should be even higher, which leads to Question 2) of how to produce the nearly Compton-thick obscuration beyond on large scales.
While strong obscuration ($N_{\rm H}\gtrsim 10^{24}~{\rm cm^{-2}}$) in AGN usually occur on small scales within the nuclear torus ($<10$ pc), mergers can lead to Compton-thick obscuration on galactic scales \citep{Hickox_2018rev}.
For example, simulations of gas-rich major mergers by \citet{Blecha_2018} show that such strong obscuration could occur at post-merger times, when the separation of galaxy cores is $\lesssim 1-10$ kpc.
The optical image of \target shown in Figure~\ref{fig:full_spec} does exhibit a tidal tail-like feature in the north, and the absence of obviously separated bright cores (note that $1''$ $\sim$ 4.5 kpc at $z=0.287$) suggests a post-merger stage.
To better understand the potential merger event, we perform an image decomposition to investigate the host galaxy and its past evolution in the next section.

\IfFileExists{imaging.tex}{\input{imaging}}{}

\section{Discussion}
\label{sec:discuss}

Thus far, we have performed different model fits to the absorption spectrum of \target.
In this section, we discuss the interpretation of our analyses in a broader context, including the physical origin of the absorber and the comparison with other broad-line sources with Balmer absorption.

\subsection{Physical origin of the absorbers}

The case of \target provides a unique opportunity to study the physical origin of the absorber based on both single-epoch spectral measurements and multi-epoch dynamical arguments.
As we have shown, to produce strong Balmer absorption in both stationary and varying absorbers, high gas densities are required. This in turn implies small effective thickness of the absorber, which we found to be $\Delta l = N_{\rm H}/n_{\rm H}\lesssim 10^{-4}$ pc.
The non-variation of the stationary absorber further implies a total physical scale of $\gtrsim 10^{-3}$ pc and a filling factor of $\epsilon _f\lesssim 10^{-1}$.
%Therefore, a low filling factor and/or stratified self-shielding structure is likely needed for both absorbers, especially the stationary one to keep it stable.
%An independent constraint on the physical scale of the varying absorber comes from the time interval.
%The physical scale of the varying absorber can also be constrained with the time interval between two observations.
%Taking the $\sim 8$ months rest-frame interval between the two DESI observations as the maximum crossing timescale, and taking a characteristic inflow velocity of $V_{\rm absorber}=|dv|+\sqrt{2}v_0$, the crossing scale length is $l_{\rm cross}<10^{-3}$ pc, compatible with the effective thickness.
In contrast, the physical scale of the dynamic absorber need to be smaller than the crossing lengths scale of $l_{\rm cross}<10^{-3}$ pc, which is also consistent with our derived values based on absorption lines in Section~\ref{sec:vary_abs}.
As a result, the variability nature of \target allows us to actually probe extremely small scale clouds through their dynamics.
A similar investigation based on X-ray spectroscopy by \citet{Maiolino_2010} found a dense obscuring medium on scales of $10^{-5}$ pc in the BLR of NGC 1365, which the authors interpreted as extended structures from BLR clouds and inferred a lifetime of a few months for each BLR cloud.
The similar physical and time scales involved in \target suggest an origin within the BLR, or at least on a comparable scale for the absorbers.

Given a bolometric luminosity of $L_{\rm bol}\approx 9\,L_{5100}\approx 10^{45.7}$ \ergs \citep{sternlaor_2012}, one expects a characteristic BLR size of roughly 0.07 pc, where the velocity of the Keplerian motion is roughly 2,700\,-\,4,200 \kms (where the uncertainty mainly comes from the line decomposition given the presence of the absorption; see Table~\ref{tab:target_properties}).
In comparison, the velocity offset of the absorber is $dv\sim 400-1,000$ \kms and the characteristic velocities are $V_{\rm absorber}=|dv|+\sqrt{2}v_0\approx$ 1,200 \kms for the stationary absorber and 1,900 \kms for the varying absorber, if they are associated with gas inflows.
To block the LOS, instead of Keplerian motions,
the absorber velocities are more likely related to the cloud motions perpendicular to the orbital plane or pure radial inflows, which could have velocities of $\sim 1/2\,v_{\rm Keplerian}$ or $\sim 1/4\,v_{\rm Keplerian}$ \citep{Osterbrock_1978,Gaskell_2009}, which can roughly match our measurements.

The non-orbital motions of the clouds can produce LOS obscuration if the inclination angle is $i>{\rm arctan(R/H)}$ (where H and R are scale heights of the clouds and R is the typical orbital radius; $i\approx 40^\circ$ according to the model of \citealp{Osterbrock_1978}).
In this case, \target represents a special viewing angle of ordinary QSOs, which, however, does not explain the rarity of absorption-line QSOs, whose number is only $\sim 10\%$ at low redshift compared to the optically selected QSO sample \citep{Trump_2006}.
The Balmer absorbed QSOs are even rarer. Within the DESI DR1 \texttt{QSO} sample at $z<0.5$ with $\rm S/N(H\alpha)>3$, the detection rate is only 0.03\% based our selection in Section~\ref{sec:data}.
In addition, the inferred column density of the major optical absorber is insufficient to extinguish X-ray emission in \target.
As we discussed in Section~\ref{xray}, while this might be explained with a kinematically distinct absorber with a high column density and a low covering factor, it might also imply a large scale obscuration with lower density gas occurs during the post-merger stage.
The merger might also have led to enhanced gas inflow traced by the absorption lines, suggesting that \target could trace an active feeding phase rather than simply a special viewing angle.

For absorbers with a BLR origin, their scale heights and the covering factors need to be larger than those of the dusty torus, otherwise one should see a typical Type 2 AGN rather than an absorbed Type 1 AGN.
This is also supported by the moderate dust attenuation we inferred.
As a result, \target appears as an AGN with a bloated BLR structure, possibly related to the active inflow seen in the spectra.
Another order-of-magnitude comparison can be made for the mass inflow rate and the accretion rate of the BH.
The mass inflow rate is given by
\begin{equation}
    \dot{M}_{\rm inflow}=4\pi C_{f,\rm ave}r_{\rm inflow}^2\dot{N_{\rm H}}\mu m_{\rm H},
\end{equation}
where $C_{f,\rm ave}$ is the covering factor average over the $4\pi$ angle (can be different from the LOS $C_f$), $r_{\rm inflow}$ is the characteristic inflow radius, $\dot{N_{\rm H}}$ is the inflow rate of gas column density, $\mu \approx1.22$ is the mean molecular weight, and $m_{\rm H}$ is the mass of a hydrogen atom.
Taking a BLR origin with $r_{\rm inflow}\sim 0.07$ pc and $\dot{N_{\rm H}}\sim 10^{22}~[{\rm cm^{-2}}]/0.67~{\rm yr}$ traced by the dynamical absorber, we obtained $\dot{M}_{\rm inflow}\sim C_{f,\rm ave}\times 9~M_\odot~{\rm yr^{-1}}$.
Meanwhile, assuming a radiative efficiency of $\eta=0.1$, the mass accretion rate of the BH in \target based on single-epoch measurements is $\dot{M}_{\rm accretion}\approx 1.5~M_\odot~{\rm yr^{-1}}$.
Interestingly, if the inflow is directly feeding the BH, the average covering factor would be $C_{f,\rm ave}\sim 15\%$, close to the LOS value we measured.
{Combining the derived $C_{f,\rm ave}$ for \target with the occurrence rate of $\sim 0.03\%$ for the LOS Balmer-absorbed QSOs in DESI DR1 implies an intrinsic occurrence rate of $0.03\%/15\%=0.2\%$ for the Balmer-absorbed QSOs.
This is significantly lower than the BAL fraction \citep{Trump_2006} and even $10\times$ lower than the intrinsic FeLoBAL fraction at $z\lesssim 2$ inferred by \citet{Dai_lobalfraction_2012}.
If the small fraction corresponds to a small timescale, this could indicate a short phase of fast accretion after a gas-rich merger, which happened before the QSO entering the blowout phase with strong outflows.
Still, we emphasize that this interpretation is currently based on a single target + a small sample without properly accounting for sample variance and completeness, which we will investigate in future work.
}

The relative locations of the stationary and varying absorbers are also important.
Compared to the stationary absorber, the varying absorber shows a higher velocity offset.
Given the low-ionization nature of the varying absorber (i.e., the lack of variability in the \hei absorption), it seems to suggest that it originates in a location further away from the accretion disc compared to the stationary absorber.
As a result, should both absorbers trace an inflow, the implied inflow is slowing down since the inflow velocity decreases with decreasing radius.
%If the varying absorber actually arises in a location closer to the accretion disc, the lack of variability in the \hei absorption cannot be explained.
Such a velocity structure is unexpected for free falling clouds in the gravitational potential of the BH.
Instead, this could imply a steep density profile if the mass inflow rate [$\propto \rho (r)v$] is relatively stable over time.
The additional variation in the dynamic absorber between the two 2022 exposures shown in Figure~\ref{fig:exposure_var} suggests that the absorber might trace a continuous gas inflow that was stably building up the column density over several months.

\begin{table}
        \centering
        \caption{Derived column densities from some metal absorption lines in the dynamic absorber. Due to the weakness of metal absorption lines compared to Balmer lines in \target, we assumed a single absorber and tied the kinematics of all metal absorption lines to those of major Balmer absorption trough at $dv\sim 1,000$ \kms (see Table~\ref{tab:abs_params}).
        }
        \label{tab:met_abs_dyn}
        \begin{tabular}{l c c}
            \hline
            \hline
            Absorption line & Lower level & $\log$ column density [$\rm cm^{-2}$] \\
            \hline
            \mgii$\lambda \lambda 2796,2803$ & ${3s\,^2S_{1/2}}$ [ground] & $14.34^{+0.17}_{-0.18}$ \\
            \caii$\lambda \lambda 3934,3968$ & ${4s\,^2S_{1/2}}$ [ground] & $13.58^{+0.10}_{-0.12}$ \\
            \feii$\lambda 4233$ & ${b\,^4P_{5/2}}$ & $<15.02~(1\sigma)$ \\
             &  & $<15.29~(2\sigma)$ \\
            \feii$\lambda 5169$ & ${a\,^6S_{5/2}}$ & $14.83^{+0.09}_{-0.10}$ \\
            \hline
        \end{tabular}
\end{table}

{Another useful tracer for the physical origin of the absorber is the chemical abundance pattern.}
In the spectrum of \target, there are metal absorption lines from \mgii, \feii, and \caii. 
These ions have similar ionization potentials (IPs), which are $\rm IP(Mg^0)=7.6$ eV,  $\rm IP(Fe^0)=7.9$ eV, and $\rm IP(Ca^0)=6.1$ eV, and $\rm IP(Mg^+)=15$ eV,  $\rm IP(Fe^+)=16$ eV, and $\rm IP(Ca^+)=12$ eV.
%$\rm Fe^0$ (8 eV) and $\rm Ca^0$ (6 eV) and $\rm Fe^+$ (16 eV) and $\rm Ca^+$ (12 eV), 
Thus, in principle, one can infer the relative abundance of these elements based on their absorption lines.
However, the optical \feii\ absorption lines observed in \target generally trace excited states at $\gtrsim 2$ eV above the ground state, and thus they cannot be easily converted to the total column density of $\rm Fe^+$.
We therefore investigate the relative abundance of Mg/Ca through the UV \mgii\ absorption and \caii\ K and H absorption.
From the varying spectrum of 2022/2021, we obtained $\log N({\rm Mg^+})/N({\rm Ca^+})=0.77\pm 0.20$.
This fit was done by fixing the kinematics of the \mgii\ and \caii\ absorption to those of the major redshifted Balmer absorption component, {and our derived column densities are listed in Table~\ref{tab:met_abs_dyn}}.
For reference, the solar value is $\rm \log(Mg/Ca)_\odot = 1.26$ \citep{grevesse2010}.
In the ISM, Ca is typically more depleted onto dust grains than Mg and Fe \citep[e.g.,][]{Dwek_1980}.
The subsolar Mg/Ca of the varying absorber indicates that the absorber is nearly dust-free and lies within the dust sublimation radius, which is $r\approx 0.38$ pc assuming a dust sublimation temperature of 1,500 K.
The relative enhancement of Ca compared to Mg could further imply recent enrichment from Type Ia supernovae \citep{Kobayashi_2020}.
For the stationary absorber, since \mgii\ is not fully covered by the DESI spectrum (see Figure~\ref{fig:full_spec}), we cannot reliably retrieve the Mg abundance.
{Future follow ups in the rest-frame UV regime of \target would help to determine both the \mgii\ abundance as well as the \feii\ abundance through \feii\ absorptions that directly trace the ground state of $a\,^6D$.}

%From the varying spectrum of 2022/2021, we obtained $\log N({\rm Ca^+})/N({\rm Fe^+})=-1.24\pm 0.15$ from measurements of \caii\ K and H and \feii$\lambda 5169$.
%For reference, the solar value is $\rm \log(Ca/Fe)_\odot = -1.16$ \citep{grevesse2010}.
%If the environment of the absorber is dust-free, the subsolar Ca/Fe might be an indication of recent enrichment from the Type Ia supernovae, where Fe is more enriched than Ca \citep{Kobayashi_2020}.
%In the ISM, Ca is typically more depleted onto dust grains than Fe \citep[e.g.,][]{Dwek_1980}.
%The near-solar Ca/Fe of the varying absorber indicates that the absorber is nearly dust-free and lies within the dust sublimation radius, which is $r\approx 0.38$ pc assuming a dust sublimation temperature of 1,500 K.
%, this might be an indication that the absorber lies close to or within the dusty torus as low-ionization BLR gas \citep{netzer1983,Gaskell_2007}.
%Assuming a dust sublimation temperature of 1,500 K, the dusty torus would lie at $r\approx 0.38$ pc for \target.
%Meanwhile, for the stationary absorber probed by the 2021 spectrum, we found $\log N({\rm Ca^+})/N({\rm Fe^+})=-1.64\pm 0.05$, where we set $0.95 < C_f < 1$ when fitting \caii.
%This seems to suggest that the stationary absorber, at least its low-ionization part, is more dusty than the varying absorber, such that Ca is more depleted onto dust than Fe.

{Finally}, we note that the gas inflow is not the only plausible physical scenario.
As shown by \citet{Starkey_2023} and Huang et al. submitted, ripple- or pillar-like structures might exist on the AGN accretion discs, which could explain reverberation mapping results for accretion disc sizes.
The ``pillars'' could be produced by local heating from massive stars embedded in the accretion discs \citep{chenyixian_2024}.
While these pillars overall undergo Keplerian motions, when viewed from an angle, they produce occultation of the inner radiation, which can manifest as temporary absorption in emission lines (Huang et al. submitted).
This picture would qualitatively predict cyclic motion of the dynamic absorber and a potential change of the kinematics in the future (e.g., from redshift to blueshift), which can be tested with further monitoring of the absorption spectrum.
The chemical abundances of the pillars traced by the absorption could also be different compared to those traced by emission lines, due to the pollution by massive stars \citep{Huang_2023}.
We will investigate different theoretical models in more detail in future work.

\subsection{Comparison with other Balmer-absorbed systems}

\begin{figure}
    \centering
    \includegraphics[width=\columnwidth]{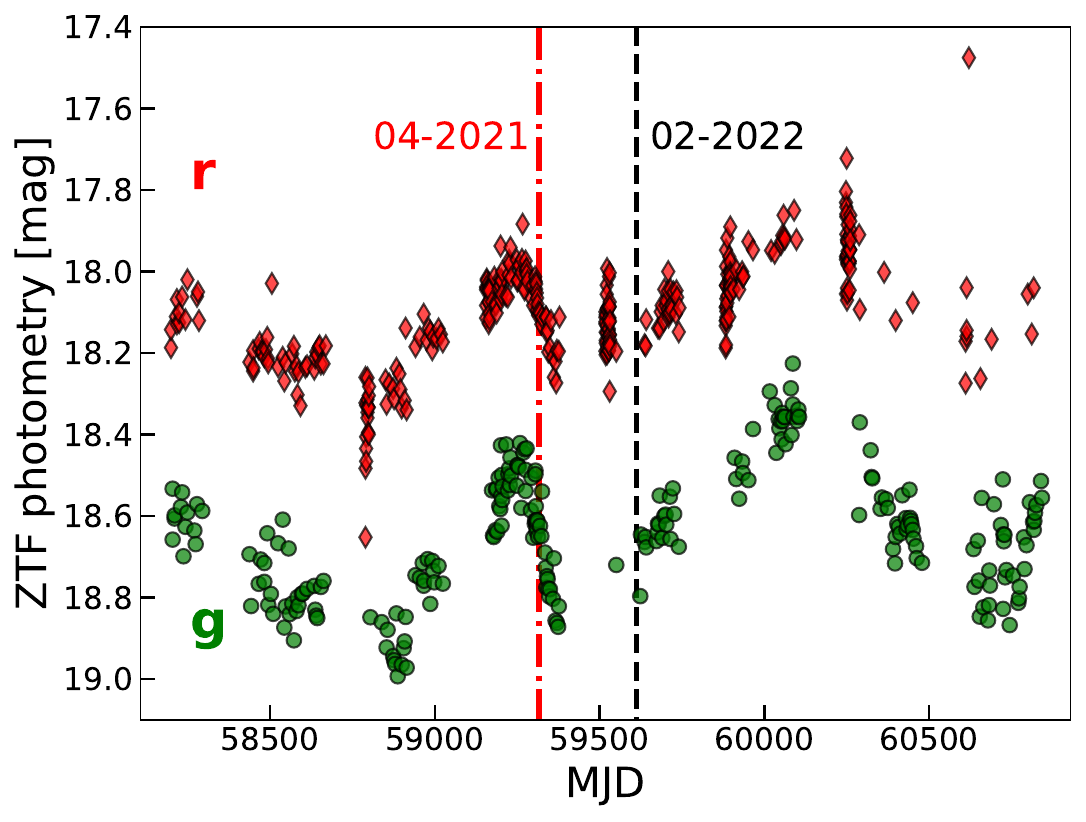}
    \caption{Broad-band light curves from ZTF for \target, where the epochs of DESI spectroscopic observations are marked with vertical lines.
    On 2021 and 2022, \target was undergoing continuum dimming and continuum brightening, respectively, although the continuum brightnesses at the two epochs happened to be consistent.
    }
    \label{fig:ztf}
\end{figure}

As we have mentioned, there are different categories of AGN that show significant Balmer absorption in observations.
With the case study of \target presented here, an immediate question is how it fits into the broader context of Balmer-absorbed AGN.

One of the well studied classes of Balmer-absorbed AGN is found within the FeLoBAL class, as discussed by \citet{Leighly_2025}.
Similar to \target, FeLoBALs with Balmer absorption often show evidence of differential absorption (i.e., higher order Balmer lines showing higher column densities).
The Balmer absorption is interpreted to be associated with radiation driven outflows.
For loitering FeLoBALs, which are those with low accretion rate ($\lambda _{\rm Edd}\lesssim 0.3$) and low inferred outflow velocities ($V_{\rm outflow}<2,000$ \kms), the inferred outflow launching radius is on subpc scales and thus can be close to or even within the BLRs.
Notably, \target would fall into the category of loitering FeLoBALs based on its accretion rate and absorber velocities.
However, a key difference is the apparently inflowing gas traced by the absorber in \target rather than the outflowing gas typical for FeLoBALs.
The overall large ($>1$ pc) outflow launching radii of FeLoBALs would indicate no absorption variability on short timescales for most of these sources, although no systematic studies on absorption variability have been carried out for Balmer-absorbed FeLoBALs.
Regarding the broader class of BAL QSOs, several studies found evidence for short-to-long ($10^{-2}-10^1$ yr) term variability in UV absorption lines for a sample of BALs \citep[e.g.,][]{Barlow_1993,Lundgren_2007,Gibson_2010,Capellupo_2011,Capellupo_2012,Capellupo_2013,Grier_2015,Wheatley_bal_civ_variability}.
Since the dominant gas obscuration tracers in these studies are high-ionization lines including \civ\ and Si\,{\sc iv}, the authors interpret the change as due either to gas motions or to a change in the ionization state of the absorber.
In the passing cloud scenario, the inferred location of the clouds are consistent with regions inside BLRs \citep{Capellupo_2013}.
In our case of \target, due to the distinct kinematics of the varying absorption that traces a change in the neutral gas obscuration, and the similarity between the continuum fluxes at the two epochs, it is more likely that the variation is driven by passing clouds rather than a change of ionization.
The X-ray weakness is another key characteristic shared by \target and FeLoBALs.
Despite obvious LOS gas obscuration,
according to hard-band X-ray observations of some BAL QSOs, intrinsic X-ray weakness rather than photoelectric absorption better explains their observations \citep[e.g.,][]{Leighly_2007,luo_2013,luo_2014}.
In the case of \target, as we discussed in Section~\ref{xray}, it is more consistent with LOS gas obscuration rather than intrinsic weakness, although deeper X-ray follow up is needed to precisely determine the obscuring gas column density.

%\redtxt{discuss Weimin's works}
As we have mentioned, variability of Balmer absorption in individual QSOs has been reported by \citet{Shixiheng_balmerqsovar_2017} and \citet{Sunluming_balmerqsovar_2017}.
Different from \target, these QSOs again show blueshifted absorption possibly tracing outflowing gas, and the authors interpret the variability as due to either a change in the ionization state of the gas or due to transverse motions of the gas.
Notably, \citet{Shixiheng_balqso_2016} and \citet{Zhouhongyan_naturebalqso_2019} studied a sample of QSOs with redshifted Balmer absorption, where they interpret the absorption as tracing gas inflows potentially feeding the BHs, although no time-domain observations of the absorption lines are present in these works.

More recently, \citet{Park_desilrd_2026} presented several Balmer absorbed AGN selected from DESI DR1, which also include \target.
These sources generally show narrower Balmer absorption compared to BALs.
One of the particularly interesting sources is an AGN at $z=0.219$, named ``BAQ1'' by \citet{Park_desilrd_2026}.
The Balmer absorption is not only strong in BAQ1, but also shows evidence of time variability.
By analyzing the archival SDSS spectrum (2005) and DESI DR1 spectrum (2021) of BAQ1, \citet{Shangguan_baq_2026} noticed weak but significant variability around the absorption trough of \ha in BAQ1.
The fact that the varying fluxes are centered around the absorption through indicates that the variability is primarily driven by a change in the gas column density similar to \target.
The only difference is that in the case of BAQ1, the variability is much more subtle, which might indicate either a lower covering factor of the varying absorption and/or a much smaller column density change.

Another case of varying Balmer absorption, as we mentioned, is the nearby Seyfert 1 NGC 4151 during 1997\,-\,2000. As shown by \citet{hutchings2002}, blueshifted Balmer, \hei, and \feii\ absorptions are visible and varying over years.
The velocity offset of \hei\ is clearly smaller than the Balmer absorption, which would again imply some stratified outflow structures.
Also noted by \citet{hutchings2002} is the fact that the outflow velocity observed in NGC 4151 is correlated with the brightness of the continuum emission, which could be an indication of the radiation driven origin.
In the case of \target, DESI's fiber-based spectroscopy shows little continuum luminosity difference between the 2021 and 2022 spectra, although it is possible that there is a time lag between the absorption and the continuum. Indeed, as shown in Figure~\ref{fig:ztf}, the broad-band photometric monitoring by the Zwicky Transient Facility \citep[ZTF,][]{ztf} reveals a continuum dimming and subsequent brightening between 2021 and 2022 for \target.
On 02-2022, when the last DESI epoch was observed, the continuum roughly recovered to the 2021 level.
If the absorber is tracing inflows, the overall continuum dimming between 2021 and 2022 might imply inefficient radiative feedback that allows additional inflows.

Finally, we compare \target with the mysterious broad-line population dubbed LRDs.
LRDs were discovered by the \textit{James Webb Space Telescope} \citep[\jwst,][]{jwst0,jwst1} in recent years, and they were found to be the dominant population of broad-line emitters at $z>2$ compared to classical QSOs \citep[e.g.,][]{Maiolino_jadesagn_2024,Kocevski_lrd_2024,akins_lrd_2024,ma_lrd2d26_2025,juodzbalis_jadesagn_2025}.
As revealed by the NIRSpec spectrograph \citep{jakobsen2022}, LRDs also show Balmer absorption with optical depth ratios deviating from theoretical values, where stellar atmosphere-like physics is proposed to explain the deviation \citep{deugenio_irony}.
The typically inferred column densities of LOS gas obscuration in LRDs are much higher, reaching $\log N_{\rm H}\gtrsim 10^{24}~{\rm cm^{-2}}$ to explain the strong Balmer breaks and optical continuum features \citep{Inayoshi_maiolino_2025,ji_lrdbreak_2025,degraaff_lrd_2025,naidu_lrd_2025} as well as large Balmer decrements \citep{ji_lord_2025,yanzu_2026}.
Within the LRD population, both blueshifted and redshifted absorptions are seen \citep{Matthee_2026}, and even individual LRDs can show both blueshifted and redshifted absorptions \citep{deugenio_qso1_2025,linxiaojing_locallrd_2025,ji_lord_2025}, suggesting strong differential obscuration or optical depth effects on small physical scales potentially similar to the case of \target.
It has been suspected that the absorption in LRDs occurs on scales of $\sim 10^{-3}$ pc \citep[e.g.,][]{linxiaojing_locallrd_2025}.
Until now, very few LRDs have been reported to exhibit significant variability \citep[see e.g.,][for a recent investigation]{Liu_twinklelrd_2026}, although \citet{lin_lrdvar_2026} argue that the apparent lack of variability might be due to the limitation of current observations. 
Sources that have been reported to show significant variability include two high-redshift lensed LRDs with variations around their Balmer emission over rest-frame timescales of years \citep{ji_lrdbreak_2025,Zhang_lrdcross}, and one high-redshift LRD with tentative variation in its \ha\ absorption \citep{deugenio_lrdoutflow_2025}.
{With \target, we can estimate whether the same level of absorption variability seen in this QSO can be detected in observations of LRDs at high redshift. For example, assuming \target is observed at a typical redshift of $z=5$ as LRDs for two epochs and with the spectral resolution matching G395M grating of \jwst/NIRSpec, the variability of \target in \ha absorption would become insignificant ($<3\sigma$) if the ratio between the flux density of \ha emission line peak and the root mean square (RMS) noise is $<12$.
The variability of Balmer-absorbed QSOs can thus provide insights for future monitoring of LRDs' variability.
We note that this estimation is only based on a single target, and we will test the detectability of variability over a larger QSO sample in future work.
}

%Although time-domain constraints on the absorber variability of LRDs are still very limited, the current observational evidence suggests that LRDs tend to have stationary absorbers, 
If the absorbers in LRDs are truly much more stable compared to Balmer-absorbed QSOs, this might be related to the lack of strong outflows in LRDs given their relatively low intrinsic luminosities and metallicities \citep{trefoloni_feii_2024}.
Alternatively, physics similar to stellar photospheres and supernova shells might modulate the variability behavior \citep{Zhang_lrdcross,jockel_smssne_2026,Naidu_matthee_2026}.
With the recent discoveries of low-redshift LRDs at $z<1$, more time-domain investigations become feasible \citep[e.g.,][]{ji_lord_2025,line_lrd2_2026,lin_lrdvar_2026}, and future follow ups might help to verify the intrinsic variability of absorption lines in LRDs, or the lack thereof.
The X-ray weakness is another key property of LRDs, and recent models propose supper-Eddington accretion as a plausible explanation \citep{madau_superedd_2024,Inayoshi_agnvar_2024,lambrides_superedd_2024}, although gas obscuration could contribute to the X-ray weakness as well \citep{Maiolino2024_Xrays}.
Meanwhile, the usually symmetric broad emission lines with very extended wings observed in LRDs indicate either electron scattering dominated line transfer \citep{Rusakov_escattering_2025} or stratified Keplerian motions \citep{Scholtz_2026,Madau_scholtz_2026}.
The line broadening mechanisms could modulate the variability timescales.
For example. \citet{Sneppen_lrd_2026} argue that the line variability in the electron scattering dominated scenario cannot be faster than the diffusion timescale proportional to the scattering optical depth, which can be on the order of $\gtrsim 1$ yr.
The line profiles of LRDs are different from the case of \target, which might explain why while the absorbers in both systems are likely on small physical scales, their time-domain behaviors appear different.
We will present a more detailed comparison between spectroscopic features of LRDs and Balmer-absorbed QSOs in future work.

\subsection{Comparison with high-redshift AGN}

If \target truly traces a transition phase before the QSO blowout, it would be interesting to compare its properties with the general population of high-$z$ AGN discovered by \jwst over the past few years.
As shown by \citet{hainline_lrd_2024} and \citet{Geris_2026}, not all broad-line emitters at high $z$ are LRDs. The broad-line emitters that have blue colors throughout rest-frame UV and optical are now referred to as ``Little Blue Dots'' \citep[LBDs,][]{Brazzini_lbdlrd_2026,Geris_2026}.
The relation between LRDs and LBDs has remained debated, and scenarios involving evolution or viewing angle effects have been proposed \citep[e.g.,][]{Matthee_2026,madau_lrdlbd}.

Based on observations of broad Balmer lines, the inferred single-epoch virial BH masses of both LRDs and LBDs are generally $M_{\rm BH}\sim 10^{7-8}~M_\odot$, and \target would be at the high mass end compared to this mass range.
A key observational result for high-$z$ AGN discovered by \jwst is their overmassiveness compared to their hosts, where $M_{\rm BH}/M_{\star}$ is usually found to be a few per cent and sometimes even on the order of 1, significantly higher than the values in local AGN \citep[e.g.,][]{harikane2023,Maiolino_jadesagn_2024,juodzbalis_dormantagn_2024,juodzbalis_jadesagn_2025}.
In the case of \target, the host galaxy is clearly resolved with HSC imaging, and the implied single epoch BH-to-host mass ratio is roughly 2\,-\,5 per cent, making \target as overmassive as the high-$z$ population.
While many high-$z$ AGN do not show a resolved host in \jwst/NIRCam imaging, \citet{Rinaldi_2026} show by mocking high-$z$ observations with a $z=2$ source that the host surface brightness can simply be too low to be observable.

We note that the virial mass estimations for high-$z$ AGN have recently been questioned, and \citet{Rusakov_escattering_2025} suggest that the actual BH masses are $M_{\rm BH}\lesssim 10^6~M_\odot$, although \citet{Scholtz_2026} and \citet{Madau_scholtz_2026} argue that the virial mass estimations are still robust.
A recent kinematic mass measurement made by \citet{juodzbalis_specast_2025} with spectroastrometry and kinematic modelling for an LRD at $z=7$ supports the virial mass estimation.
Measuring BH masses independent of the single-epoch virial method in more systems will be useful to verify the overmassiveness of these AGN.

In addition to the BH mass, \target also shows X-ray weakness similar to high-$z$ AGN.
Both LRDs and LBDs are extremely weak in X-rays, with non detection in deep \chandra fields implying $\kappa _{2-10}>100$ \citep{Maiolino2024_Xrays}. While \target is detected in X-rays, the large observed $\kappa _{2-10}\approx 250$ is comparable to or higher than the general limits of individual high-$z$ AGN discovered by \jwst.
Notably, such an X-ray similarity was also suggested by \citet{Leighly_2025} between FeLoBALs and high-$z$ AGN.

Finally, a notable difference between \target and many high-$z$ AGN is the broad-band variability.
As shown in Figure~\ref{fig:ztf}, the variability amplitude of \target in ZTF bands can reach 0.2\,-\,0.4 mag, whereas the multi-epoch NIRCam constraints place $<0.1$ mag variability for high-$z$ AGN \citep{kokubo_harikane2024}.
However, by using a sample of local analogs and mocked observations, \citet{lin_lrdvar_2026} argue that the variability of LRDs might be underestimated with the current cadence of \jwst observations.
It would thus be interesting to compile more epochs of high-$z$ observations and compare the light curves of high-$z$ AGN with local sources such as \target.

\section{Conclusions}
\label{sec:conclude}

In this manuscript, we present dedicated analyses of a QSO, \target, at $z=0.287$, which shows time variability in its absorption lines including those from hydrogen Balmer transitions between 2021 and 2022.
The QSO has a dust-corrected bolometric luminosity of $L_{\rm bol}=10^{45.7}$ \ergs, and is characterized by a single epoch BH mass of $M_{\rm BH}\sim 10^{8.2-8.7}~{\rm M_\odot}$ and an Eddington ratio of $\lambda _{\rm Edd}\sim 0.14-0.38$.
The unusual spectral variability provides a probe into the gas structures on BLR scales in this QSO.
We summarize our conclusions as follows.
\begin{itemize}
    \item The optical spectrum of \target is characterized by two absorbers, one stationary absorber that remains unchanged over 10 months in the observed frame, and one dynamic absorber that appears during the same time interval.
    Both absorbers show redshifted absorptions indicative of inward gas motions along the LOS, with the stationary absorber having a velocity offset of $dv\approx 500$ \kms from the systemic redshift, and the main component of the dynamic absorber having $dv\approx 1,000$ \kms.
    A further absorption variation occurred between the two exposures taken in 2022 separated by 2 months, which has consistent kinematics with the main dynamic absorber with $dv\approx 1,000$ \kms.
    \item By performing spectral fitting with partial covering models, we found that column densities of excited neutral hydrogen of $N_{\rm H(n=2)}\sim 10^{14-15}~[{\rm cm^{-2}}]$ are needed to produce the Balmer absorption in both absorbers.
    Meanwhile, the stationary absorber has a high LOS covering factor of $C_f\approx 1$ and the main component of the dynamic absorber has $C_f\approx 0.2$.
    The dynamic absorber might have an additional blueshifted component at $dv\approx -300$ \kms, with $C_f\approx 0.08$ but strongly saturated.
    The blueshifted absorption is only visible in the $n>3$ Balmer lines, which might indicate that the absorber is co-spatial with BLR clouds, such that lines produced in the inner regions of the BLR are more obscured.
    We further examined the possibility of differential obscuration where the continuum and lines are obscured differently.
    The differential obscuration is statistically not preferred for the stationary absorber, but might occur in the dynamic absorber and suggest that the line covering factor is lower than the continuum covering factor.
    \item Using \cloudy photoionization models, we found that a typical total hydrogen column density of $N_{\rm H}\sim 10^{22}~{\rm cm^{-2}}$ and a density of $n_{\rm H}\gtrsim 10^{8}~{\rm cm^{-3}}$ are required to reproduce the Balmer absorption in both absorbers.
    This implies effective physical scales of $\lesssim 10^{-4}$ pc for the absorbers, smaller than the crossing length scale of $\sim 10^{-3}$ pc within the time interval.
    The non-varying nature of the stationary absorber might imply a lower filling factor compared to the dynamic absorber.
    The physical scales of the absorbers are compatible with moving clouds or gas inflow within the BLR, where the dynamic absorber traces the neutral layer at a possibly larger radial distance compared to the stationary absorber.
    This explains why the dynamic absorber does not have strong \hei$\lambda 3889$ absorption due to its lower ionization and/or shielding by inner gas clouds.
    In addition, we inferred a radial distance of $r_{\rm abs}\approx 0.07$ pc from the accretion disc for the stationary absorber, consistent with the characteristic size of the BLR inferred from the single-epoch method.
    If the absorbers trace gas inflows that feed the accreting BH, the implied $4\pi$-averaged covering factor is $C_{f,{\rm ave}}\approx 0.15$, close to the LOS covering factor of the dynamic absorber.
    This implies that the absorption spectrum observed in \target might not be simply due to a viewing angle effect, but tracing an actual feeding process during an active accretion phase.
    The additional absorption variation occurred in 2022 implies a change of 10\% in the $N_{\rm H}$ of the main dynamic absorber over 2 months.
    This indicates that the dynamic absorber likely traces a relatively continuous inflow that gradually increased the gas column density over several months.
    Other physical scenarios such as occultation by inhomogeneous disc structures are also possible, and further monitoring is needed to constrain different models.
    \item The \cloudy photoionization models we computed further predict the presence of a small Balmer break caused by the dynamic absorber.
    There exists a flux density break in the varying spectrum matching \cloudy models, although more epochs of follow-up are needed to verify its identity.
    If confirmed, this is the first direct evidence that gas obscuration within a BLR origin can lead to the formation of a Balmer break in QSOs.
    \item With 10 ks integration by the \textit{Einstein Probe} mission, we found an X-ray flux of $F_{\rm 2-10~keV}=1.0^{+0.3}_{-0.4}\times 10^{-13}$ \ergscm and a luminosity of $L_{\rm 2-10~keV}=2.0^{+0.6}_{-0.7}\times 10^{43}$ \ergs. The X-ray SED is consistent with strong LOS absorption with a 67\% confidence interval of $\log N_{\rm H}=23.6^{+0.2}_{-0.2}~[{\rm cm^{-2}}]$. The column density of gas probed by X-rays is nearly $10\times$ of that probed by optical absorption lines. 
    We considered three possible scenarios, including 1) an overall enhanced gas obscuration at the current epoch, corresponding to an average net gas inflow rate of $\dot{N_{\rm H}}\sim 10^{23}~{\rm cm^{-2}\,yr^{-1}}$; 
    2) a warm absorber inside the BLR that provides extra obscuration for the hot corona;
    3) galactic-scale gas obscuration driven by a recent gas-rich merger, which is dominated by low-density gas and thus is not traced by Balmer absorption but seen in the X-rays.
    Scenario 1) is disfavored by the non-detection from \textit{eROSITA} during 2019\,-\,2020.
    Scenario 2) is supported by the consistent column density derived from the blueshifted absorption as part of the dynamic absorber.
    Scenario 3) is also plausible given the tidal feature seen in the HSC-SSP image of \target indicative of a post merger stage.
    Future X-ray and UV-optical spectroscopic follow ups will help disentangle the above scenarios.
    \item The host galaxy of \target is spatially resolved in the HSC-SSP imaging.
    With image decomposition, we found a stellar mass of $\log(M_\star/M_\odot)\sim 10$.
    The light distribution of the host galaxy is clearly asymmetric.
    We further identified extended, diffuse emission and clump structures likely related to tidal features and star-forming clumps.
    These structures suggest a post gas-rich merger scenario with induced star formation and possibly induced gas inflows that trigger the AGN.
    This is consistent with the gas inflow interpretation for the Balmer absorbers in \target.
    \item {Compared to the high-$z$ AGN population discovered by \jwst,
    \target shows a similar ratio of $M_{\rm BH}/M_\star$ and X-ray weakness. While Balmer absorption is frequently seen in high-$z$ sources, no significant variability has been confirmed, which, however, might be limited by the S/N and the spectral resolution of current observations.
    It would be interesting to further compare the multi-epoch properties of high-$z$ AGN and local Balmer-absorbed QSOs in future observations.
    }
\end{itemize}

As a concluding remark, this case study of the varying absorption spectrum of the QSO \target offers a unique LOS view of the motions of gas on BLR scales.
Further time-domain follow ups as well as statistical studies of a sample of such QSOs will help to verify whether \target truly represents a special evolutionary phase of QSOs, where gas inflows driven by gas-rich galaxy mergers feed the central BHs.

\section*{Acknowledgements}
%We thank the anonymous referee, whose thoughtful suggestions improved the clarity of this work.
The authors thank Weimin Yuan and Jingwei Hu for comments on a draft of this manuscript.
This work is based on data obtained with the \textit{Einstein Probe}, a space mission of the Chinese Academy of Sciences, in collaboration with the European Space Agency, the Max-Planck-Institute for extraterrestrial Physics (Germany), and the Centre National d’tudes Spatiales (France).
This research used data obtained with the Dark Energy Spectroscopic Instrument (DESI). DESI construction and operations is managed by the Lawrence Berkeley National Laboratory. This material is based upon work supported by the U.S. Department of Energy, Office of Science, Office of High-Energy Physics, under Contract No. DE–AC02–05CH11231, and by the National Energy Research Scientific Computing Center, a DOE Office of Science User Facility under the same contract. Additional support for DESI was provided by the U.S. National Science Foundation (NSF), Division of Astronomical Sciences under Contract No. AST-0950945 to the NSF’s National Optical-Infrared Astronomy Research Laboratory; the Science and Technology Facilities Council of the United Kingdom; the Gordon and Betty Moore Foundation; the Heising-Simons Foundation; the French Alternative Energies and Atomic Energy Commission (CEA); the National Council of Humanities, Science and Technology of Mexico (CONAHCYT); the Ministry of Science and Innovation of Spain (MICINN), and by the DESI Member Institutions: \url{www.desi.lbl.gov/collaborating-institutions}. The DESI collaboration is honored to be permitted to conduct scientific research on I’oligam Du’ag (Kitt Peak), a mountain with particular significance to the Tohono O’odham Nation. Any opinions, findings, and conclusions or recommendations expressed in this material are those of the author(s) and do not necessarily reflect the views of the U.S. National Science Foundation, the U.S. Department of Energy, or any of the listed funding agencies.

XJ, FDE, IJ, and RM acknowledge ERC Advanced Grant 695671 ``QUENCH'' and support by the Science and Technology Facilities Council (STFC) and by the UKRI Frontier Research grant RISEandFALL.
DJW also acknowledges support from STFC (grant code ST/Y001060/1).
RM acknowledges funding from a research professorship from the Royal Society.

%\redtxt{[add your acknowledgements here]}

\section*{Data Availability}

The DESI data used in this work are publicly available at \url{https://data.desi.lbl.gov/doc/releases/dr1/}.
Archival photometric data can be queried from \url{https://vizier.cds.unistra.fr/vizier/}.
Analysis results of the optical and X-ray observations will be shared on reasonable request to the corresponding author.

% WARNING
%-------------------------------------------------------------------
% Please note that we have included the references to the file aa.dem in
% order to compile it, but we ask you to:
%
% - use BibTeX with the regular commands:
   \bibliographystyle{mnras} % style aa.bst
   \bibliography{ref} % your references ref.bib
%
% - join the .bib files when you upload your source files
%-------------------------------------------------------------------

\appendix

\section{SED fitting}
\label{appendix:sed}

\begin{figure*}
    \centering
    \includegraphics[width=\columnwidth]{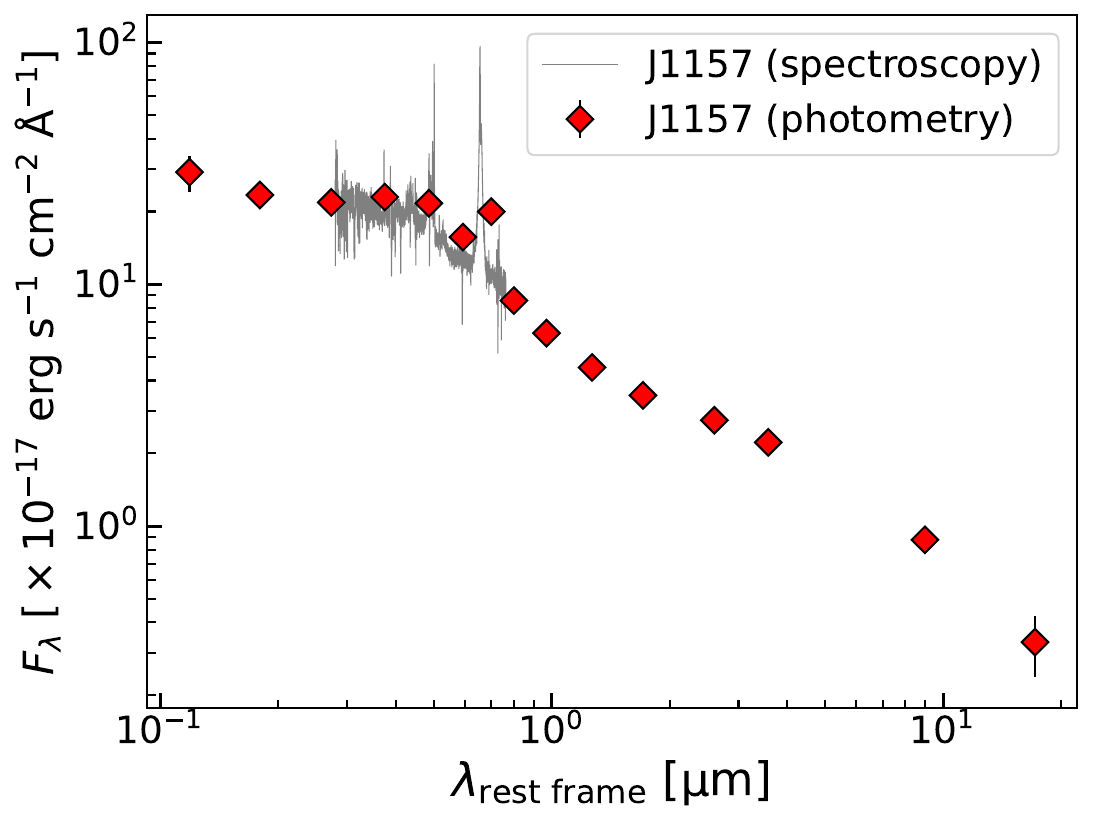}
    \includegraphics[width=\columnwidth]{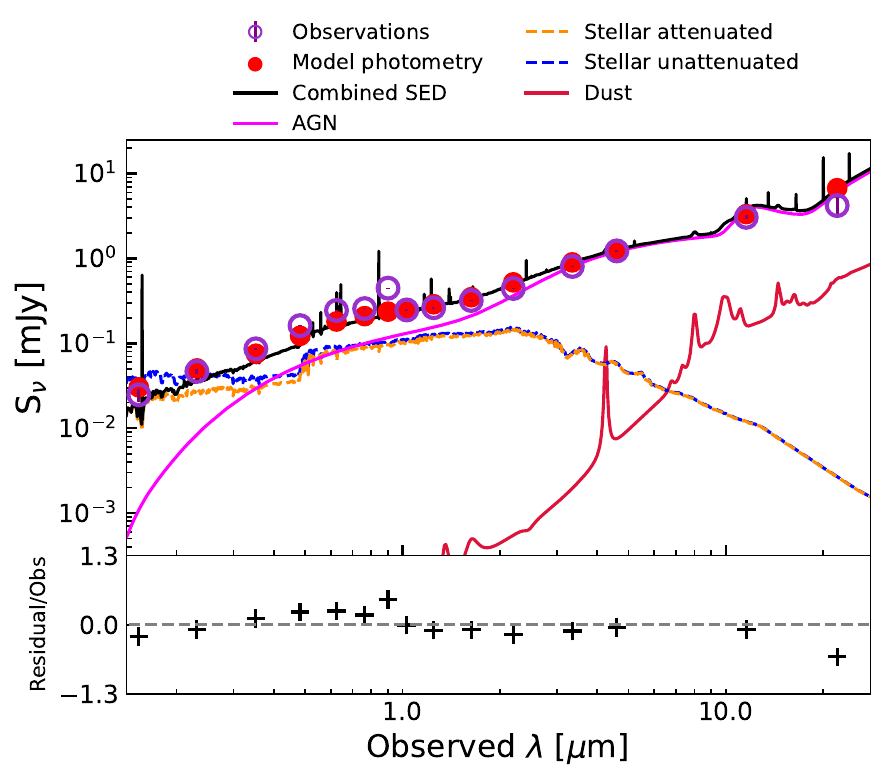}
    
    \caption{Archival SED from $GALEX~FUV$ to $WISE~W4$ for \target, where a clear spectral slope flattening in the UV compared to the optical is visible in the left panel.
    The right panel shows the SED fitting results from \textsc{cigale}, which yield a moderate attenuation of $A_{\rm V}\approx 1$ mag similar to the nebular attenuation inferred from the broad-line Balmer decrement (Table~\ref{tab:target_properties}).
    The SED fit attributes the UV continuum to stellar populations. However, given the presence of strong \feii, \hei, and Mg\,{\sc ii} UV absorptions, where Mg\,{\sc ii} absorption shows clear variability (Figure~\ref{fig:uv_spec}), the SED fitting result needs to be taken with caution and spectroscopic follow up in the UV is needed.
    }
    \label{fig:sed_vizier}
\end{figure*}

In this appendix, we describe the SED fitting of the broad-band photometry for \target.
The photometry we used was queried from the web-based tool \textsc{VizieR} \citep{vizier}.
We queried FUV and NUV data from \textit{GALEX} \citep{galex}, $u',g',r',i',z'$-band data from SDSS \citep{york2000}, 
Y, J, H, K-band data from UKIDSS \citep{ukidss}, and W1, W2, W3, W4-band data from \textit{WISE} \citep{wise}.
The full FUV to NIR SED is shown in the left panel of Figure~\ref{fig:sed_vizier} together with the DESI spectrum, which shows a clear flattening of the continuum from optical to UV.

We used the code \cigale \citep{cigale} to perform SED fitting in order to treat the IR dust emission with energy conservation.
We modeled the AGN component with the SKIRTOR model \citep{skirtor} and assumed an SMC extinction curve \citep{gordon2003}.
The stellar populations are modeled with \citet{bc03} assuming a \citet{Salpeter1955} initial mass function (IMF) and a delayed-$\tau$ star formation history.
The stellar dust emission adopts the \citet{Dale_2014} model with a variable slope of $\alpha=1.3-4$.
The \citet{calzetti2000} attenuation curve is used to describe reddening of the stellar populations, and the ratio between the stellar-  and nebular- dust attenuation is set to $\rm E(B-V)_\star /E(B-V)_{\rm gas}=0.44$ following \citet{calzetti2000}.

The right panel of Figure~\ref{fig:sed_vizier} shows the best-fit model, where the ``AGN'' component includes both the disc and the dusty torus emission, and ``dust'' refers to emission by dust absorbing stellar light.
The model predicts that the AGN light dominates the IR with $A_{\rm V}=0.93\pm 0.02$ mag and a bolometric luminosity of $\log L_{\rm bol}=45.06\pm 0.02$ [\ergs].
The stellar light dominates the UV and has $M_\star \approx 10^{10}~M_\odot$.
As we mentioned in Section~\ref{sec:data}, the caveat of this interpretation is that the UV has \feii\ and \mgii absorptions that are varying with time, which cannot have a stellar origin.
Meanwhile, if the UV spectrum is stellar dominated, the variability should be diluted compared to the optical spectrum.
However, as shown in the varying spectrum, the UV exhibits significant variability, either from a continuum origin or by bound-free absorption of hydrogen.
More time-domain spectroscopic follow ups will help disentangle the AGN and stellar contributions in the UV.

\section{Posterior distributions of fitted parameters for the varying absorber}
\label{appendix:posteriors}

\begin{figure*}
    \centering
    \includegraphics[width=\linewidth]{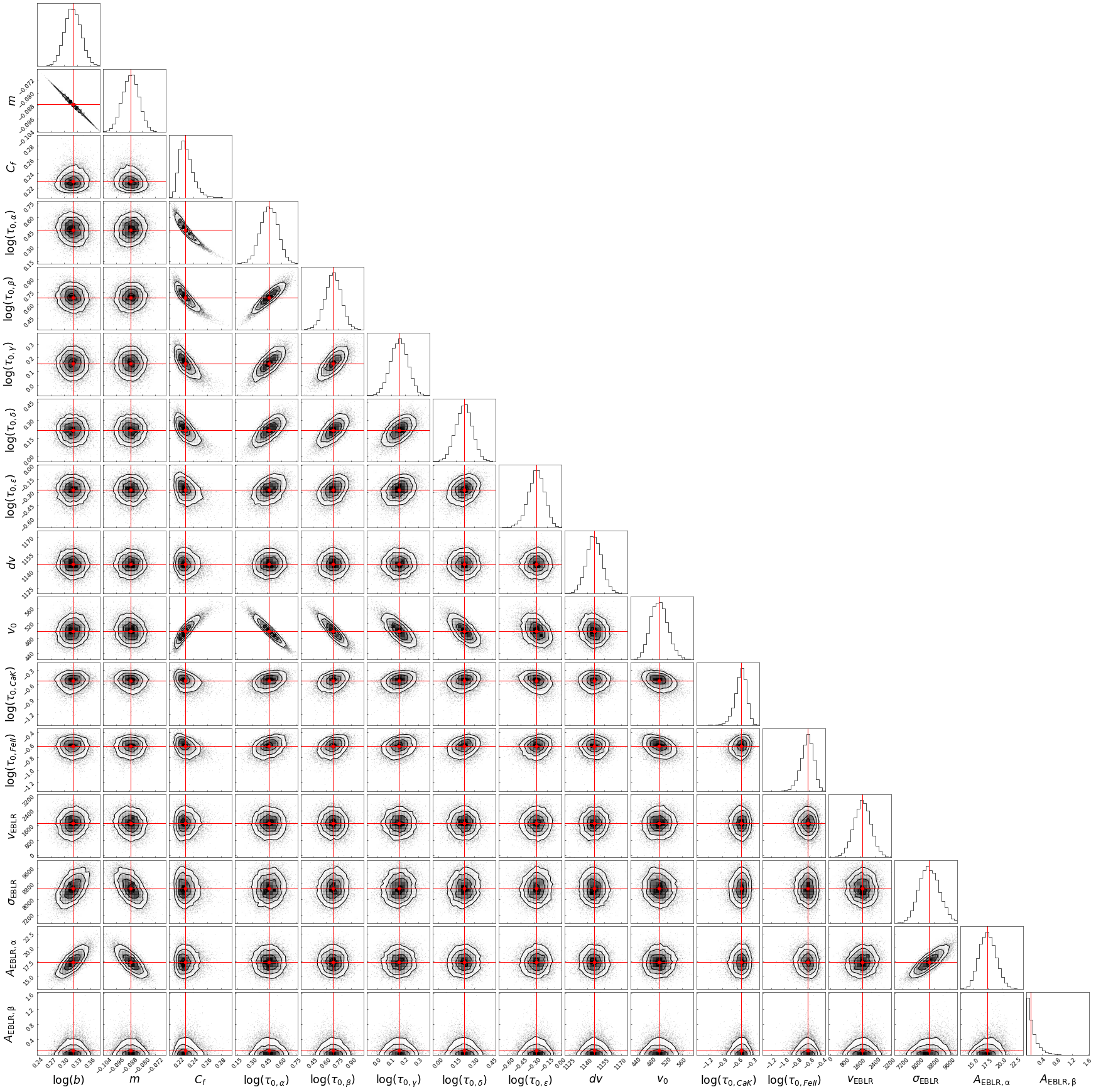}
    \caption{Posterior distributions of the parameters for the partial covering model shown in the left panel of Figure~\ref{fig:single_abs}. The red lines mark the median values, which we use as the best-fit values.
    }
    \label{fig:var_abs_corner}
\end{figure*}

\begin{figure*}
    \centering
    \includegraphics[width=\linewidth]{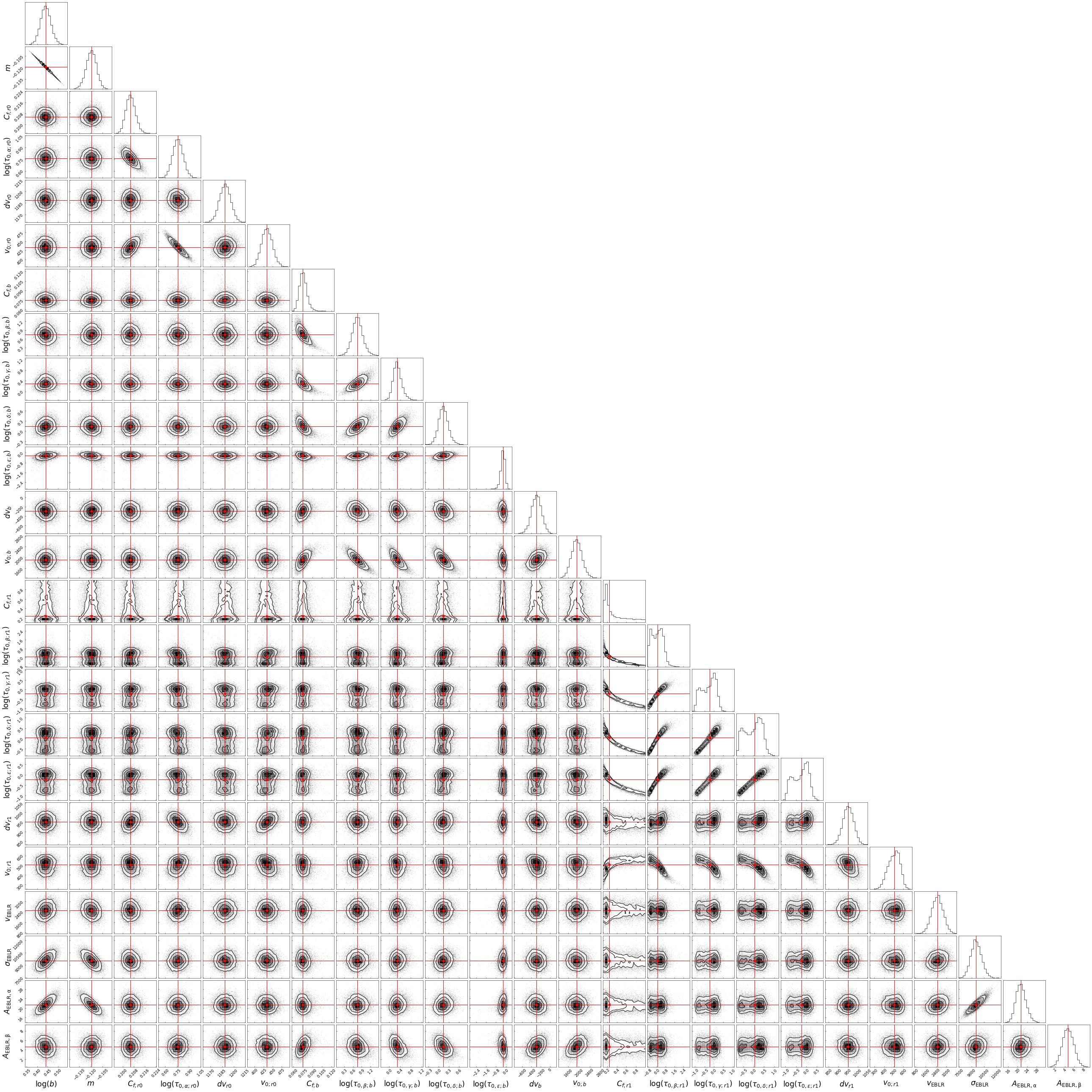}
    \caption{Posterior distributions of the parameters for the partial covering model shown in Figure~\ref{fig:vel_com}, where three absorbers with independent kinematics are considered. The red lines mark the median values, which we use as the best-fit values.
    }
    \label{fig:var_mulabs_corner}
\end{figure*}

In this appendix, we present the posterior distribution of the fitted parameters for the partial covering model that describes the varying absorber shown in Figures~\ref{fig:abs_var} and \ref{fig:single_abs}.
Figures~\ref{fig:var_abs_corner} and \ref{fig:var_mulabs_corner} show the posterior distributions for the single absorber model and the multi absorber model, respectively.

For the single absorber model, there are a total of 16 parameters: $m$ and $b$ describe the continuum as a power law of $F_\lambda = b\lambda ^m$; $C_f$ is the covering factor; parameters of $\tau _{0,i}$ describe line-center optical depths of different lines, including \ha, \hb, \hg, \hd, \heps, \caii\ K, and \feii$\lambda 5169$; $dv$ describes the velocity offset of the absorber in \kms; $v_0$ is the velocity width of the absorber in \kms; $v_{\rm EBLR}$ and $\sigma _{\rm EBLR}$ describes the velocity and velocity dispersion in \kms for a potential extremely broad component in \ha and \hb; $A_{\rm EBLR, \alpha}$ and $A_{\rm EBLR, \beta}$ are the amplitudes in $10^{-15}$ \ergscm\ for the extremely broad components in \ha and \hb, respectively.

The multi absorber model further separates the absorption into three components, including an absorber traced by \ha, which produces higher order absorptions based on theoretical oscillator strengths, and two other absorbers only present in Balmer lines with upper levels of $n>3$ with free optical depths.
The best-fit model has two redshifted absorbers (one with \ha and one without \ha) and one blueshifted absorber (without \ha). 
The two redshifted absorbers fit the main absorption trough in the varying spectrum.
One of the redshifted absorber is less well constrained by the data as shown by the extended tails in the distribution of $C_{f,r1}$ and optical depths.

\section{Continuum-only absorption}
\label{appendix:contabs}

\begin{figure}
    \centering
    \includegraphics[width=\columnwidth]{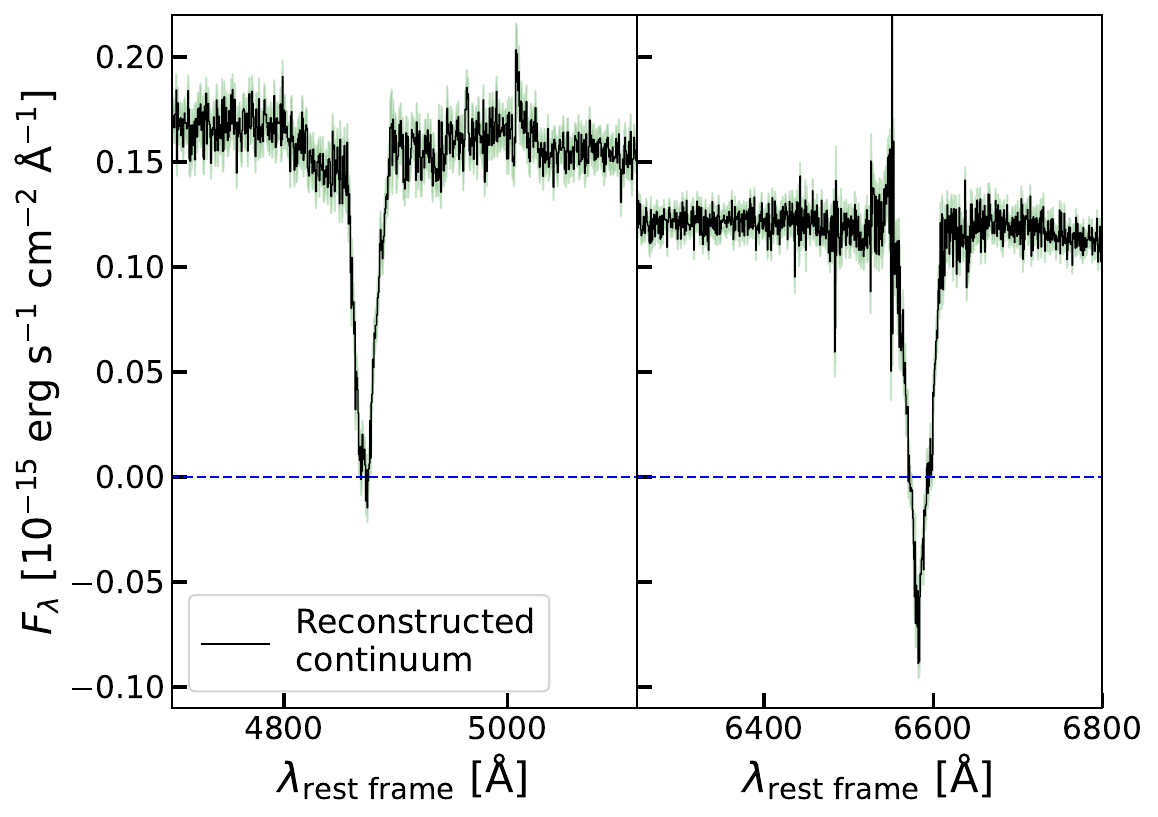}
    \caption{Reconstructed continuum for the 2022 epoch assuming that only the continuum is absorbed in \target.
    The continuum is made by adding the difference spectrum ($2022-2021$) to the best-fit continuum model for the 2021 epoch assuming continuum-only absorption.
    The shaded region represents $1\sigma$ measurement uncertainties.
    The left and right panels show spectral regions around \hb and \ha, respectively.
    The reconstructed continuum shows clear absorption that goes negative, which is unphysical.
    This implies that both broad lines and the continuum should be absorbed.
    }
    \label{fig:reconstruct_cont1}
\end{figure}

In this appendix, we present the reconstructed continuum for \target at the 2022 epoch if only the continuum is absorbed.
In Section~\ref{sec:vary_abs}, we present two model fits where broad lines and the continuum are absorbed together or differently.
The continuum-only absorption model for the 2021 epoch is shown in panel (b) of Figure~\ref{fig:stationary_fit}, from which we can reconstruct the continuum model for 2022.
Taking the best-fit continuum model at 2021, $F_{\rm \lambda,cont.model;2021}$, and taking the observed flux difference between 2022 and 2021, $\Delta F_{\rm \lambda}=F_{\rm \lambda,cont.obs;2022}-F_{\rm \lambda,cont.obs;2021}$, the inferred continuum model for the 2022 epoch under the assumption of the continuum-only absorption is given by $F_{\rm \lambda,cont.model;2022}=F_{\rm \lambda,cont.model;2021}+\Delta F_{\rm \lambda}$.
The reconstructed continuum model around \hb and \ha is shown in Figure~\ref{fig:reconstruct_cont1}.
It is clear that the absorption goes negative near the core, which is unphysical for any absorption.
As a result, it is likely that both broad lines and continuum are absorbed.

\section{Density constraint from FeII absorption}
\label{appendix:feii_abs}

\begin{figure}
    \centering
    \includegraphics[width=\columnwidth]{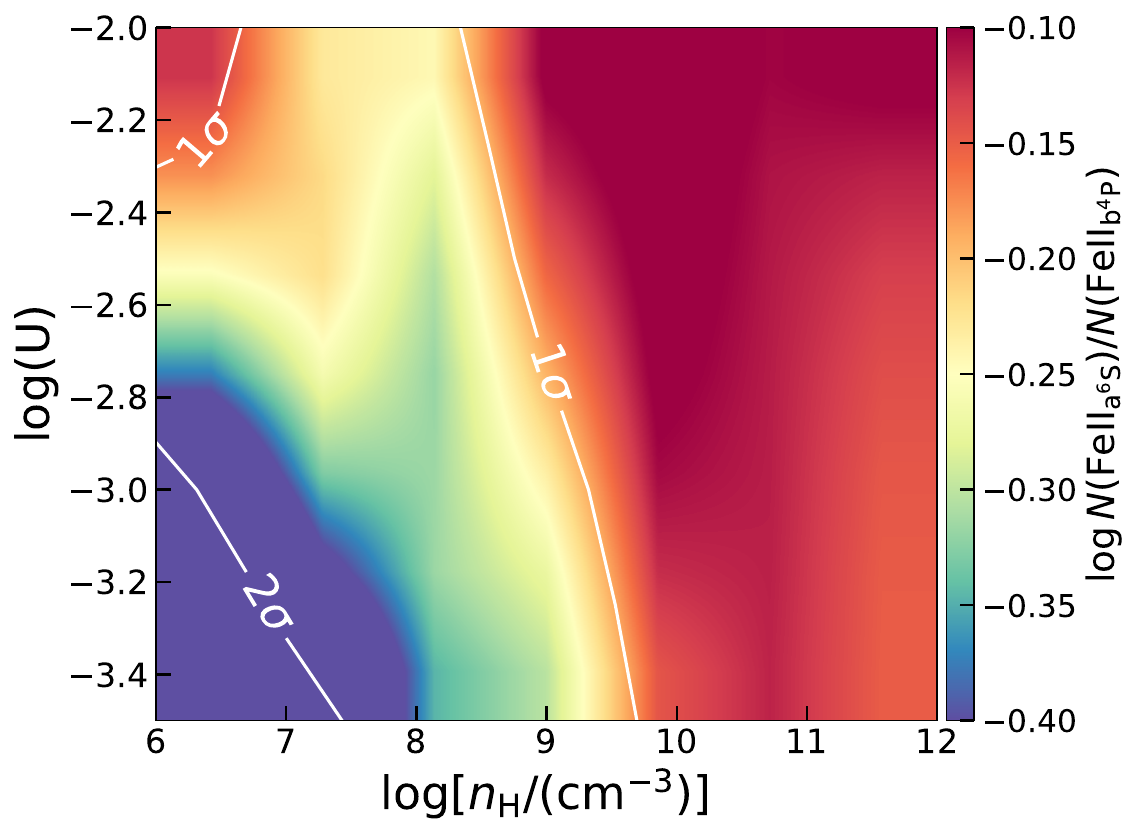}
    \caption{
    Constraints on $n_{\rm H}$ and $\log U$ of the dynamic absorber based on the column densities of different \feii\ energy levels.
    The background shows \cloudy models color coded according to the log of the \feii\ column density ratio.
    While contours show the lower limit we measured from the dynamic absorber at different significance.
    The $3\sigma$ limit is outside the range of the plot.
    Although these limits potentially point towards high gas densities, the constraints are relatively loose compared to \hei\ as shown in Section~\ref{sec:vary_abs}.
    }
    \label{fig:denlgu_feii}
\end{figure}

In this appendix, we present density constraint for the dynamic absorber based on the optical \feii$\lambda 5169$ and \feii$\lambda 4233$ absorption.
As we mentioned in Section~\ref{sec:vary_abs}, \feii$\lambda 4233$ is not detected in the varying absorption spectrum, which yields an upper limit on the column density of the $b^4P_{5/2}$ level.
To derive the plausible range of density based on these lines, we checked \cloudy models with the same parameter space listed in Table~\ref{tab:cloudy_models}, but limiting $\log U\leq -2$ and truncating the models at a column density of $N_{\rm H}=10^{22.5}~{\rm cm^{-2}}$.

Figure~\ref{fig:denlgu_feii} shows a comparison between our measurements and \cloudy models.
Even at $2\sigma$, the lower limit of the column density ratio between the $a\,^6S_{5/2}$ (traced by \feii$\lambda 5169$) and $b\,^4P_{5/2}$ (traced by \feii$\lambda 4233$) levels is not constraining compared to \hei shown in Section~\ref{sec:vary_abs}.
Overall, the constraints are consistent with high densities for the dynamic absorber.

%\IfFileExists{imaging.tex}{\input{imaging}}{}

% Don't change these lines
\bsp	% typesetting comment
\label{lastpage}
\end{document}

%% file: imaging.tex
\section{Host galaxy}\label{sec:host}

%\textcolor{red}{Section Under Construction}
As shown by the HSC-SSP image in Figure~\ref{fig:full_spec}, the host galaxy of \target is clearly resolved spatially, with substructures including blue, likely star-forming clumps and a tail possibly indicative of a recent galaxy merger.
Notably, strong obscuration of X-rays in AGN is frequently found in post galaxy mergers \citep{Hickox_2018rev}.
In this section, we performed an image decomposition to investigate the properties of the host galaxy.

To measure the morphology of \target, we model multi-band HSC-SSP \textit{grizy} imaging data 
%(Section~\ref{sec:data}) 
and match empirical PSF cutouts.
HSC-SSP is a wide-field multi-band imaging survey with the Subaru 8.2m telescope.
The image cutouts we used are from HSC-SSP DR3 Wide, which include data collected between March 2014 and January 2020 in \textit{grizy} bands with a limiting magnitude of $\sim 26$ mag and typical seeing of $\sim 0.\!\!''7$ \citep{Aihara_hscssp}.
The data are reduced with the pipeline \textsc{hscPipe v8}.
For details on the data reduction, we refer the readers to \citet{Aihara_hscssp}.
%{\color{red} [describe hsc data]}

\begin{figure}
  \centering
  {\phantomsubcaption\label{f.hsc.gz.a}
   \phantomsubcaption\label{f.hsc.gz.b}
   \phantomsubcaption\label{f.hsc.gz.c}
   \phantomsubcaption\label{f.hsc.gz.d}
   \phantomsubcaption\label{f.hsc.gz.e}
   \phantomsubcaption\label{f.hsc.gz.f}
   \phantomsubcaption\label{f.hsc.gz.g}
   \phantomsubcaption\label{f.hsc.gz.h}
   \phantomsubcaption\label{f.hsc.gz.i}
   \phantomsubcaption\label{f.hsc.gz.j}}
  \includegraphics[width=\columnwidth]{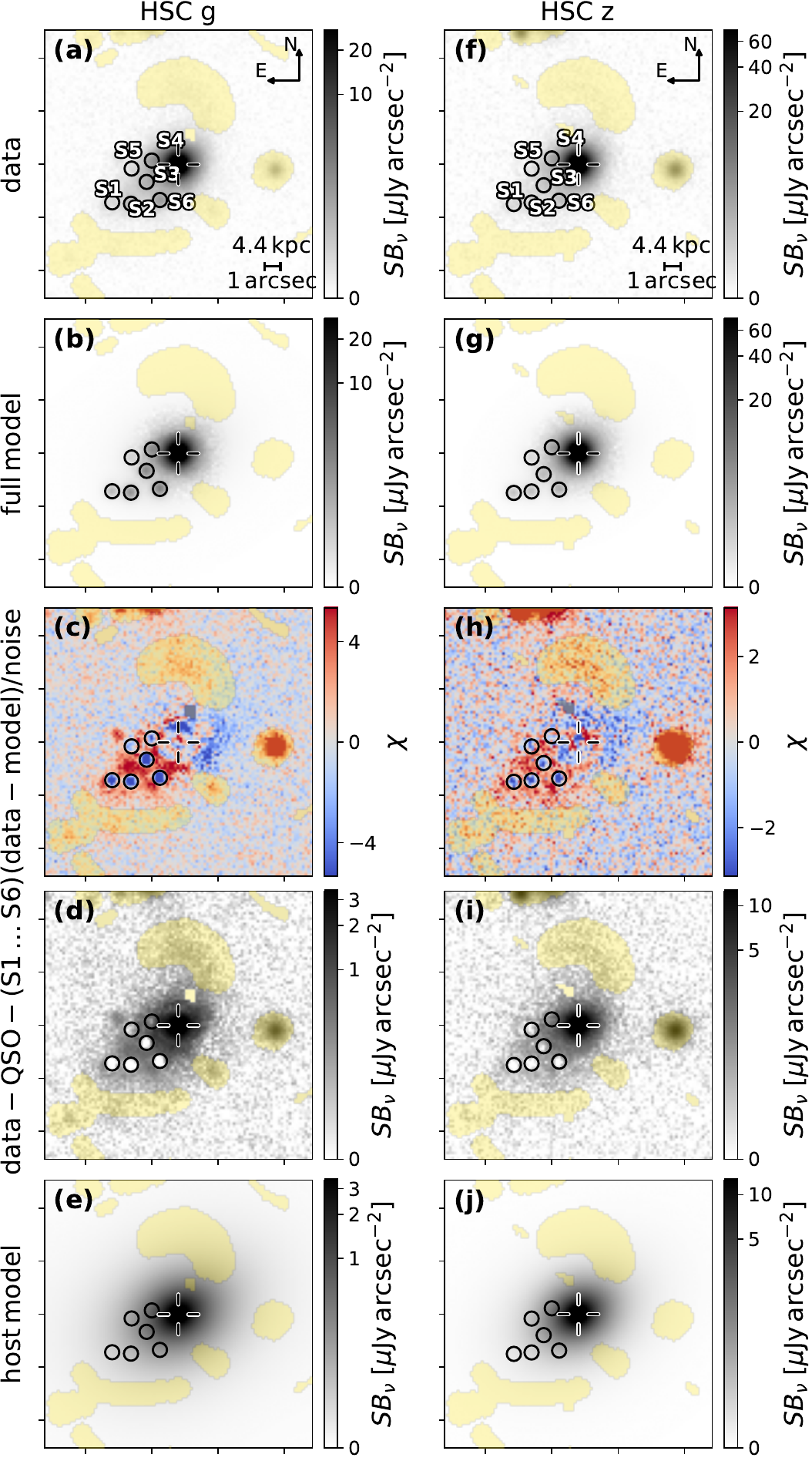}
  \caption{
  Light-profile modeling of HSC $g$ (left) and $z$ (right). From top to bottom,
  the rows display the data, the fiducial \textsc{pysersic} model (consisting of
  a QSO, host galaxy, and six additional point sources), the $\chi$ map, the
  empirical host view (subtracting all model point sources from the data), and
  the host model. Yellow shaded regions were masked and are not fit. Panels~\subref{f.hsc.gz.c} and~\subref{f.hsc.gz.h} show
  that the model overpredicts/underpredicts the data to the north west/south east;
  this is likely associated with irregular light distribution from a recent
  merger.
  }
  \label{f.hsc.gz}
\end{figure}

We used the Bayesian software \textsc{pysersic} \citep{pasha+miller2023}, optimized with standard methods as described in, for example, \citet{deugenio_lrdoutflow_2025}.
Each imaging band was fit separately, but we used the
marginalized posterior probabilities from some bands as probability priors on other bands, as described below.
Before the fit, we masked bad detector pixels, any detected source not part of the main segmentation around \target,
and manually selected foreground or interloper
regions, including the spatially extended emission to the north-west (yellow shaded
regions in Figure~\ref{f.hsc.gz}).
We model \target with a S\'ersic profile (representing the host galaxy) plus
a point source (representing the QSO). In addition, we used six more point
sources, indexed S1 to S6, representing as many sources located in the south-east
(particularly visible in the $g$- and $r$-band cutouts; see Figure~\ref{f.hsc.gz.a}
for the $g$-band imaging).
We determined the number of extra point sources iteratively, starting from a fit
without any extra point source, and adding S1--S6 sequentially at the location
of the largest residuals in $\left| \chi \right|$ from the previous fit, until
the $BIC$ stopped decreasing substantially ($\Delta{BIC}>-10$;
we do not allow extra point sources within $0.\!\!''5$ from the QSO).
We completed this selection procedure for $g-$ and $z-$band
independently; since these two bands do not agree on the number and location of
the necessary point sources, we adopted the union of the two solutions as the
fiducial setup.
We remark that while S1--S6 are modelled as point sources, we do not claim them
to be spatially unresolved; rather, a point-source parametrization is necessary to avoid the
optimizer using the additional degrees of freedom to model any of S1--S6 as
extended and elongated sources that are degenerate with the host galaxy.
A constant sky background completes the scene.

A further complication to our modeling is that the $r$- and $i$-band images carry saturation
flags at the location of the quasar; while the images are not visibly affected
by saturation, these flags may indicate that the detector entered the non-linear
regime, causing unreliable flux and/or point-source degradation, even without
reaching full well.
For the $i$-band in particular, the model parameters are very sensitive to whether
or not we mask the pixels flagged as saturated; in no case the fit parameters
look acceptable. 
For this reason, we use a fiducial tied-geometry model: the
QSO centre has a strong prior from the unsaturated bands, while the host
centre, S\'ersic index, and position-angle are tied to the best-determined
\(z\)-band solution. We then refit all five bands with the appropriate HSC
bad-pixel masks, and report this latter fit as the fiducial model.
We also tested the robustness of this method by refitting $g$-band using the
$r$-band saturated-pixel mask, finding that the fit results are not strongly
affected, provided we use the aforementioned priors.

For brevity, we show only the $g$- and $z$-band analysis
(Figure~\ref{f.hsc.gz}). Both bands display a structured `red-to-blue' residual
pattern in the $\chi$ maps; this is due to faint, spatially
extended and asymmetric emission, mirrored by the arch to the north of quasar.
To estimate the importance of this feature, we compare
the sum of the absolute value of the residuals to the fluxes of the QSO and of the host models, all
calculated inside a circle of radius $3''$. In this manner, we find that the extended
emission accounts for no more than 3.5 (13.5) per cent of the QSO (host galaxy) flux in $g$,
and 4.1 (11.5) in $z$. Here the upper limits are because we sum the absolute residuals, instead of a signed sum. Our choice is more conservative and avoids the possible bias of negative residuals indicating overestimated source flux, which, if ignored, would double-count the source strength.
Taking a seeing of roughly $1''$ from the DESI catalog and an on-sky fiber diameter of $1.\!\!''5$, the imaging model predicts that less than 10 per cent of the light within the DESI fiber is from the host galaxy, validating our assumption that the DESI spectra are dominated by the QSO.

While the scene is complex, and the fit is unstable, all models and bands
require two components to fit the data, which we interpret naturally as the
extended host galaxy and the point-source QSO. The QSO and host fluxes, and
the host physical properties are unstable across adjacent photometric filters
and different masking choices. However, the fits can be stabilized by using
physically motivated probability priors, which reduce the variations in, for example,
the galaxy half-light radius to within a factor of two.
The spatially extended emission with embedded clumps is unlikely to arise from
a chance alignment: first, due to the number of clumps (at least six) over a
few square arcseconds. Second, because the clumps occupy a region of diffuse
light that connects to the northern arc (Figure~\ref{f.hsc.gz.a}
and~\subref{f.hsc.gz.c}). Our interpretation is that both the diffuse emission
and clumps are merger debris and star-forming knots, pointing to a recent
gas-rich merger -- a natural candidate for triggering the AGN with induced gas inflows.

To estimate the stellar mass of the host, we use the $g-i$ color calibration
from \citet{taylor+2011}, which assumes a delayed-exponential star-formation
history and \citet{chabrier2003} IMF.
To calculate the $k$-correction we use the spectroscopic redshift and a simple
linear interpolation, plus the $2.5 \log (1 + z)$ term for filter stretch, but
without accounting for intervening passive evolution between the epoch of
observation and the median redshift from \citet{taylor+2011}.
With the adopted cosmology, we infer $\log (M_\star/\mathrm{M_\odot}) \simeq
10.0$. As an independent test, we use the mass--size relation of
star-forming galaxies from \citet{vanderwel+2014}. For the appropriate redshift
bin at $z=0.25$, this relation predicts $R_\mathrm{e} = 4.78$\,kpc, in agreement
with the fiducial $z$-band host radius; the $g$- and $i$-band measurements are
instead 1.6-$\sigma$ deviations.
The inferred stellar mass is equivalent to $\log (M_\star/\mathrm{M_\odot}) \simeq
10.2$ if assuming instead \citet{Salpeter1955} IMF, comparable to $\log (M_\star/\mathrm{M_\odot}) \simeq
10.1$ inferred from the \textsc{cigale} SED fitting in Appendix~\ref{appendix:sed}, although the latter uses the full FUV-MIR SED without imaging decomposition.

Overall, the emerging picture is that of a typical star-forming galaxy that
recently underwent a strong merger, triggering both star formation and
the AGN.